\newif\ifarxiv\arxivtrue

\ifarxiv
\documentclass{ar-1col_bland}
\usepackage{changepage}

\usepackage[total={30.5pc,47pc},centering]{geometry}
\else
\documentclass{ar-1col}
\fi

\usepackage{xspace}
\usepackage[comma]{natbib}
\usepackage{url}

\usepackage{biblio}
\usepackage{amssymb}
\usepackage{amsmath}
\usepackage{outlines,hyperref,bm}
\usepackage[normalem]{ulem}
\usepackage{subcaption}

\usepackage{url,color}

\newcommand{\photoz}{photo-$z$}

\jname{in press}
\jvol{AA}
\jyear{2022}

\begin{document}

\markboth{Newman \& Gruen}{Photometric Redshifts}

\title{Photometric Redshifts for Next-Generation Surveys}

\author{Jeffrey A. Newman$^1$ and Daniel Gruen$^{2}$
\affil{$^1$Department of Physics and Astronomy and PITT PACC, University of Pittsburgh, Pittsburgh, PA 15260, USA; email: janewman@pitt.edu}
\affil{$^2$Ludwig-Maximilians Universit\"at, Faculty of Physics, University Observatory, Scheinerstr.~1, 81679 Munich, Germany; email: daniel.gruen@lmu.de}}

\begin{abstract}
Photometric redshifts are essential in studies of both galaxy evolution and cosmology, as they enable analyses of objects too numerous or faint for spectroscopy.  The Rubin Observatory, Euclid, and Roman Space Telescope will soon provide a new generation of imaging surveys with unprecedented area coverage, wavelength range, and depth.  To take full advantage of these datasets, further progress in photometric redshift methods is needed.  In this review, we focus on the greatest common challenges and prospects for improvement in applications of photo-$z$'s to the next generation of surveys:
\ifarxiv
\begin{adjustwidth}{0pt}{0.222 \linewidth}
\fi
\begin{itemize}
\item{Gains in \emph{performance} -- i.e., the precision of redshift estimates for individual galaxies -- could greatly enhance studies of galaxy evolution and some probes of cosmology.}
\item{Improvements in \emph{characterization} -- i.e., the accurate recovery of redshift \emph{distributions} of galaxies in the presence of uncertainty on individual redshifts -- are urgently needed for cosmological measurements with next-generation surveys.}
\item{To achieve both of these goals, improvements in the scope and treatment of the samples of spectroscopic redshifts which make high-fidelity photo-$z$'s possible}{} will also be needed. 
\end{itemize}
\ifarxiv
\end{adjustwidth}

\fi

For the full potential of the next generation of surveys to be reached, the characterization of redshift distributions will need to improve by roughly an order of magnitude compared to the current state of the art, requiring progress on a wide variety of fronts. We conclude by presenting a speculative evaluation of how photometric redshift methods and the collection of the necessary spectroscopic samples may improve by the time near-future surveys are completed.
\end{abstract}


\maketitle

\tableofcontents



\newpage

\section{INTRODUCTION}

\label{sec:intro}

The advent of large-format CCD mosaic cameras makes it possible to obtain deep imaging of large fractions of the sky.  By measuring fluxes through
multiple filters, near-future projects\footnote{E.g., the Vera C. Rubin Observatory's Legacy Survey of Space and Time \citep{LSSTScienceBook}, the Nancy Grace Roman Space Telescope \citep{Spergel2015}, and the Euclid mission \citep{Euclid}}
will obtain high signal-to-noise but limited-spectral-resolution information on billions of objects. Distances based on redshift estimates will be essential for interpreting these observations. We can only feasibly obtain this redshift information from imaging data alone; we thus must determine {\it photometric redshifts}, also  referred to as ``photo-$z$'s''.  

\ifarxiv
\begin{marginnote}
\entry{Take-away}{Distances based on photometric redshifts enable the inference of many properties from imaging data, key for studies in both galaxy evolution and cosmology.}
\end{marginnote}
\fi

A broad range of extragalactic studies rely on photometric redshifts. Given redshift estimates, intrinsic physical properties of galaxies can be inferred from
their observed spectral energy distributions.
Photo-$z$'s thereby enable studies of how the demographics of galaxies have changed over time, 
constraining models of galaxy evolution. 
Photo-$z$'s are also frequently used to select objects of interest for follow-up spectroscopy (e.g., galaxies 
or quasars whose properties are consistent with extremely high redshifts), enabling large imaging samples to be mined for rare objects.  They also are  vital for identifying the host galaxies of transient sources.   

Photometric redshifts likewise are necessary for many methods of constraining cosmological models.  Probes of cosmology generally rely on determining the relationship between observable quantities and redshift.
As we analyze deeper and wider data sets for more and more stringent tests of cosmological models, requirements on photometric redshift methods steeply increase.

A number of recent works have described the large variety of photometric redshift methodologies available to the community, including evaluation of their performance under more or less idealized conditions \citep{Sanchez2014,Tanaka2018,Salvato2019,Schmidt2020,Desprez2020,Brescia2021}. In this review, we instead focus on common challenges that any approach to photometric redshifts must account for, rather than the methods themselves. We concentrate upon evaluating the needs for the new generation of deep, wide-field imaging surveys that will come online in the 2020s -- the ``Stage IV'' dark energy surveys in the classification of the Dark Energy Task Force \citep{DETF} -- in light of the results of the current-generation, ``Stage III'' surveys. 

These needs can be broadly divided into the twin goals of performance and characterization.  Throughout this review, we will use the term \textbf{performance} to refer to the ability to predict the redshift of an \textit{individual galaxy} precisely; i.e., with small uncertainty when compared to its true redshift.  \textbf{Characterization}, in contrast, will refer to the ability to constrain the properties of the \textit{redshift distribution of an ensemble of galaxies} -- e.g., the mean redshift or its higher moments -- as for many high-precision cosmology measurements it is that distribution which we need to know well. 

\begin{marginnote}
\entry{Performance}{The ability of a photometric redshift algorithm to deliver higher-precision redshift estimates for individual galaxies.} 
\end{marginnote}

In the remainder of \S\ref{sec:intro} we will describe how the performance and characterization of photo-$z$'s impact both galaxy evolution and cosmological studies.  In \S \ref{sec:methods}, we outline the principles underlying recent photometric algorithms, providing context for the issues discussed in this review.  \S \ref{sec:spectroscopy} describes the needs for spectroscopic redshift measurements for large samples of objects to improve both photo-$z$ performance and characterization.  In \S \ref{sec:openissues} we describe a variety of open issues, highlighting areas where future work would be valuable.  Finally, in \S \ref{sec:vision}, we attempt to forecast how photo-$z$ methods might ultimately evolve to optimize both performance and characterization.

\begin{marginnote}
\entry{Characterization}{The recovery of the redshift distributions of ensembles of galaxies, including their \emph{mean} redshifts, variances, and higher-order moments.}
\end{marginnote}

\ifarxiv
\begin{marginnote}
\entry{Take-Away}{This review focuses on common challenges and strategies for applications of photometric redshifts to the next generation of deep, wide-area imaging surveys.}
\end{marginnote}
\fi

\subsection{Performance and Characterization of Photometric Redshifts}

\label{sec:definitions}

Here we first describe fundamental limitations to both the performance and characterization of photometric redshifts. We then discuss the impact of these sources of uncertainty on different science cases.  A photometric redshift method could be extremely well-characterized despite its poor performance, or vice versa; applications of photometric redshifts across different subfields will place widely varying requirements on each aspect. 

\textbf{Limits to the potential performance} of redshift determination are set by the quality of data collected on a galaxy. In the case of spectroscopy, the flux obtained from an object is measured as a function of wavelength with a resolution ($R = \frac{\lambda}{\Delta \lambda}$) typically ranging 
from 100 to $>30,000$. When multiple features (e.g., absorption or emission lines) are detected in an object's spectrum, its redshift may be determined 
securely, as any possible pair of significant features in a spectrum has a unique wavelength ratio.  Even when individual features are not found at high 
signal-to-noise ratio (SNR), comparisons to spectral templates may enable a determination of redshift from 
the combination of many weaker lines.  If the features in an object's spectrum are relatively narrow (with wavelength 
extent not much larger than the instrumental resolution) and SNRs are sufficiently high, the redshift of an object may be determined via spectroscopy with an uncertainty that is well below $R^{-1}$.

\begin{figure}[t]
\centering
\begin{subfigure}{.63\textwidth}
  \centering
  \includegraphics[width=\linewidth]{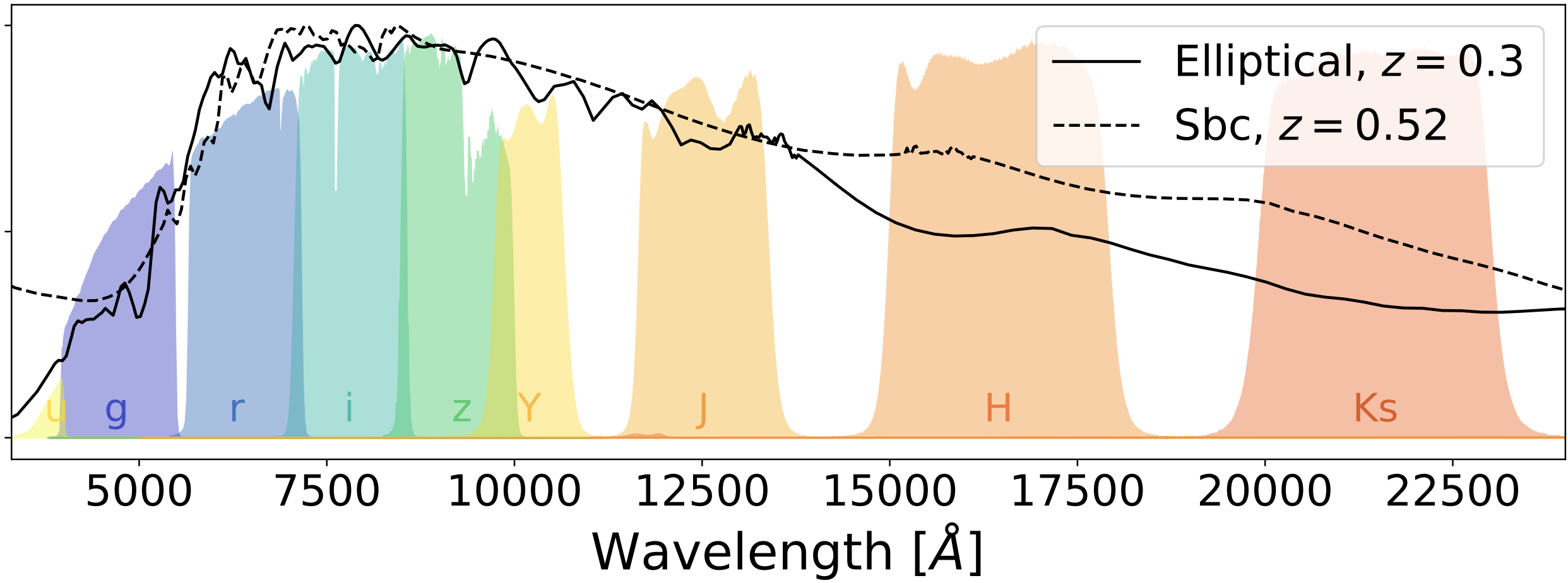}
\end{subfigure}%
\hspace{0.02\textwidth}
\begin{subfigure}{.33\textwidth}
  \centering
  \includegraphics[width=\linewidth]{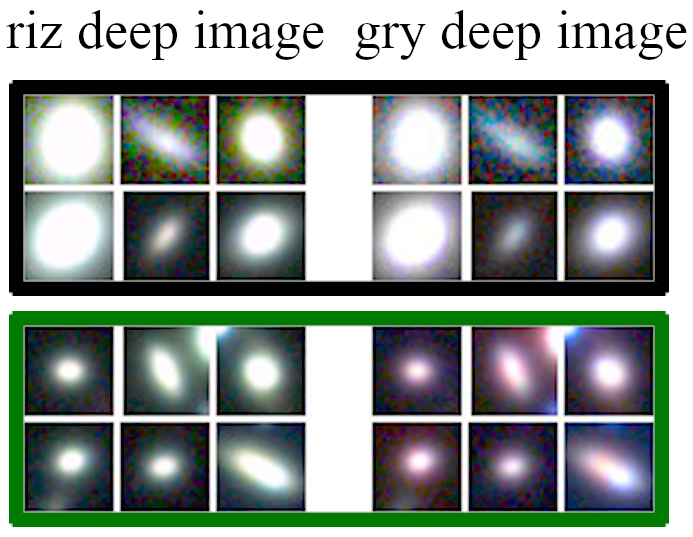}
\end{subfigure}
\caption{\emph{Left:} Given flux information in only a few broad bands, degeneracies between different combinations of intrinsic rest-frame spectral energy distributions and redshift are common, as illustrated by templates for an elliptical galaxy at $z=0.3$ and a spiral galaxy at $z=0.52$ that are virtually indistinguishable if only $riz$ photometry is used (as may be optimal for ground-based lensing analyses, \citet{Sheldon2020}). The resulting redshift PDF $p(z)$ would have to be bimodal, with the relative weight of the two modes determined by a model for the galaxy population. Figure adapted with permission from \citet{Buchs2019}. \emph{Right:} A sample of galaxies with similar $riz$ color-magnitude \citep{desdr2}. These come from two sub-populations that are distinguishable using additional colors, such as $g-r$ and $r-y$ (black and green boxes, showing HSC images, cf. \citealt{hscdr}); the populations have different redshifts (black: $z~0.85$, green: $z~0.50$). Color coding of boxes corresponds to that used in Figure~1 in \citet{Myles2021}. 
}
\label{fig:galaxy_zoo}
\end{figure}

Photometric redshifts rely on measurements of integrated fluxes within a set of filters, with resolutions of at most $R\sim 50$ when many narrow bands are used \citep[e.g.,][]{Molino2014, Marti2014,spherex}. 
More commonly, present and near future large area surveys such as the Dark Energy Survey \citep{DES2016}, the Hyper Suprime-Cam SSC Survey \citep{Aihara2018}, the Kilo Degree Survey \citep{Hildebrandt2021}, and the Rubin Observatory LSST \citep{LSSTScienceBook} use broad-band filters with $R<10$. As a result of the low effective spectral resolution provided by this imaging, fewer spectral features which may constrain redshift are available, and each of those features provides weaker constraints due to the poorer wavelength localization. Additionally, it is rare that one can identify multiple, well-localized spectral features at such low $R$, leading to ambiguity in their identification. This can lead to  degeneracies between multiple fits to a galaxy's spectral energy distribution (SED) -- i.e., its observed flux as a function of wavelength -- that further increase the uncertainty of redshift estimates (cf. the left panel of \autoref{fig:galaxy_zoo}).  For instance, in many cases it can be difficult to distinguish whether a strong jump in a galaxy's SED corresponds to the 4000~\AA~break at a lower redshift, or the Lyman break at a higher $z$ \citep{Stabenau2008}.   

In the best cases, for high-quality, uniform photometry, either with a large number of narrow-band filters 
or for object classes with uniform intrinsic spectra that have strong, broad spectral features, photometric redshift errors $\sigma_z \sim 0.01(1+z)$  (where $z$ is 
the redshift of the object) have been obtained \citep[e.g.,][]{Pandey2021}. A similar performance has been achieved for broader samples using narrow-band photometry \citep{Alarcon2021}. However, photometric redshifts for diverse galaxy populations estimated from few broad-band filters are necessarily more uncertain, with $\sigma_z$ from current surveys commonly $\sim 0.05 
(1+z)$ or larger.
\ifarxiv
\begin{marginnote}
\entry{Take-Away}{The performance of photo-$z$ algorithms is limited by having measurements in only a few, broad, noisy photometric bands -- but that trade-off enables studies of large samples of faint galaxies.}
\end{marginnote}
\fi

Photometric redshifts are vital despite their inferior precision because they presently are the only available option for estimating redshifts for the samples of large numbers of faint galaxies required by many current science cases. 
At an extreme, one could imagine using a 5000-fiber spectrograph (the equivalent of the most highly-multiplexed among the current options, the DESI instrument \citep{DESIInstrument}) on a 10m telescope to obtain spectroscopy of the LSST "gold sample" of 4 billion galaxies with $i<25.3$ that are expected to be used for weak lensing and large-scale-structure measurements \citep{Ivezic2019}. Extrapolating from past campaigns \citep{Newman2013}, it would require nearly 16,000 years of continuous integration time under clear, dark conditions to obtain spectra of the same signal-to-noise (S/N) for this entire sample as the DEEP2 survey attained for objects with $i<22.5$. Furthermore, at that S/N many faint objects (10-25\%) do not yield secure spectroscopic redshift measurements, a failure rate that is much higher than the catastrophic error rates in the best photo-$z$ algorithms.

\textbf{The probability density function (PDF) of redshift,   $p(z)$}, provides the best representation of the result of a photometric redshift algorithm, as the resulting distributions may be highly non-Gaussian or multimodal. Three different flavors of PDFs are commonly used in the literature, though they are not always clearly distinguished: \label{pz}
\begin{itemize}
    \item \textbf{Bayesian posteriors}, with $\int_{z_0}^{z_1} p(z|\mathrm{photometry}) \, dz$ interpreted as the \textit{degree of belief} that the redshift of a given galaxy is between $z_0$ and $z_1$. These can properly incorporate any uncertainties in the underlying model (e.g., the prior of a template fitting scheme, or the limited training sample), broadening posteriors accordingly.. 
    \item \textbf{Frequentist probabilities}, with $\int_{z_0}^{z_1} p(z|\mathrm{photometry}) \, dz$ interpreted as an estimate of the fraction of times the true redshift of one out of an ensemble of galaxies with indistinguishable photometry is between $z_0$ and $z_1$. These $p(z)$'s assume a fixed model for the galaxy population - they cannot marginalize over model uncertainties, as that procedure is inconsistent with the frequentist formalism.
    \item \textbf{Likelihoods} that interpret $p(z)$ as $p(\mathrm{photometry}|z)$ for a fixed template chosen in some way; e.g., for the best-fitting among a set of templates.
\end{itemize}
\begin{marginnote}
\entry{Redshift Probability Density Function (PDF)}{A function whose integral between two  limits corresponds to the probability that the redshift lies in that range: $p(z)$.}

\entry{Likelihood}{The probability of obtaining the actually observed values as a function of the assumed model parameters (including redshift): $p({data} | {model})$.}

\entry{Prior}{A function describing the relative probability of a given set of model parameters, a key ingredient in Bayesian inference: $p({model}$.}

\entry{Redshift Posterior}{A redshift PDF derived multiplying the likelihood of the observations by the assumed prior probabilities for model parameters: $p({model} | {data})$.}

\end{marginnote}
Bayesian posteriors are required in schemes that allow the data of all galaxies to inform the model (see \S \ref{sec:bhm}). Frequentist approaches are particularly appropriate for applications where the $n(z)$ of a set of galaxies is estimated with a fixed model; e.g., directly from the spectroscopic redshift distribution of a set of reference galaxies (cf. \S \ref{sec:stage3}). 

A key distinction is that the stack (i.e., the sum over several galaxies) of frequentist $p(z)$'s, is meaningful. It provides an estimate of the ensemble's redshift distribution under the assumption of the fixed model, as the frequentist $p(z)$ corresponds to the fraction of times a given object should be found at a particular redshift, such that the expectation number in a redshift bin from a sample must be a simple sum. However, in the Bayesian definition, likelihoods (not posteriors) must be combined to infer overall redshift distributions, as summing individual posteriors that marginalize over the same model does not properly account for the effect of model uncertainty on the inferred redshift distribution \citep{Malz2021}. A variety of statistics derived from PDFs, including the redshift corresponding to the maximum likelihood or the maximum posterior probability, the expectation value of redshift evaluated across the full PDF, or the expectation calculated only on the highest peak \citep{Dahlen2013} have been used as single ``point'' estimates for the redshift of a galaxy in the past, although their use in cosmological studies has been waning due to the recognition that full information on redshift distributions is required already for present surveys and will continue to be needed for future applications.

\ifarxiv
\begin{marginnote}
\entry{Take-Away}{Photo-$z$'s are best reported and interpreted as PDFs. Frequentist and Bayesian approaches differ.}
\end{marginnote}
\fi

\vspace{10pt}

\textbf{The characterization of photometric redshifts is limited} by our incomplete knowledge of the galaxy population. For instance, the samples of galaxies with spectroscopic redshifts used for this characterization may systematically miss some populations or include objects with incorrect redshift measurements.  In forward-modeling approaches, the set of model galaxy templates and associated prior probability distributions used may not be capable of describing the full ensemble of galaxies in the Universe, leading to biased results.  Imperfect characterization of statistical or systematic error distributions in the photometric data can likewise distort the redshift distributions that result from such methods. Furthermore, the limited size of deep spectroscopic samples will reduce the fidelity with which redshift distributions can be constrained. 

A connected issue is the difficulty of confidently estimating uncertainties in the characterization of redshift distribution. In the absence of accurate knowledge about the population of galaxies one is trying to estimate the redshift for, error estimates often have to remain approximate. Worse, imperfections in photometric redshift methodology can cause slight but relevant errors in the estimated ensemble redshift distributions that are impossible to know $a priori$, and difficult to evaluate on existing data.

These issues do not represent fundamental limitations to the potential power of photometric redshifts -- there is no reason we could not, in principle, formulate and constrain a model of the galaxy population, or understand photometric data, in a way that is sufficient to characterize photometric redshifts at any level required, given sufficiently constraining data (including spectroscopy). This is a major benefit of photo-$z$'s for science cases that do not require individual galaxy redshifts. So long as the distribution of redshifts is known well and with minimal bias, strong constraints may be obtained. This is particularly the case for cosmological studies using a small number of redshift bins, such as for weak gravitational lensing or angular galaxy clustering measurements.

However, this implies that those applications have extremely stringent requirements for the characterization of redshifts. If the mean, width, or perhaps even full shape of the estimated redshift distribution is not close to what one would obtain from measuring the actual redshifts of the objects in that sample, significant biases can result. Given the increasing precision of cosmological experiments, the characterization of photometric redshift distributions has become perhaps the leading systematic uncertainty in these analyses.
\ifarxiv
\begin{marginnote}
\entry{Take-Away}{The scarcity of suitable deep spectroscopy, combined with stringent requirements, makes photometric redshift characterization a leading challenge and source of systematic uncertainty.}
\end{marginnote}
\fi

\subsection{Applications of Photometric Redshifts}

\subsubsection{Galaxy Evolution}

Studies of galaxy evolution which rely upon photometric redshifts vary in their sensitivity to photometric redshift performance and characterization.  We illustrate these dependencies by considering three major applications: subdivision of samples according to their redshifts; measurements of the abundances of objects as a function of their properties and redshift (as in luminosity function studies); and studies which rely on measurement of galaxy clustering or environmental densities.  We focus on applications in the not-too-distant future, when systematic uncertainties in modelling galaxy evolution should remain substantial.  If this limitation is overcome and we wish to extract as much information as possible from the data, requirements on the characterization of photometric redshifts for galaxy evolution studies will become much more stringent, resembling the needs for cosmological studies (q.v. \S \ref{sec:cosmo_requirements}).

\textbf{The subdivision of objects according to their redshifts} can be used to study evolution of galaxy demographics over time 
\citep[e.g.,][]{Finkelstein2015} or to select targets in a particular $z$ range for spectroscopic surveys \citep{Mclure2018, Takada2014}.  There is a trade-off between contamination from objects at other redshifts (i.e., what fraction of objects selected are not in the desired redshift range) versus the completeness of samples selected to be at a given $z$ (i.e., what fraction of objects truly at that redshift are included).  More stringent selections will reduce contamination, but will also cause some objects that are in the desired range to be missed.  

Improving the photometric redshift performance for typical objects will decrease contamination rates by causing fewer objects to scatter into a redshift bin due to errors, while simultaneously improving completeness by reducing the number of objects that scatter  out.  
In contrast, if their prevalence is low, catastrophic outliers (i.e., objects for which the photometric redshift is far from the spectroscopic redshift, on a non-Gaussian tail of the error distribution) will have limited impact, contributing to incompleteness and contamination at levels proportional to their rates. If the rates at which outliers occur is known well, corrections for them can be included in analyses.  In the case of targeting for spectroscopic surveys the impact of outliers is further reduced, as better redshift measurements will show that such objects do not belong in the sample.
\ifarxiv
\begin{marginnote}
\entry{Take-Away}{For most current galaxy evolution studies, the \emph{performance} of photo-$z$'s is the most important factor.}
\end{marginnote}
\fi
In most cases errors in characterization will have limited impact on these analyses.
For instance, a small additive bias in redshifts 
will have only minor effects on how observed samples are interpreted via galaxy evolution models at all but the lowest $z$. \begin{marginnote}
\entry{Catastrophic outliers}{Objects for which the photometric (or spectroscopic) redshift is far from the true redshift, corresponding to a non-Gaussian tail of the error distribution.}
\end{marginnote}

In some analyses integrals over the redshift PDF for an object are used to divide up samples \citep[as in][]{Finkelstein2015}; in that case, inaccuracies of those PDFs can affect binned analyses.  For instance, one can consider a sample where photometry may be consistent with either redshift $z \sim  2$ or $z \sim 6$ due to confusion between the 4000 Angstrom and Lyman alpha breaks, which will lead to bimodal (two-peaked) redshift PDFs for each object.  In that case, if the relative amount of probability assigned to each of these redshifts is incorrect, the abundance of $z = 6$ galaxies could be badly mis-estimated. 

\textbf{Measuring distributions of galaxy properties} introduces additional complexity. Photo-$z$'s have been used to measure the distribution of galaxy luminosities or stellar masses (commonly referred to as a luminosity function or mass function, respectively), as in \citet{Wolf2003} and \citet{Bundy2017}.  In these applications, modeling uncertainties are substantial, due to our limited knowledge of how to relate the star formation history of a galaxy to the observed light from it; for instance, changing the assumed initial mass function of stars can alter inferred stellar masses by $\sim$0.5 dex \citep{Courteau2014}.  

In these applications, the gains from improving photo-$z$ performance are generally modest, as luminosity or mass functions are often measured in bins which are broader than photo-$z$ uncertainties (e.g., $\Delta z = 0.2$ or $0.5$).  The effects are larger at the bright/high-mass end of the luminosity/mass function, as the propagation of distance errors into the inferred luminosity will alter the shape of the exponential cutoff of the Schechter function substantially \citep[to first order, it will be convolved with a Gaussian kernel determined by this propagation; cf.][]{Santini2015}, resulting in an Eddington-like bias \citep{Eddington1913}.  The effects are weaker for fainter objects, as convolution with a Gaussian leaves a distribution unchanged when its second and higher derivatives are negligible.  However, where incompleteness becomes large that condition will no longer hold, and photometric redshift errors can again bias results \citep{Sheth2007}.

Catastrophic (i.e., large and non-Gaussian) photometric redshift errors will primarily affect the bright end of the luminosity function at higher redshifts.  Since faint objects are common but bright ones are rare, if a luminous, higher-$z$ object is erroneously placed at low redshift there is little impact, as then it will have a low inferred luminosity and be outweighed by the numerous faint objects that are truly at that $z$.  However, if a faint lower-$z$ object is falsely assumed to be at high redshift, it will have a high inferred luminosity; as a result, contaminants can dominate samples at the bright end.  

However, studies of luminosity and mass functions are not very sensitive to overall biases in photometric redshifts; a 1\% error in the mean redshift of all objects in a sample would alter their inferred stellar masses by less than 0.01 dex, in contrast to systematic uncertainties of 0.1-0.5 dex.  Nevertheless, characterization of the rate and distribution of catastrophic outliers can be important if their effects are to be removed to enable studies of the bright end of the luminosity function.

The final category of galaxy evolution measurements where photometric redshifts have played an important role is measurements of the \textbf{clustering (or environments) of galaxies as a function of their properties}.  In contrast to the previous cases, for such studies improving typical photometric redshift performance will have a large effect.

As a simple illustration of this, one can consider counting the number of objects within a cylinder with length in the redshift direction $\Delta z$ and projected comoving radius $r_p$ about some object.  The count of objects within the cylinder can be used as a measure of local overdensity \citep{Cooper2005}, and is equivalent to a measurement of the mean projected correlation function within the cylinder, $<w_p(r_p)>$,  integrating to a maximum separation $\pi_{max} \equiv \Delta z / 2$.  In the Poisson-dominated regime (common for environment measures), the fractional uncertainty on the density within the cylinder will be $\frac{\sigma(n)}{n} = \frac{1}{\sqrt{n \pi r_p^2 \Delta z dl/dz}} \propto \frac{1}{\sqrt{\Delta z}}$, where $n$ is the mean comoving density and $dl/dz$ is the derivative of comoving distance with respect to redshift.  However, the number of objects truly associated with a cylinder -- the signal which one desires to measure -- remains fixed. In the simplest case, where $\Delta z$ is large compared to photometric redshift errors, the signal-to-noise of overdensity measurements will be proportional to $\frac{1}{\sqrt{\Delta z}}$; if objects scatter out of the cylinder due to photo-$z$ uncertainties, the S/N will only get worse.  Improving photometric redshift performance enables smaller redshift windows to be used without losing physical pairs of objects, reducing $\Delta z$ and increasing the signal-to-noise accordingly.
\ifarxiv
\begin{marginnote}
\entry{Take-Away}{Photo-$z$ \emph{performance} strongly affects the signal-to-noise ratio of clustering and environment studies.}
\end{marginnote}
\fi

Catastrophic outliers will cause systematic biases in the inferred clustering strength.  If a fraction $f_{\rm outlier}$ of photo-$z$'s are far from their true redshifts, correlation function and overdensity measurements will be reduced by a factor of $(1-f_{\rm outlier})^2$, so large catastrophic outlier rates can cause clustering to be badly underestimated.   Outliers will also cause the effective density of a sample (generally used as a constraint in halo model interpretations of clustering) to be overestimated by a factor of $(1-f_{\rm outlier})^{-1}$.  For analyses not to be systematically biased as a result, it is necessary either for the outlier rate to be known and corrected for, or for outlier rates to be marginalized over in analysis (as in \citealt{Zhou2021}), at the cost of degraded constraints on other quantities.  

As for luminosity functions, biases in inferred photometric redshifts have only a modest effect on environment and clustering measurements, as differences in redshift between pairs of galaxies will remain unchanged. Instead, the impact is to alter the mean redshift of the samples for which clustering has been measured.  Given current limitations on modeling galaxy evolution, small biases in effective redshift should be subdominant to other systematics in the near future.  

Accurate characterization of the uncertainties in individual redshift estimates, or particularly having accurate photo-$z$ PDFs for individual objects, is beneficial for clustering analyses.  If the redshift PDF is well-known, measurements can directly utilize the range of possible redshifts of each object, rather than relying only on (for instance) calculating angular correlation functions within fixed redshift bins.  This can be exploited to maximize the SNR of measurements.  For instance, \citet{Zhou2021} replaced each object with a large number of samples from its redshift PDF and measured the number of pairs within a fixed redshift window around each, eliminating the loss of pairs due to bin edges while taking PDF information  fully into account.  However, if errors (or PDFs or outlier rates) are mis-estimated, inferences based on clustering measurements can be systematically biased.  Fitting for additional parameters characterizing the degree to which errors are over- or under-estimated can be used to account this effect at the cost of degraded constraints on other parameters, as in \citet{Zhou2021}.   

\subsubsection{Cosmology}
\label{sec:cosmo_requirements}

The principal objective of observational cosmology is to test predictions for the expansion history of the universe and the 
growth of structure across time. Measurements based upon the cosmic microwave background, the distance-redshift relation 
with supernovae and baryonic acoustic peaks in the clustering of galaxies, the growth of structure observed 
through the clustering of galaxies, gravitational lensing, and through galaxy clusters broadly agree. Jointly they indicate that overall 
deviations of expansion and structure growth from a $\Lambda$CDM \emph{standard model} of cosmology are at most 
at the level of a few percent. The next steps of this research program thus must advance into a regime of 
highly precise and accurate measurements to significantly challenge $\Lambda$CDM predictions with data. 
Present and future experiments have been designed to reduce statistical uncertainties on cosmological measurements beyond the current state of the art. Assuming successful data collection, the outcomes from these experiments are almost certain to be limited by systematic uncertainties.

For this reason, the requirements on photometric redshifts for the purpose of cosmology are quite different from those for galaxy evolution, and largely shared among different 
probes. Redshift affects the observables predicted by a cosmological model -- e.g., the amplitude of large-scale matter density fluctuations, the number density of galaxy clusters, or a cosmological distance measure -- typically via an integral over the redshift distribution $n(z)$. As a consequence,  reporting frequentist $p(z)$ estimates for individual galaxies or sets of galaxies, stacking them, and marginalizing over uncertain characterization by repeating the whole procedure at varied model parameters \citep[e.g.][]{Stoelzner2020, Cordero2021} can be well-matched to these applications.

The relevant integrals change by of order unity per unit change in mean 
redshift. To improve upon the current few-percent-level tests of $\Lambda$CDM 
predictions, the mean redshifts of observed samples will need to be known to similarly high accuracy. This includes accurately characterizing the tails of the redshift distribution of photometric samples associated with catastrophic outliers.
Characterization of  higher order moments of the redshift distribution of samples is likely to be a secondary 
limiting factor. In addition to the stringent requirements placed upon the characterization of photometric redshifts, in some instances photo-$z$ performance will affect the signal-to-noise ratio of cosmological measurements greatly. 
We consider the requirements on photometric redshifts for each of the major imaging-based probes of cosmology below.
\ifarxiv
\begin{marginnote}
\entry{Take-Away}{Cosmological studies primarily require exquisite \emph{characterization} of photo-$z$'s.}
\end{marginnote}
\fi

\begin{itemize}
\item In \textbf{weak gravitational lensing}, the angular diameter distances of lensed galaxies, determined 
from their redshift by the cosmological model, set the amplitude of lensing signals \citep[for a recent review, see][particularly their sections 2.8 and 3.2]{Mandelbaum2018}. For weak gravitational lensing, only large ensembles of galaxies will yield useful signal-to-noise ratio; subdividing samples into a small number of minimally-overlapping redshift bins is sufficient to capture most information.  The assignment of galaxies to redshift bins benefits from improvements to photo-$z$ performance, but due to the relatively shallow increase of lensing efficiency with source redshift, the gain in constraining power is comparatively modest \citep{Hu1999}. However, extremely stringent characterization of the redshift distribution of the ensemble of galaxies assigned to each bin is required for both present and future experiments to reach their goals.

\item In \textbf{strong gravitational lensing} \citep[e.g.,][]{Treu2010}, photometric redshifts can aid in the identification of potential lens systems, as well as in the study of foreground galaxies which contribute additional lensing effects and influence the inferred properties for a given system; these applications will benefit weakly from improvements to photo-$z$ performance. Photometric redshift estimates of the multiply imaged background galaxies are likewise useful for the reconstruction of lens geometries. Commonly, follow-up spectroscopy will be needed for cosmological analyses of strong lensing, so photo-$z$ characterization requirements are minimal. 

\item The \textbf{clustering of galaxies} is a probe both of the amplitude of density fluctuations (when joined with lensing; see, e.g., \citealt{Baldauf2010}) and of the scale of baryonic acoustic oscillations (BAO). 

Large-scale-structure measurements can  be performed either by simply measuring angular clustering in photometric redshift slices \citep[e.g.,][]{Sanchez2011,Carnero2012,DESY3BAO} or, for more sensitive results, by using photometric redshift estimates (or PDFs) for individual objects \citep[e.g.,][]{Padmanabhan2007,Seo2012, Zhou2021}.

The observed amplitude of angular clustering will depend directly on the redshift distribution of the galaxy sample. As was the case for clustering-based studies of galaxy evolution, increasing the performance of photo-$z$'s will improve signal-to-noise  for  cosmological large-scale-structure studies. Characterization of the width of the ensemble redshift distribution is particularly important for minimizing systematic uncertainties in measurements of the clustering amplitude \citep[e.g.,][]{Cawthon2021}, while characterization of the mean redshift will be more important for constraints on the BAO distance scale.

\item For \textbf{clusters of galaxies} \citep[e.g.,][]{Allen2011}, the expected abundance of clusters and the relationships of observables to the intrinsic physical properties of a cluster both depend on redshift.  However, these should all be relatively slow functions of $z$, and photometric redshifts for clusters tend to be very well determined (due both to their red galaxy populations and the ability to average photo-$z$'s from many objects), so that improvements to photo-$z$ performance will have limited effect on cosmological inference for clusters selected based on their gas properties. Photo-$z$ performance is however a crucial factor for optically-selected cluster samples, where objects are selected as overdensities in the three dimensional distribution of galaxies. Uncertainties in the photometric redshifts of \emph{individual galaxies} will set the $\Delta z$ scale over which foreground or background objects may affect optical observables for a given cluster (such as richness, total flux, etc.). Photo-$z$ performance thus directly impacts the measured distribution of richnesses and the mass limit down to which physical clusters can be distinguished from the random galaxy background.  The angular clustering of clusters depends on their redshift distribution (as was true of galaxy clustering), requiring uncertainties in cluster photometric redshifts to be accurately characterized for applications that depend on that quantity.  Calibrations of cluster masses based upon weak lensing measurements will have stringent requirements on the characterization of the redshift distributions of background objects, much as for other weak lensing measurements \citep{DESC_SRD}.

\item For analyses of \textbf{photometric supernovae} used to constrain the distance-redshift relation without spectroscopic follow-up, imaging alone must be used to determine redshifts \citep{Linder2019}. In this case performance must be sufficient to help classify supernova type, with the important distinction that for these analysis extreme photo-$z$ outliers can be rejected based upon the observed brightnesses of supernovae regardless of the host galaxy photometry.  Accurate characterization of photo-$z$'s will be needed to use such supernovae in cosmological analyses. Additionally, photometric redshifts can be used to select suitable targets for spectroscopic follow-up that are likely to be supernovae of the desired type Ia (as opposed to other types of transients); this places only weak requirements on performance, however.
\end{itemize}

A broad assessment of requirements on the level of accuracy of the characterization of photometric redshifts is presented in \citet{DESC_SRD}, which tested the impact of both biases and mis-estimations of photometric redshift errors on cosmological measurements with the Rubin Observatory LSST.  This study found that for cosmological measurements combining weak gravitational lensing and large-scale structure, the mean redshift of each tomographic bin must be known to $\delta z < 0.001(1+z)$ by the end of the survey for systematic uncertainties in the dark energy equation of state not to exceed statistical uncertainties.  Similarly, the photometric redshift scatter $\sigma_z$ must be known to better than $\delta \sigma_z < 0.003(1+z)$.  Requirements on the characterization of the redshifts of the lensed objects behind galaxy clusters used to calibrate cluster masses are similar: biases must be below $\delta z < 0.001(1+z)$ and uncertainty in scatter $\delta \sigma_z < 0.005(1+z)$.  These requirements are all extremely stringent and will be challenging both to meet and to demonstrate that they have been definitively achieved.  The ultimate impact of photo-$z$ characterization on constraints based on strong lensing or supernova brightness measurements is much weaker, however, as for those probes analyses of only the subset of objects with spectroscopic measurements are already likely to be systematics limited; they will thus not be major drivers of photometric redshift requirements for the upcoming imaging surveys.  We note that this study considered only a simple Gaussian model of photometric redshift distributions without outliers, but in real applications it is likely that higher moments of the redshift distribution, not only mean and variance, must be characterized stringently.
\ifarxiv
\begin{marginnote}
\entry{Take-Away}{Weak lensing, large-scale-structure, and cosmology studies with upcoming surveys all require characterization at the 0.1\% level.}
\end{marginnote}
\fi

\section{OVERVIEW OF PHOTOMETRIC REDSHIFT METHODS}

\label{sec:methods}

The idea that broad-band photometry of galaxies could be used to constrain their redshift goes back almost 60 years \citep[e.g.,][]{Baum1962,Koo1985,Loh1986}. Since then, two families of methods have commonly been considered separate - one based on comparing observations to galaxy spectral energy distribution templates, and one based on empirical relations of photometry to redshift found, often by means of machine learning, on samples of galaxies with known redshift. This dichotomy, while useful in characterizing methods, is somewhat superficial. All photometric redshift methods can be interpreted in the same context of Bayesian statistics, of inferring the posterior probability (or some related statistic) of redshift given observational data \citep{Budavari2009}. The model of galaxy templates or the sample of reference galaxies with known redshift are part of the prior - along with other implicit assumptions made in the respective method \citep{Schmidt2020}. 
\ifarxiv
\begin{marginnote}
\entry{Take-Away}{Photometric redshift methods can be categorized by how they use prior information, including training samples and SED templates.}
\end{marginnote}
\fi

\subsection{Template-Based Methods}

\label{sec:templates}

The family of methods often described as \emph{template-based} perform inference based upon an \textit{a priori} model of the range of galaxy SEDs that exist. 
These codes commonly construct PDFs for the redshift of a galaxy via an application of Bayes' theorem, following \cite{Benitez2000}.  The posterior probability for the redshift $z$ given a set of observed fluxes, $F$, $p(z | F)$ can be determined from an equation:

\begin{equation}
p(z | F) = \frac{\int  p(F | z,t,O) p(z,t,O) \,  dt \, dO}{p(F)}.
\label{eqn:posterior}
\end{equation}

Here the prior, $p(z,t,O)$, represents the joint probability of a given combination of template $t$, redshift, and potential extra parameters such as luminosity or apparent magnitude in some band, absent any other information about the individual object of interest; the choice of extra parameters used varies amongst implementations. If the templates are not distributed over a continuous space,  the integral contains a discrete sum. The likelihood, $p(F | z,t,O)$, corresponds to the probability of getting the particular values of the  fluxes in each band observed for the object of interest, assuming a set of values of the redshift, template type, and any additional parameters.  For Gaussian errors, the likelihood will be proportional to $\exp{-\chi^2/2}$, where the $\chi^2$ value is computed as $\sum[F_{\rm observed,i} - F_{{\rm predicted}(z,t,O),i}]^2/\sigma_i^2$.  Here $F_i$ corresponds to the $i$th element of either the observed flux vector or the flux vector predicted for a given set of parameters $z,t,O$ and $\sigma_i$ is the uncertainty in the $i$th observed flux. 

As long as the model for galaxy SEDs is considered fixed, the resulting $p(z)$ can be interpreted as either a Bayesian or frequentist estimate. Some template methods report the posterior probability of redshift inferred from the observed fluxes, $p(z | F)$; others instead provide the likelihood $p(F|z,t,O)$ without incorporating the prior term $p(z,t,O)$. Care must thus be taken to interpret outputs correctly.

Commonly applied methods that use a $\chi^2$-based likelihood include LePhare \citep{Arnouts1999,Ilbert2006}, BPZ \citep{Benitez2000}, ZEBRA \citep{Feldmann2006} and EAZY \citep{Brammer2008}. They, and other template-based methods, differ primarily in their choice of: 
\begin{itemize}
\item \textit{What observables are used to predict redshifts;} e.g., a set of fluxes (LePhare, ZEBRA), or instead a set of colors (i.e., differences between magnitudes between photometric bands) which may be combined with a magnitude-dependent redshift prior (BPZ, EAZY). One could imagine  exploiting morphological parameters in priors as well. We note that using magnitudes or magnitude-derived colors has the disadvantage of discarding the information present in negative flux measurements; dropping such measurements entirely can result in biased inference.
\item \textit{What set of templates to use,} e.g.,~ones derived from spectroscopic observations (\citealt{Coleman1980,Kinney1996} as used~in BPZ or \citealt{Connolly1995}) or synthetic spectra based upon stellar population synthesis models (e.g., \citealt{Bruzual1993,Bruzual2003} in LePhare). Variants use best-fitting linear combinations of templates at each redshift (as in EAZY) or iteratively update the initial template set to better match observed colors (e.g., ZEBRA).
\item \textit{Whether to multiply the likelihood $p(F | z,t,O)$ by the prior probability, $p(z,t,O)$,} which some methods chose to not do (LePhare), others do using a redshift/luminosity prior (EAZY) or redshift/type/magnitude prior (BPZ) derived from training data, and others with an iteratively adjusted prior (ZEBRA).
\item \textit{How to marginalize over templates:} formally, to calculate the posterior redshift distribution $p(z | F)$ one must integrate the multi-dimensional posterior on the right-hand side of equation \ref{eqn:posterior} over the template dimension $t$ (a process known as marginalization).  However, some codes (e.g., EAZY) instead approximate the marginalized likelihood $p(F | z)$ by $\exp{-\chi^2(t_{\rm best},z)/2}$,  where $t_{\rm best}$ is the template (or combination of templates) which provides the best fit for a given $z$, replacing the integral by its value for only the highest-likelihood template at each $z$, ignoring other templates which may also be consistent with the photometry.
\item \textit{What quantity to report,} whether the full redshift posterior (BPZ, ZEBRA) or single ``point'' values such as the combination of template type and redshift that have the maximum likelihood (e.g., LePhare, ZEBRA); the redshift which has the maximum posterior probability (e.g., EAZY); or the expectation value of redshift $\int z p(z) dz$ (e.g., BPZ, EAZY). 
\end{itemize}

A strength of template-based methods is that they use an informative prior, the underlying model of the full galaxy population, to infer the redshift posterior from the noisy information from an individual galaxy. The importance of this prior, however, makes template-based methods subject to a number of potential problems:
\begin{itemize}
\item \textit{The template set is incomplete.} Since no two galaxies have exactly the same SED, no finite set of templates can fully describe a population. When more degrees of freedom are given to the set of templates, conversely, unphysical solutions or biases can result.
\item \textit{The prior is wrong.} Even with a complete and sufficiently flexible set of templates, the priors on the abundance, redshift, and luminosity of galaxies resembling a template may not be accurate. Especially when photometry is noisy and only available in a few bands, it can be consistent with multiple combinations of template and redshift, making the photometric redshift estimate highly sensitive to the priors used.
\ifarxiv
\begin{marginnote}
\entry{Take-Away}{Incomplete, incorrect, and/or inflexible models for the galaxy population currently limit template-fitting redshift performance at levels of $|\langle\Delta  z\rangle/(1+z)|>0.01$.}
\end{marginnote}
\fi
\item \textit{The data do not inform the model.} In the limit where the templates and priors describing the galaxy population are already accurately known, separating their determination from the estimation of individual galaxy posterior PDFs, as done in most template-fitting codes, is an appropriate choice. However, such accurate knowledge does not exist about the general galaxy population. Photometric datasets could themselves be used to constrain models of the set of template SEDs needed and prior probability distributions for redshift and type.
\end{itemize}

Template-based methods have addressed these concerns in a variety of ways: e.g., by using flexible and/or optimized template sets (as in EAZY) or by adjusting templates and priors using the ensemble of galaxy data (as in ZEBRA). These and other recent developments bring them closer to the Bayesian Hierarchical methods described in Section~\ref{sec:bhm}.

\subsection{Machine Learning Methods}

Empirical methods for photometric redshift estimation find a relation between galaxy observables (e.g., fluxes and errors) and statistics related to the redshift. 
Most current methods use machine learning techniques, which rely on training samples of galaxies whose observables and known redshifts are taken as samples from the desired relation. Methods can be distinguished by:
\begin{itemize}
\item The \textit{training sample} of galaxies with known redshift, as well as the information about the training sample that will be used to predict redshifts (the ``features'' used for prediction, in machine learning parlance). Some methods only utilize galaxy colors (or flux ratios between bands) to predict redshift, while others incorporate the magnitude or flux in at least one band as a separate feature used for prediction, and some methods use full pixelized images.  The selection of objects in the training sample and appropriate reweighting as a function of their properties can be used to reduce biases or the impact of sample variance on redshift characterization.

\item The \textit{quantity they are trained to optimize}. Early methods commonly were optimized to minimize the variance of a point estimate of the redshift of a galaxy given its observables \citep[e.g., ANNz][]{Collister2004}. Many newer approaches aim to provide an estimate of the PDF of redshift instead \citep[e.g., TPZ, ANNz2][]{CarrascoKind2013,Sadeh2016,deVicente2016}. Approaches differ (and sometimes are ambiguous) in how exactly the resulting PDF should be interpreted (i.e., which of the types of PDFs described in \S \ref{sec:definitions} is being produced by a given algorithm). 

\item Further \textit{assumptions or choices} that affect the estimation of the target quantity. For instance, some methods choose to divide the training sample of galaxies with spectroscopic redshifts into subsets by distinguishable properties \citep[e.g., cells in self-organizing maps][]{Masters2015,Buchs2019}. Other methods define a neighborhood in photometric space over which reference galaxies are used to estimate the redshift of an object with given photometry \citep[e.g., DNF, DIR, CMNN][]{deVicente2016,Hildebrandt2017,Graham2018}, and potentially also a functional form (or, equivalently, machine learning architecture) relating photometry to redshift that is fitted within that neighborhood \citep[e.g., GPz, ANNz2, DNF][]{Almosallam2016,Sadeh2016,deVicente2016}.  
\end{itemize}

Each of these characteristics can lead to limitations on performance or characterization:
\begin{itemize}
\item \textit{The training sample is not a superset of the target sample.} When the sample of galaxies for which photometric redshifts are needed occupy regions of observable or physical-property space that are not also populated by the training sample, empirical models that are not built upon physical knowledge about galaxies cannot be assumed to successfully extrapolate.  For example, due to the expense of spectroscopy for faint galaxies, most objects with spectroscopic redshift measurements are much brighter than the objects for which photo-$z$'s are desired.
\item \textit{The training sample is not representative of the target sample.} A more treacherous case is when the training sample covers the distribution of the target sample in the space of observables that are available for both, but is non-representative in additional dimensions that are not known for the target sample.  For instance, spectroscopy may fail to deliver secure redshifts for objects of some types or at some redshifts while succeeding for other objects with similar colors, biasing training.
\item \textit{The training sample contains faulty entries.} Errors in the redshifts or photometric observables assigned to the training sample will generally lead directly to systematic errors in the estimated photo-$z$'s.
\item \textit{The trained quantity does not match the science}; for instance, in many cases a science analysis requires the full distribution of redshifts of a sample of galaxies, but the output of a photo-$z$ algorithm may be a single point estimate of redshift or some other quantity that does not allow the full distribution to be reconstructed accurately. 
\item \textit{Further choices introduce bias.} Even seemingly reasonable analysis choices -- e.g., to estimate photo-$z$'s through nearest neighbors in photometry space, or simplifying the treatment of noise -- can be shown to introduce significant biases in mean redshift \citep[e.g.,][]{Wright2020}.
\end{itemize}
\ifarxiv
\begin{marginnote}
\entry{Take-Away}{Insufficient training samples or analysis choices that do not match the science case commonly limit empirical redshift methods at levels of $|\langle\Delta z\rangle/(1+z)|>0.008$ \citep{Pasquet2019,Hayat2020}. }
\end{marginnote}
\fi

Whether a method is appropriate depends on the science case -- e.g., the target quantity that must be estimated, the level of performance needed, and whether PDFs are needed for individual objects or if instead only overall redshift distributions are required. Tomographic weak gravitational lensing analyses, for instance, need the full redshift distribution for a sample; point estimates are not suitable due to the non-linear dependence of lensing strength on redshift, making the tails of PDFs important. Determining these distributions will require a sufficiently-representative reference sample of nearly outlier-free and precise redshifts to train. {For this application, so long as the same selections can be performed on both the training and target sets of galaxies, it is sufficient to estimate the combined redshift distribution of all objects within a set of bins in parameter space, enabling the compression of continuous photometric information into a discrete number of subsets.}  

\subsection{Methods Employed by Recent Surveys}
\label{sec:stage3}
Cosmological analyses from recent surveys such as DES, HSC, and KiDS have generally responded to the above concerns by not assuming the output of any particular classical template-fitting or empirical photo-$z$ method will be correct. Where they did use such methods, they calibrated the result with a comparison to a reference catalog of spectroscopic \citep{Hildebrandt2019} or high-quality photometric \citep{Hoyle2018,Hikage2019} redshifts. Where they did not, they developed custom approaches designed to produce histograms of those reference redshifts re-weighted to represent the lensing source galaxy sample, designed to reproduce their redshift distribution $n(z)$ with minimal bias and variance \citep{Hikage2019,Wright2020,Myles2021}. By design, the result of these methods are frequentist $p(z)$'s for individual galaxies or, when stacked, $n(z)$'s.

The uncertainties in the mean redshift resulting from applying these characterization methods under idealized conditions to recent Stage III\footnote{The current generation of imaging dark energy experiments -- including DES \citep{DES2016}, HSC \citep{hscdr}, and KIDS \citep{Hildebrandt2021} -- all are classified as ``Stage III'' surveys in the scheme of the Dark Energy Task force \citep{DETF} based upon the level of constraints on the dark energy equation of state and its evolution that they will achieve.  The Vera C. Rubin Observatory's LSST, Nancy Grace Roman Space Telescope, and Euclid mission are classified as Stage IV experiments. } dark energy experiments have been estimated to range from 0.004 to 0.05 with earlier methods \citep{Hoyle2018,Hildebrandt2019,Wright2020} and from 0.001 to 0.006 in the most recent studies \citep{Wright2020,Myles2021},
\ifarxiv
as illustrated in the Figure within the margin.
\fi

Comparing to the requirements described in \S \ref{sec:cosmo_requirements}, one finds that the methods used for recent surveys are thus in principle capable of characterizing redshift distributions at a level approaching the requirements for future, Stage IV experiments. However, these estimated  characterization uncertainties all \emph{exclude} the effects discussed in \S \ref{sec:characterization}, whose impact can be several times larger.
\ifarxiv
\begin{marginnote}
\entry{Characterization of photometric redshifts in ideal conditions, by method}{\includegraphics[width=3.5cm]{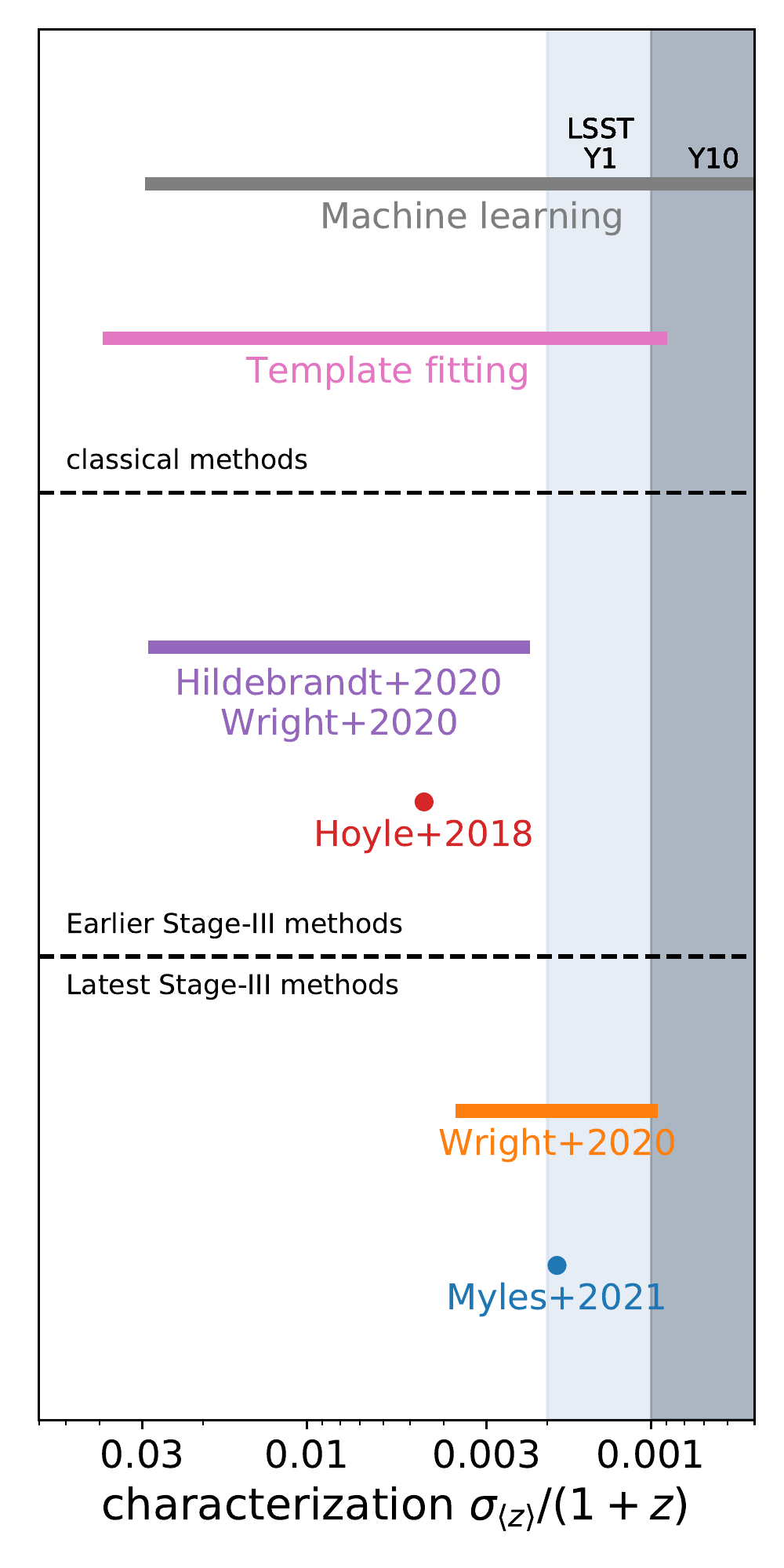}}
\label{fig:char_error_margin}
\end{marginnote}
\fi

\subsection{Bayesian Hierarchical Methods}
\label{sec:bhm}

Photometric redshift methods are necessarily hierarchical, as the term is used in the statistics literature; i.e., there exist at least two levels of parameters, one set describing the properties of an individual galaxy, and one set describing (either via a reference sample or sets of templates and priors) the distribution of properties of the ensemble of galaxies. Methods can be distinguished by how they treat the parameters of the latter sort. 
Most commonly at present, the model of the underlying population of galaxies remains fixed while photometric redshift inference is performed. One may choose a template prescription or train an empirical method based upon a sample of trusted redshifts and photometry, and then estimate the redshift of each target galaxy given that input set of information. Some methods allow a limited degree of feedback from the inference of redshift distributions to the parameters describing the galaxy ensemble; examples include the BPZ or ZEBRA methods described in \S \ref{sec:templates}, or the combination of SOMPZ, 3sDIR, and WZ used in DES Y3 \citep{Myles2021,Gatti2021}. 

Bayesian Hierarchical Methods take this idea to its limit by performing a simultaneous, joint Bayesian inference to constrain  parameters at both levels: i.e., determining posterior probability distributions both for the redshifts of individual galaxies and for parameters that describe the properties of the broader population of galaxies. In the context of photometric redshift estimation, such methods were first introduced by \citet{Leistedt2016}, who used a hierarchical model to constrain the underlying distributions of template types and redshifts while at the same time computing PDFs for the redshifts of each individual object using a mock data set. These underlying distributions correspond to the prior $p(z,t)$ used within a Bayesian photometric redshift method; by inferring $p(z,t)$ from the data itself, uncertainties in the prior can be propagated into the final redshift PDFs for each object. \citet{Leistedt2019} extended this method to also allow the set of rest-frame SED templates used to be modified via hierarchical inference. A different approach is taken by \citet{Sanchez2019,Alarcon2020}, who instead infer a prior probability distribution for the density of galaxies in the \emph{observed} high-dimensional color space, rather than in the space of intrinsic properties, via a hierarchical inference process. 

Another variant of hierarchical models, forward-modeling,  has also been successfully applied to data \citep[e.g.,][]{Herbel2017,Tortorelli2021}. In these methods, a parametric model for the galaxy population is used to simulate galaxy images and/or catalogs. The model parameters, along with other cosmological and nuisance parameters, are constrained via Markov Chain Monte-Carlo methods to resemble the observed distributions of galaxy properties, including the measured redshifts of objects with spectroscopy. In addition to constraints on model parameters, hierarchical methods can simultaneously provide photo-$z$ PDFs for individual objects or redshift distributions for ensembles of galaxies, marginalizing over the values of those parameters.  

Hierarchical and forward-modeling approaches are still at early stages of development, but hold significant promise for addressing many of the challenges discussed in this review. They can, in principle, overcome the limitations posed by incomplete training sets or inaccurate templates or priors, so long as the models used are sufficiently general to encompass the underlying reality without providing so much freedom that redshifts are poorly constrained.
\ifarxiv
\begin{marginnote}
\entry{Take-Away}{Methods that inform a model for the galaxy population with all collected data have promise for addressing current limitations of photo-$z$ algorithms.}
\end{marginnote}
\fi

\section{IMPROVING PERFORMANCE AND CHARACTERIZATION VIA SPECTROSCOPY}

\label{sec:spectroscopy}

Modern photometric redshift methods are dependent upon having a set of objects whose redshifts are securely known. Most directly, cosmological analyses of current-generation (Stage III) surveys have estimated redshift distributions of galaxy samples by simply using a weighted histogram of the redshifts in reference samples (cf.  \S \ref{sec:stage3}). In machine 
learning-based techniques, samples of objects with known redshifts and photometry provide the training data used to optimize algorithms for estimating photo-$z$'s.  Template-based 
methods are less directly dependent upon having redshift measurements available; however, the best-performing algorithms today use such samples to 
optimize the libraries of galaxy spectral energy distributions used to compute likelihoods \citep{Crenshaw2020}; to optimize photometric passband throughput 
curves and zero-points \citep{Ilbert2006}; to refine error models \citep{Brammer2008}; and/or to develop redshift priors \citep{Benitez2000}.  Sets of secure redshift measurements are also fundamental to testing for (and, if necessary, correcting) any systematic errors in photometric redshift measurements; if uncorrected, 
such errors can far exceed random errors in precision cosmology measurements \citep{DESC_SRD}.  
\ifarxiv
\begin{marginnote}
\entry{Take-Away}{There can be no photometric redshifts of high quality without spectroscopic redshifts of high quality \emph{and} quantity.}
\end{marginnote}
\fi

In this section, we describe the needs for external redshift measurements to improve photo-$z$ performance and characterize redshift distributions, and lay out the scope of the problem 
for future imaging surveys.  The limited availability of secure spectroscopic redshifts for objects as faint as those to which photo-$z$ algorithms will be applied in the near future may be a major stumbling block for photometric redshift methods. 

\subsection{The Twin Needs for Spectroscopy}

One application of spectroscopic samples is to develop (in the case of machine-learning methods) or 
optimize (for template-based methods) photometric redshift algorithms, {\it reducing random uncertainties in redshift estimates for individual objects}. In these cases, a set of objects with precision redshift measurements is used to improve the \textbf{performance} of algorithms.  This application of spectroscopy is referred to as ``training'' in \citet{Newman2015}.  

For template-based methods, when sufficiently large training samples are available, we should be able to refine the underlying model used arbitrarily well, in which case photo-$z$ errors should be determined only by photometric uncertainties and not be degraded by our limited knowledge of the intrinsic SEDs of galaxies, the system used to obtain photometry, etc.  For machine learning algorithms, a perfectly-trained algorithm will have fully determined the mapping from observed properties to redshifts; the performance of photo-$z$ algorithms will then be limited only by the information contained in the photometry itself.

However, for many precision studies (particularly in cosmology),  photometric redshifts for individual galaxies do not need to be highly precise to constrain the quantities of interest. Instead, it is only requirements on the \textbf{characterization} of redshift distributions that are extremely stringent, as discussed in \S \ref{sec:cosmo_requirements}.

This characterization is generally performed using samples of objects with spectroscopic redshift measurements, which may be used to estimate redshift distributions directly when weighted to match photometric samples (this application is referred to as ``calibration'' in \citealt{Newman2015}).  

In work to date, the same basic spectroscopic samples are frequently used both to train photometric redshift algorithms and to characterize their results \citep[e.g.,][]{Hikage2019,Wright2020,Myles2021}.  However, for upcoming imaging surveys, it will be very challenging to obtain low-error-rate sets of redshifts with minimal systematic incompleteness down to the magnitude limits of samples that will be used for cosmological measurements (cf. \S\S \ref{sec:incompleteness} and \ref{sec:outliers}).  Large, deep samples are already systematically incomplete at $i=22.5$, whereas Rubin Observatory cosmology samples will extend to $i=25$ or greater, and Roman Observatory will utilize deep IR-limited samples that are even more challenging spectroscopically.
In that case, small, deep but incomplete samples may be used to improve the performance of photometric redshift algorithms, but approaches that are less sensitive to incompleteness must be used for characterization (cf. \S\S \ref{sec:bhm}, \ref{sec:xcorr}, and \ref{sec:vision}); for instance, much larger but shallower samples of secure redshifts can characterize distributions by exploiting correlations from large-scale-structure \citep{Newman2008}.

We note that the term ``spectroscopy'' should be interpreted broadly in this section.  For improving photo-$z$ performance, at least, useful information may be obtained from extremely-high-quality, many-band photometric redshifts which are available in some limited areas of sky (as those from e.g.~\citealt{Laigle2016,Alarcon2021}).  However,  many-band photo-$z$'s exhibit much larger catastrophic outlier rates and redshift errors than higher-resolution spectroscopy does.  This is a consequence of the limited spectral resolution of the information available from many-band surveys and the poorer signal-to-noise compared to broadband imaging, as well as the limited deep data available for training the many-band photo-$z$'s. The lower robustness of many-band redshifts will generally limit their utility for precision characterization. Slitless and prism spectroscopy exhibit similar characteristics to many-band imaging, providing less-secure redshifts for larger samples derived from lower-resolution spectral information.

\subsection{Spectroscopy for Improving Photometric Redshift Performance}

For all training-based methods of determining photometric redshifts, it is necessary to have a set of objects for which both the 
properties that will be used to make predictions and the quantity one wishes to predict (in this case, the redshift) have been measured in the same way as for the galaxies to which algorithms will be applied.  

Since the relationship between color and redshift is complex, it requires many samples to fully map it. A machine 
learning algorithm trained with only a sparse set of objects with spectroscopic redshifts may have to rely on information from galaxies with very different properties (color, brightness, $z$, etc.) to predict the redshift of a given galaxy, and prediction errors will be correspondingly degraded.  This degradation will be even worse in regions of parameter space where training redshifts are unavailable, in which case algorithms will extrapolate (often nonlinearly) from objects with systematically different properties.  The left and center panels of \S \ref{fig:sparsesampling} illustrate the loss of information when training samples become sparse.

In contrast, given a very large and representative input sample the ability of a training-based 
algorithm to predict redshift will be limited only by the uncertainties in the fluxes provided in inputs and the intrinsic scatter in the mapping from noiseless photometry to 
redshift; at that point, results should not improve as sample sizes get larger.  However, how fast this transition occurs will depend upon the algorithm used to predict photometric redshifts, the quantity that is being estimated (e.g., the full $p(z)$ as opposed to point estimates of redshift), and the photometric data it is estimated from (cf. \S \ref{sec:morphology}).   Methods which effectively interpolate between members in the training sample in a manner which takes more advantage of the underlying, simpler structure of the distribution of galaxies in the space of 
rest-frame spectral energy distributions should be more effective at predicting redshifts for galaxies outside their training set.  Scalings of photometric redshift 
errors when several standard machine learning algorithms are applied to the mock LSST dataset of \citet{Graham2018} are illustrated in Figure 1 of \citet{Newman2019}.  Errors scale with training set size approximately as $\sigma_z \sim \sigma_{\infty}(1 + a N_{\rm training}^{-0.4})$, where $\sigma_z$ is the RMS scatter between a
photometric redshift estimate and true redshift, $\sigma_{\infty}$ is the scatter that would be obtained with an infinite, perfect training set, $N_{\rm training}$ is the 
number of objects in the training set, and $a$ is a constant which depends upon the algorithm used.  The reason for the observed power-law exponent remains unknown.  In machine learning methods applied to this dataset to date, improvements in errors generally become slow beyond sample sizes of 20-30,000.

Deep-learning-based algorithms which utilize pixel-level information in galaxy images to predict redshift require larger training samples to reach their optimum performance; in the most recent algorithms, the core scatter does not plateau until training samples comprise $>100,000$ objects, and catastrophic outlier rates continue to fall even for training samples of $>400,000$ galaxies \citep{Dey2021b}.  However, such methods are unlikely to yield large photo-$z$ performance improvements for the faint, poorly-resolved galaxies from next-generation surveys whose redshifts are most difficult to measure spectroscopically.  If we therefore set aside the ambition to apply deep learning methods at faint magnitudes, a practical goal for near-optimal performance from near-future photo-$z$ algorithms would be to obtain redshifts for a sample of 20,000-30,000 objects in total, spanning the flux range of the samples that will be used for cosmological studies.
\ifarxiv
\begin{marginnote}
\entry{Take-Away}{Roughly 30,000 deep spectroscopic redshift measurements are needed to optimize \emph{performance } of photo-$z$ algorithms in near-future surveys.}
\end{marginnote}
\fi

\subsection{Spectroscopy for Characterizing Redshift Distributions}

\label{sec:spec_characterization}

As we shall see, if our goal is to characterize redshift distributions rather than optimize the performance of photometric redshift algorithms, one coincidentally obtains a very similar estimate of the sample size required.
We note, however, that characterization of redshift distributions for precision cosmology measurements with future imaging surveys will require spectroscopic samples with very high redshift success rates (99\% or higher);  very low incorrect-redshift rates ($<0.1\%$); \textbf{and} minimal sample/cosmic variance (cf. \S \ref{sec:sample_variance}), in order to ensure validity of results (assuming  characterization through a simple reweighted redshift histogram, as is done in current analyses).  

No deep redshift survey to date has approached the required levels of completeness (i.e., the fraction of targets that yield extremely-secure redshifts) needed for direct characterization of photometric redshifts for future cosmology surveys, as will be discussed in \S \ref{sec:incompleteness}. However, new instruments and strategies could change this situation, so it is desirable for samples designed to improve photometric redshift performance to be also capable of fulfilling our characterization needs if the necessary high success rates are achieved. If the redshift estimation process including its failure modes can be forward-modelled accurately, higher failure rates may still be tolerable for characterization purposes. 

A number of theoretical works have explored strategies for the characterization of redshift distributions, and have each determined that sample sizes of 20-30,000 should be sufficient whether simple Gaussian errors or more complex scenarios including catastrophic outliers are considered \citep{Ma2008,Bernstein2010,Hearin2010}.  The fact that improvements in errors for machine learning methods begin to become slow beyond this sample size appears to occur purely by  coincidence.
\ifarxiv
\begin{marginnote}
\entry{Take-Away}{Only if the deep spectroscopic samples collected for training reach unprecedentedly low or well-understood incompleteness and outlier rates, or if characterization methods greatly improve, can the stringent requirements for future dark energy studies be met.}
\end{marginnote}
\fi

\begin{figure}[t]
\centering
\includegraphics[width=\textwidth]{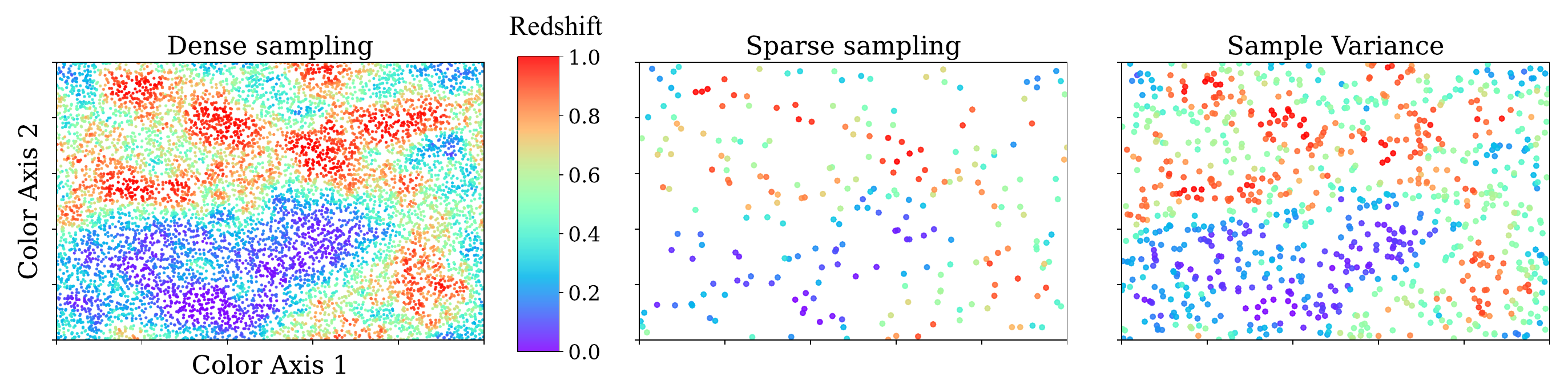}
\caption{Illustration of the impact of limitations on spectroscopic training sets on the ability to map out relations between color and redshift.  In this toy model, the X and Y coordinates could correspond to measures of galaxy colors in different bands or to a dimensionality-reduced transformation of color space into two dimensions (such as that produced by a self-organizing map).  The relationship between galaxy redshift and its position in this color space is determined by evaluating a Gaussian random field; redshifts ranging from 0 to 1 are mapped onto colors ranging from indigo at the lowest $z$ to red at the highest, as shown by the color bar.  When large spectroscopic training sets that densely sample the color space are available, as in the panel at left, machine learning methods can easily predict the redshift at any location in this space by interpolation or determination of the distribution of training redshifts in a local neighborhood.  When spectroscopic training samples are sparse, as in the middle panel, the relationship of color to redshift will be only poorly determined and photo-$z$ accuracy will be degraded. Finally, in the right panel we emulate the effect of sample variance by selecting galaxies at a rate which is a periodic function of $z$. This causes some redshifts to be overrepresented in training sets and others to be underrepresented, resulting in systematic gaps in the coverage of color space as well as biased characterization of $z$ distributions due to some redshifts being favored compared to others.
}
\label{fig:sparsesampling}
\end{figure}

An additional consideration when designing spectroscopic samples for improving the performance and particularly characterization of photo-$z$'s is sample (or ``cosmic'') variance: the variation in density from one region of the universe to another due to the underlying matter density fluctuations \citep{Cunha2012}.  Deep spectroscopic surveys generally target only one or a few fields each covering only a very small area of sky; as a consequence, the volume at a given redshift is low, and density fluctuations are correspondingly large.  As a result, the redshift distribution in each field will exhibit large fluctuations (much larger than would be expected from Poisson statistics), with some redshifts being over-represented and others under-represented.  This effect is illustrated in the right panel of \autoref{fig:sparsesampling}.  Furthermore, the \textit{types} of galaxies will also vary, as the most massive quiescent galaxies will only be found in extreme overdensities, whereas bluer galaxies will comparatively favor underdensities. This will affect the characterization of any photo-$z$ method, whether training-based, direct, or Bayesian hierarchical. 

The effects of sample/cosmic variance can be mitigated in a variety of ways.
One option is to obtain spectroscopic data sets over a larger number of small but widely-separated fields \citep{Cunha2012}, as in that case the fluctuations in each field will be independent and tend to average out.  
\citet{Newman2015} propose a baseline survey for future dark energy experiments in which spectroscopy is obtained over 15 widely-separated, 20 arcminute diameter fields.  
This proposed design would produce similar total fluctuations in density as the C3R2 survey \citet{Masters2019}, 
but would require only 1.3 square degrees of sky to be sampled, rather than the $>6$ square degrees (spread over six fields) planned for the latter. 

A second option is to use spectroscopy to characterize the color-redshift relation in a photometric space that has a large number of bands \citep{Gruen2017}, possibly larger than the wide field survey itself. When the photometry alone constrains redshift well, sample variance largely manifests as a fluctuation of galaxy density in photometric space. It can thus be mitigated by reweighting according to the density of purely photometric galaxies observed in those same photometric bands, reducing the need for spectroscopic data \citep{Buchs2019}, so long as the volume and number of objects surveyed is sufficient that all cells in the photometric space are well-characterized. At the same time, as the number of photometric bands increases, spectroscopic incompleteness should manifest as a variation of success rates across photometric space \citep{Masters2015} and its impact can thus be better isolated and potentially reduced.

\section{MAJOR CHALLENGES FOR NEXT-GENERATION PHOTOMETRIC REDSHIFTS}

\label{sec:openissues}

A great deal of progress has been made on improving both the performance and characterization of photometric redshift algorithms in recent years.  However, upcoming large imaging surveys will only be able to reach their full promise if further advancements are made on a variety of fronts.  In this section, we focus on areas where current algorithms fall short and the potential to make gains is clear, serving as possible areas of focus for research in the near term. We do not discuss one open issue, the measurement of photometric redshifts for objects with (often time-variable) emission from an active galactic nucleus, as that is reviewed in detail within \citet{Salvato2019}. 

We first consider issues that primarily affect the performance of photo-$z$ algorithms, and then describe potential sources of problems that primarily influence characterization. As future imaging-based probes of cosmology will require exquisite calibration of redshift distributions, the latter section will focus primarily on characterization requirements for such analyses. The needs for galaxy evolution work will be comparatively easier to meet.

\subsection{Challenges for Improving Photometric Redshift Performance}

\subsubsection{Interpreting ``Probability Distributions'' from Photometric Redshift Algorithms}
\label{sec:pdf}

The probability distribution functions (PDFs) produced by photometric redshift codes often do not meet either frequentist or Bayesian expectations (cf.~\autoref{pz}). A frequentist expects that the true redshift of an object, as measured e.g.~spectroscopically, does lie between $z_0$ and $z_1$ in a fraction of trials equal to $\int_{z_0}^{z_1} p(z)\,\mathrm{d}z$. Alternatively, the criterion can be expressed via the Probability Integral Transform (PIT): the distribution of the values of the cumulative distribution function for a given object, evaluated at its true redshift, should be uniform between 0 and 1. Any inference that depends upon the assumption that PDFs accord with the expectations of frequentist statistics will be biased if that definition is not fulfilled.

This failure mode is wide-spread. \cite{Dahlen2013} found that out of 11 different photometric redshift codes run on data from the CANDELS survey, all of which delivered point estimates with comparable scatter from spectroscopic redshifts, the fraction of objects whose true redshift fell within the 68.3\% confidence region of their PDF ranged from 2.5\% to 89\%, versus the expected 68.3\%. On the other hand, between 2.9\% and 97\% of the time the true redshift fell in the 95.4\% confidence region (and even when a code came close for the 68\% region the fraction within the 95\% region was badly off, as well as vice versa). 

\begin{figure}[htp!]
{\includegraphics[width=0.95\textwidth]{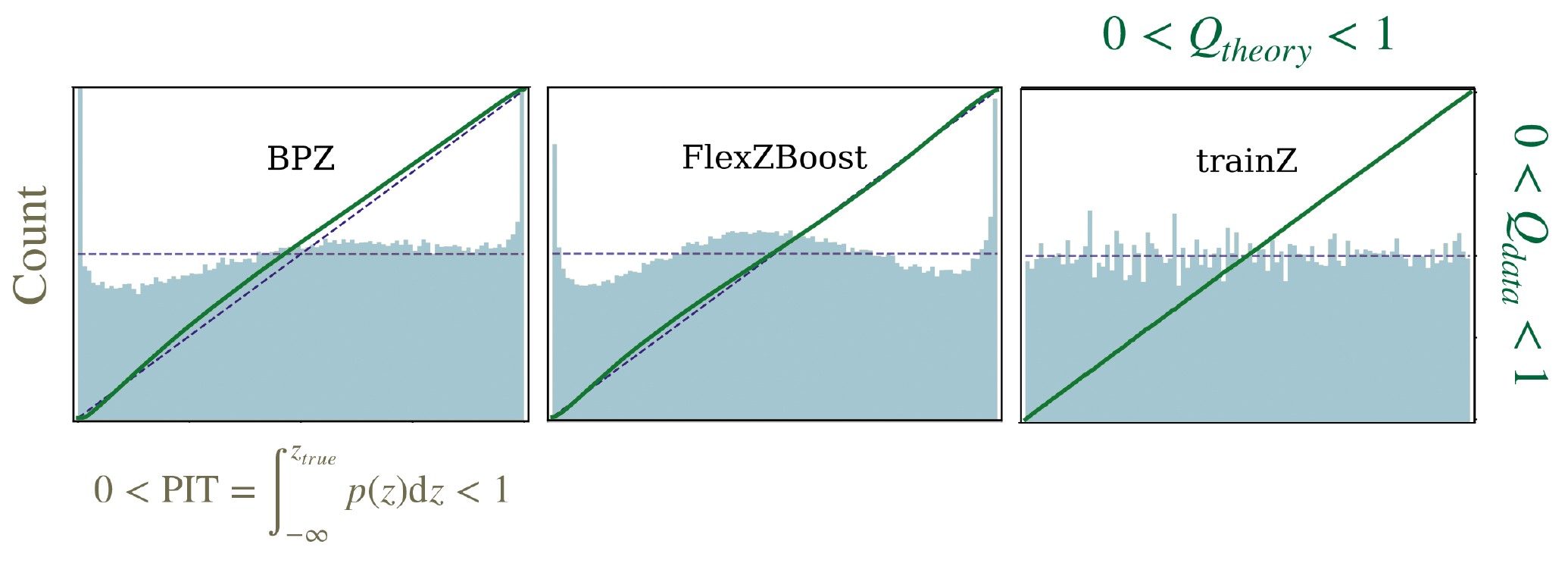}}
{\caption{Photometric redshift codes which deliver good performance generally fail to produce results which match the frequentist definition of a PDF. Plotted are results for the best-performing template-based code in the controlled tests of \citet{Schmidt2020}, BPZ \citep{Benitez2000}; the best-performing training-based code, FlexZBoost \citep{Dalmasso2020}; and a control method developed for that work which has pessimal performance at predicting redshifts for individual objects but delivers well-characterized PDFs when given ideal training sets, \texttt{trainZ}.  Blue histograms (corresponding to the gray axis labels) show the distribution of the Probability Integral Transform (PIT) statistic in the tests by \citet{Schmidt2020}, which should follow a uniform distribution for a proper frequentist PDF.  Green curves (associated with the green axis labels) show the Quantile-Quantile (QQ) plot for each method, which shows the fraction of the time that the actual redshift of an object is below a given quantile in the redshift PDF ($Q_{data}$) as a function of the chosen quantile ($Q_{theory}$); ideally, this should fall along the dashed diagonal unity line. Figure adapted with permission from \citealt{Malz_thesis}.
}
\label{fig:DC1}}
\vskip-0.15in
\end{figure}

\citet{Schmidt2020} performed a test of photometric redshift codes on simulated Rubin Observatory LSST data, in which a large and perfectly representative training set was provided for training-based algorithms, and the actual template sets used to generate photometry were made available for template-based methods. Even in this best-case scenario, all current photo-$z$ codes fell short at providing accurate PDFs in comparison to a control method. The latter, named \texttt{trainZ}, was designed to return a maximally broad yet perfectly frequentist $p(z)$, identical for each galaxy in the sample, corresponding to the histogram of the redshifts of all galaxies from the representative training set. For any other code, the deviations of the PIT distribution from the expected uniform distribution were larger by an order of magnitude or more than those of \texttt{trainZ}, as illustrated in \autoref{fig:DC1}; this degree of inaccuracy would greatly compromise cosmological inference from future surveys.  Conversely, the redshift performance for individual objects from \texttt{trainZ} would be badly insufficient for most analyses.

The reasons for these shortcomings can be either errors in the redshift PDF estimation process, among them the issues discussed in this review, or incorrect interpretation of the produced outputs of a photo-$z$ code. The latter can occur if the output of a photometric redshift code is not actually \emph{intended} to meet the frequentist definition. 
The Bayesian definition of probability as a degree of belief that a value lies in some range (cf. \S \ref{sec:definitions}) is harder to test quantitatively. However, even codes which are nominally Bayesian sometimes fail to properly marginalize probability over parameter uncertainties (e.g., template types), causing them not to match any statistical definition of a PDF.

As described in \S \ref{sec:templates}, template-based methods typically compute posterior probability distribution functions for galaxies via a simple application of Bayes' theorem. When this is done with a set of templates and priors for each that are matched to the true distribution of galaxy SEDs, with proper marginalization over all templates and redshifts and accounting for selection effects, the output should fulfill the frequentist definition of a PDF. However, when the output is calculated from (for instance) the likelihood of the best-fitting template at a given redshift, rather than marginalizing over templates, it is incorrect to interpret outputs as either frequentist or Bayesian PDFs. Even if one were to correctly marginalize over an appropriate model for templates and priors that is itself uncertain, the output could be correct in a Bayesian sense, but would not match the frequentist definition. For instance, when a set of discrete templates that are not evenly distributed within the underlying parameter space are all given equal prior probability, the result corresponds to an (incorrect) implicit prior on template type, with regions of parameter space having more templates getting extra weight in computing the PDF. It is even less clear how to properly perform marginalization when likelihoods are computed between the observed colors and the best-fit linear combination of templates, which is a common procedure. 

Although they generally do not compute either a posterior or likelihood directly, machine learning-based methods still often output a redshift PDF, potentially incorporating both measurement uncertainties and factors that contribute to the spread in redshift at fixed color. A variety of techniques exist for this \citep[see, e.g.,][]{Sadeh2016,QuantileRegressionPZ} for which there often is no expectation that they meet the frequentist criterion. Even if the loss (i.e., the quantity that a machine learning algorithm is optimized to minimize) incorporates measures of PDF-ness, that loss will be minimized across the entire training set, but may yield biases for specific subsets of the training domain even if the distribution over the full sample appears consistent with expectations. Unlike typical template-based or machine learning approaches, the techniques employed in current-generation (Stage III) analyses are constructed to return frequentist PDFs if a set of underlying assumptions are valid \S \ref{sec:stage3}, but are not necessarily consistent with Bayesian approaches.

\citet{Bordoloi2010} addressed the mismatch between photo-$z$ PDFs and the frequentist definition by remapping the PDFs of all objects according to the global PIT distribution for galaxies with known redshifts, redistributing probability for each PDF such that the PIT distribution will be uniform for the overall spectroscopic sample by construction.  This procedure corrects the individual PDFs accurately, \textit{if}  the set of objects with known redshifts is representative of the set of objects to which the corrections are applied, and if the degree of mis-specification of PDFs does not depend on object properties (e.g., brightness, intrinsic SED, or redshift).  Those assumptions do not hold in general: the spectroscopic samples used to retune the PIT will in general be biased and incomplete compared to photometric samples, and (for instance) the contributions of incorrectly estimated photometric errors and of incorrect handling of templates will have different impact for galaxies of different brightnesses or different restframe colors. In such scenarios, the overall PIT distribution for the entire spectroscopic sample may match the ideal case even while the PDFs for particular subsets of the sample are poorly calibrated (cf. \citealt{Zhao2021}).  

This can be addressed by predicting the PIT distribution at all points in parameter space via machine learning regression \citep{Zhao2021} and then correcting each object's PDF according to the PIT prediction evaluated at its location. This procedure has produced good results in initial tests \citep{Dey2021a}.
If spectroscopic samples are biased this procedure may still fail to yield accurate PDFs, however; it would be even better to develop a fundamental understanding of \textit{why} current codes fall short of the ideal, and to then develop methods constructed such that they provide well-defined PDF outputs.

\ifarxiv
\begin{marginnote}
\entry{Take-away}{Current photo-$z$ codes optimized for predicting redshifts of individual objects fail to produce outputs that fulfill the frequentist definition of a PDF; Bayesian methods have also fallen short of the ideal.}
\end{marginnote}
\fi

\subsubsection{Combining Results from Multiple Methods}

It has repeatedly been found that better photometric redshift performance can be attained by combining information from multiple photometric redshift codes than by considering only results from one, even when the same input data is used for each.  For instance, \citet{Gorecki2014} found that neural-network-based and template-based codes exhibited catastrophic redshift errors for \textit{different} objects.  As a result, requiring consistency between the results of two very different codes can produce much better performance than samples where only one code is used in isolation. More strikingly, \citet{Dahlen2013} found that even when only template codes are considered, combining results can improve performance, likely related to the use of complementary templates between the codes. In that work, the median redshift prediction from the five best-performing codes yielded both smaller scatter and smaller outlier rates than any single code achieved.  It is clear that with current methods, no single photo-$z$ code performs best for all objects, allowing performance to be improved if the strengths of different codes can all be exploited.

\begin{figure}[htp!]
{\includegraphics[width=0.95\textwidth]{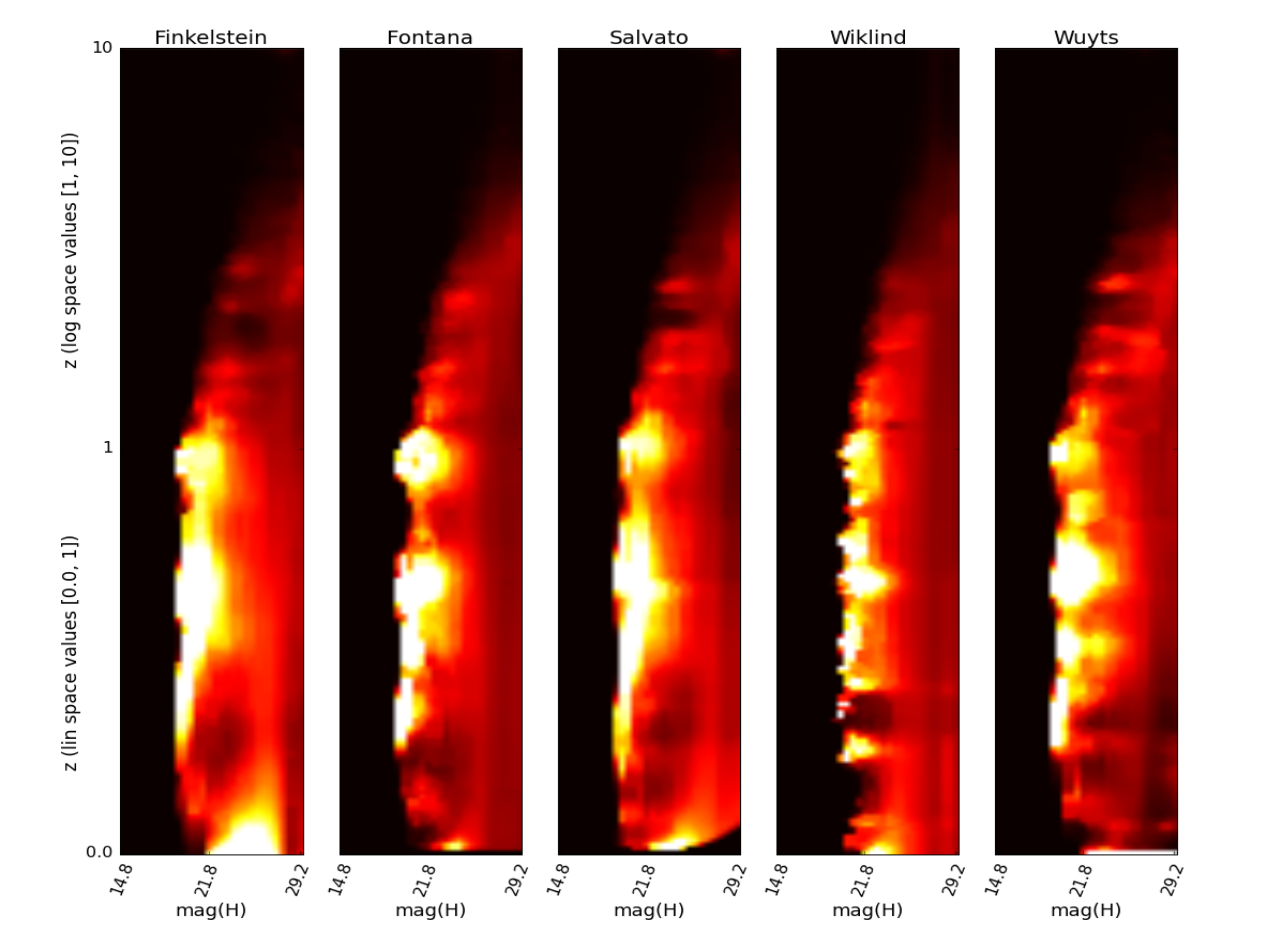}}
{\caption{Even photo-$z$ codes which yield excellent performance for galaxies with known spectroscopic redshifts clearly disagree when applied to faint objects.  Each panel shows averages of the photometric redshift PDF for objects in bins of $H$-band magnitude for five different template-based \photoz\ implementations applied to data from the CANDELS survey \citep{Grogin2011}, with lighter colors corresponding to higher average probability.  The resulting redshift distribution estimates are plotted using a linear y-axis scale for $0 < z_{phot} < 1$ and a log scale  for $1 < z_{phot} < 10$.  At the lowest redshifts, differences in the assumed priors produce dramatically different distributions.  At higher $z$ the codes do not agree on the redshifts at which overdensities of galaxies are found (which are visible as horizontal features in each panel), despite all exhibiting small scatter when compared to available samples of spectroscopic redshifts.  Figure provided by Dritan Kodra (priv.~comm.).
}
\label{fig:z-H_heatmap}}
\vskip-0.15in
\end{figure}

It is straightforward to apply a median or some other algorithm for defining a consensus value to combine point estimates for the redshift of an object.  However, it is also possible to combine posterior PDF estimates from different codes to produce a single PDF that incorporates information from each. The fact that each code's estimate is influenced by its particular (perhaps inaccurate) choices can at the same time be an opportunity for more robust results when multiple results are combined. As can be seen in \autoref{fig:z-H_heatmap}, a variety of template-based codes which all yield excellent performance when tested on galaxies with spectroscopic redshifts \citep{Kodra2019} yield PDFs which correspond to very different redshift distributions from each other for faint objects.  

\citet{Dahlen2013} presented a ``hierarchical Bayesian'' method for combining such disparate PDFs based upon techniques employed in \citet{Licquia2015};  similar methods were introduced in other contexts in \citet{Press1997}, \citet{Newman1999} and \citet{Lang2012}. This algorithm calculates a posterior probability distribution for the redshift under the assumption that an unknown fraction $f_{bad}$ of the PDFs being combined are false and contain no information.   A free parameter, $\alpha$, describes the degree of covariance between the outputs of different codes.   $\alpha=1$ corresponds to the case where there is no covariance and PDFs may be treated as statistically independent, so that the posterior PDF is the product of the input PDFs; $\alpha=1/N$, where N is the number of results combined, would instead correspond to the case where all estimates are completely redundant.

The results of the hierarchical Bayesian combination will be tighter than the input PDFs where the results all agree, so long as $\alpha > 1/N$; however, for objects for which the input PDFs disagree, the posterior PDF that results will be broader than any individual input PDF, reflecting this uncertainty.  For the CANDELS test data, the most probable value of $\alpha$ (when marginalizing over all other parameters) was $\alpha=1/2.1$; i.e., even when provided identical input photometry, the PDFs from different template-based codes behave somewhat more like independent estimates than redundant ones.

Using a hierarchical Bayesian combination of input PDFs will tend to complicate the deviations of the posteriors from meeting the statistical definition of a PDF, as different codes will contribute differently for different objects, and the choice of how to treat 'noninformative' results can substantially change the output PDF: if the fit value of $f_{bad}$ is non-zero, and if bad measurements are treated as completely noninformative (corresponding to a uniform PDF across all redshifts), the hierarchical Bayesian result will have non-zero probability at all $z$.    Kodra et al. (in prep.; cf. \citet{Kodra2019}) introduces a new method to combine photo-$z$ PDFs which avoids this problem: the use of Fr\'echet means.  Essentially, given a set of photo-$z$ PDFs for a given object, the Fr\'echet mean PDF will be the one with the smallest total distance from all the other PDFs considered, integrated over all redshifts.  It is thus a functional analog of the median (which minimizes the sum of absolute values of the deviations from a set of data values) or the arithmetic mean (which minimizes the sum of the squares); similarly, Kodra et al. find the PDF for each object which minimizes the total $\textsl{l}^1$ or $\textsl{l}^2$ norm across the results from all codes.  Since the Fr\'echet mean PDF must be one of the inputs, if the input PDFs all fulfill the frequentist definition of a PDF, so will the result of this process.  Kodra et al. have tested this method using independent sets of spectroscopic or grism redshifts in the CANDELS fields, and found that the Fr\'echet mean of the four best-performing codes that provided PDFs yielded results that come closer to meeting the frequentist definition of a PDF than the outputs of any individual code.

The fact that it is possible to realize gains in photo-$z$ perfomance by combining results from multiple methods suggests that further improvements in photo-$z$ techniques are clearly possible; ideally, only one code would be needed to produce the best results. Lacking such an ultimate code, the question remains how best to do that combination.  Ultimately, this is a choice of what to optimize for: we may desire methods that minimize the typical deviations of point estimates from their true values, reduce the frequency of outliers, present us with a conservative range of possibilities for the redshift of an object, or provide PDFs that best fulfill the statistical definition.  

Each of these goals would lead us to a different method of combination.  As long as we can define a loss function that we wish to minimize, this can be treated as a machine learning problem. There exist a variety of ``ensemble'' machine-learning methods that take as input the predictions from separate machine learning models and then output a result based on those inputs which minimizes the desired loss.  As the concept is general, the final prediction could come from simple regression methods, decision trees, or deep neural networks and still fit within this framework.  This remains an active area of machine learning research, but to date such methods have only begun to be explored in astrophysics \citep{Vilalta2017}.  

\ifarxiv
\begin{marginnote}
\entry{Take-away}{Because photo-$z$ codes make analysis choices that fall short in different ways, combining PDFs from multiple methods can provide high-performing results, but better techniques for combination are still needed.}
\end{marginnote}
\fi

\subsubsection{Improving Joint Inference of Redshift and Physical Properties from Photometry}

\label{sec:phys_props}

The observed colors of a galaxy depend not only upon its redshift, but also on its intrinsic physical properties such as its history of star formation (commonly summarized via the total stellar mass, current star formation rate, and stellar metallicity) and the amount and nature of dust attenuation along the line of sight.   Some template-based codes can exploit this fact to determine multi-dimensional posterior probability distributions for redshift and a variety of intrinsic galaxy properties simultaneously (e.g., BEAGLE \citep{Chevallard2016} and BAGPIPES \citep{Carnall2018}; see also \cite{Acquaviva2011}). We can think of these codes as not producing just a PDF for redshift, $p(z\, | \,{\rm photometry})$, but rather for the combination of redshift and a vector of galaxy properties, \textbf{G}: i.e., $p(z, \mathbf{G} \,|\, {\rm photometry})$.  A good example of such a multidimensional PDF is presented in Figure 8 of \cite{Chevallard2016}.

Such a joint probability distribution can encode everything that we can infer about the nature of a given galaxy from its observed photometry.  However, there are two current limitations on such analyses: our limited ability to predict the observed spectrum of a galaxy from its physical properties, and the extensive run time required per object for such an analysis.  We consider these separately.

Inference of physical properties from the observed SED of a galaxy will generally rely on population synthesis models, which predict the combined spectrum of a population of stars of varying masses and formation times \citep{Conroy2013}.  Given an initial mass function of stars and star formation history, the distribution of intrinsic properties of stars that should exist at a given time can be predicted from theoretical isochrones, and then the spectra of the resulting set of stars can be combined to produce a prediction for the overall SED.  We can compare the SEDs predicted for different star formation histories to the observed properties of a galaxy in order to constrain the values of its intrinsic parameters (total stellar mass, star formation rate, etc.) in either a likelihood or Bayesian framework.

However, the predictions from population synthesis models will only be as accurate as the inputs to those models are.  Observed isochrones have been characterized well and the initial mass function has been studied extensively in conditions that can be explored within our Galaxy (e.g., higher-metallicity, young stellar clusters and lower-metallicity, old globular clusters); however, extragalactic objects may fall within other regions of parameter space.  Similarly, the libraries of observed stellar spectra that may be used for population synthesis are limited both in their sampling of different stellar types -- including some populations of rare but luminous stars that can have large impact on the SED \citep{Conroy2010} -- and in their wavelength coverage.  Synthetic stellar spectra can alternatively be used, but the required modeling remains a difficult challenge;  empirical and theoretical stellar spectral libraries produce differing results \citep{Coelho2019}.  

These issues can all cause systematic biases in the inference of physical parameters from observed photometry.  However, even if population synthesis models were perfect, the computational complexity of determining the multi-dimensional joint probability distribution of galaxy properties and redshift is daunting.  In general, sampling-based techniques (Markov Chain Monte Carlo [MCMC] or related algorithms) are used to characterize distributions over this parameter space efficiently. Even so, for each object, complex population synthesis models must be evaluated many thousands of times in order to characterize the joint posterior probability distribution $p(z, \mathbf{G} \,|\, {\rm photometry})$.  With current computational resources, it would be infeasible to apply such methods to the billions of objects that will be cataloged by upcoming surveys.

There are two potential routes to make progress on this problem: we can either speed up sampling-based inference, or bypass it entirely.  Methods have already begun to take advantage of newer sampling algorithms; e.g., BAGPIPES utilizes the MultiNest nested sampling algorithm, which performs better than MCMC for degenerate or multimodal probability distributions as are commonly encountered in this application \citep{Feroz2009}.  Still greater speed improvements may be possible by substituting a deep neural network-based emulator trained to match population synthesis model calculations in place of new calculations at each sampling step.  Such emulators can take extensive time to train but evaluate extremely rapidly \citep{Kasim2020}, making them well-suited for applications where large numbers of objects will be analyzed.  

Still greater speed-ups would be possible if we avoid sampling at all: deep neural networks could instead be utilized to predict physical parameters from the observed SED directly.  This application would be similar in spirit to past work which used deep learning methods to fit models trained from simulations to strong lens observations \citep{Hezaveh2017} or gravitational wave signals \citep{George2018}.  Uncertainties on the physical parameters could then be determined by measuring the derivatives of the likelihood around the point in parameter space predicted by the machine learning analysis; neural network-based emulators of population synthesis models could be used to calculate those derivatives rapidly.  This speed advantage would come at a cost: the derivatives about the peak would provide a local approximation to the likelihood surface around it, but would not capture the effects of parameter degeneracies or secondary maxima which often occur in the inference of galaxy parameters.  Nevertheless, if we wish to characterize the physical parameters for all objects from upcoming surveys, this may prove to be the most feasible option for the near future.

\ifarxiv
\begin{marginnote}
\entry{Take-away}{Photometry can be used to jointly constrain many galaxy physical parameters along with redshift, but this will be computationally prohibitive to perform with the large samples from upcoming surveys unless methods are improved.}
\end{marginnote}
\fi

\subsubsection{Storing Multi-Dimensional Probability Distributions}

Characterizing joint probability distributions for both redshift and galaxy properties -- $p(z, \mathbf{G} \,|\, {\rm photometry})$, as we defined it in \S \ref{sec:phys_props} -- poses difficulties that go beyond just computing the PDF. If methods for measuring these joint distributions will be applied to the samples of billions of objects that will be cataloged by upcoming surveys, it quickly becomes infeasible to store the results directly.  While a variety of methods have been produced to reduce the storage needs for one-dimensional photometric redshift probability distributions \citep{Carrasco2014,Malz2018}, those methods break down in multiple dimensions. If we consider a five-dimensional grid of redshift, stellar mass, specific star formation rate (i.e., star formation rate per unit mass), metallicity, and dust extinction -- a standard set of parameters to estimate from photometry -- using a 100-element grid for each dimension would mean that $10^8$ floating point numbers are needed to store $p(z, \mathbf{G} \,|\, {\rm photometry})$ \textit{for each individual object in the catalog}, corresponding to nearly an exabyte of storage per billion objects. This level of storage would cost millions of dollars per month to host at 2021 pricesAdding additional parameters (e.g., to allow for variation in dust attenuation curves or greater diversity in star formation histories) would only make this problem worse.

It is clear that alternative methods are needed if such multi-dimensional posteriors are desired for large samples of objects.  One option is simply to store a limited number of samples from the posterior PDF for each object.  These samples would include the effects of all covariances between parameters, and can be used  to construct aggregate distributions in parameter space by combining the samples from all objects of interest (though it is worth noting that to predict aggregate distributions one should combine likelihoods, not posteriors; cf. \citet{Malz2021}).  This option is quite inexpensive to store, requiring $N_{samples} \times N_{properties}$ of numbers to be stored per object, where $N_{samples}$ is the number of samples from the posterior PDF that are stored (e.g., 10 or 100), and $N_{properties}$ is the number of different properties recorded for each sample (redshift included).  However, a limited set of samples will provide correspondingly limited fidelity in describing the overall posterior probability distribution for individual objects.

A second option is to be able to calculate posterior PDFs so quickly that storage is not needed.  This would require many orders of magnitude of speed-up compared to current algorithms.  A promising option may be to develop deep learning-based emulators of the entire posterior fitting process: although such methods would require millions of examples and considerable CPU resources to train, once that is completed they would take minimal time per object to run.
\ifarxiv
\begin{marginnote}
\entry{Take-away}{The cost of storing joint many-dimensional  PDFs for large samples would be prohibitive with current methods.}
\end{marginnote}
\fi

A third possibility would be to exploit the fact that, while the joint distribution of redshift and galaxy properties is broad, in many cases (e.g., specific star formation rate or mass-to-light ratio) the distribution of a property \textit{conditioned on the redshift and observed photometry} is quite narrow; that is, if one knows (or assumes) the value of the redshift, the quantity of interest may be determined from that value and the observed photometry with only small scatter.  In that case, we can store a one-dimensional PDF for redshift, $p(z) | {\rm photometry}$, as well as a compact description for $p(\mathbf{G}\, |\, z, {\rm photometry})$; the product of those two probability distributions will be the $p(z, \mathbf{G} \,|\, {\rm photometry})$ we desire.  As an example, if the conditional distribution at a single $z$ is well-described by a multi-dimensional Gaussian in parameter space, we might store the coefficients of polynomial fits to the parameters of that Gaussian as a function of $z$; that could be combined with $p(z) | {\rm photometry}$ to reconstruct the full multi-dimensional PDF.  The most simplified version of this would be to only store the peak position and derivatives about the peak (or, equivalently, the best-fit solution and the covariances between all parameters), as would result from the deep learning-based inference of properties described in \S \ref{sec:phys_props}.  If we wish full, high-dimensional posteriors to be stored for large numbers of objects (and not only samples from those posteriors), though, new methods will be necessary.

\subsubsection{Incorporating Morphological Information} 

\label{sec:morphology}

Most photometric redshift algorithms only utilize integrated measurements of flux or color as inputs.  However, well-resolved galaxy images contain more information than just flux.  For instance, a galaxy's morphological type exhibits strong correlations with its restframe spectral energy distribution; e.g., elliptical galaxies in the local universe exhibit de Vaucoleurs-like light profiles and red restframe colors characteristic of old, passively-evolved stellar populations.  If a spiral galaxy's image is well-resolved, the redder characteristic colors and stronger spectral breaks exhibited by its older bulge can provide increased information for a photometric redshift algorithm to exploit, even if that component is dominated by the younger, blue disk in integrated colors.  

Additionally, the observed angular size of a galaxy of fixed physical size will vary quickly with redshift at low $z$, though only slowly at high $z$; as a result, size information can constrain the possible redshift of a galaxy (e.g., a galaxy observed to be several arcminutes across could not plausibly be at a high redshift). Furthermore, the observed bolometric surface brightness of a galaxy is dimmed by a factor of $(1+z)^4$ compared to its restframe value \citep{Tolman1930,Hubble1935}.  As a result, at higher redshifts the combination of galaxy brightness and size could provide additional information about redshift; this has been exploited for photometric redshift applications previously \citep{Stabenau2008}.  However, galaxy evolution could easily obscure the surface brightness signal, making it difficult to exploit.  

\begin{figure}[t]
\centering
\includegraphics[width=0.85\textwidth]{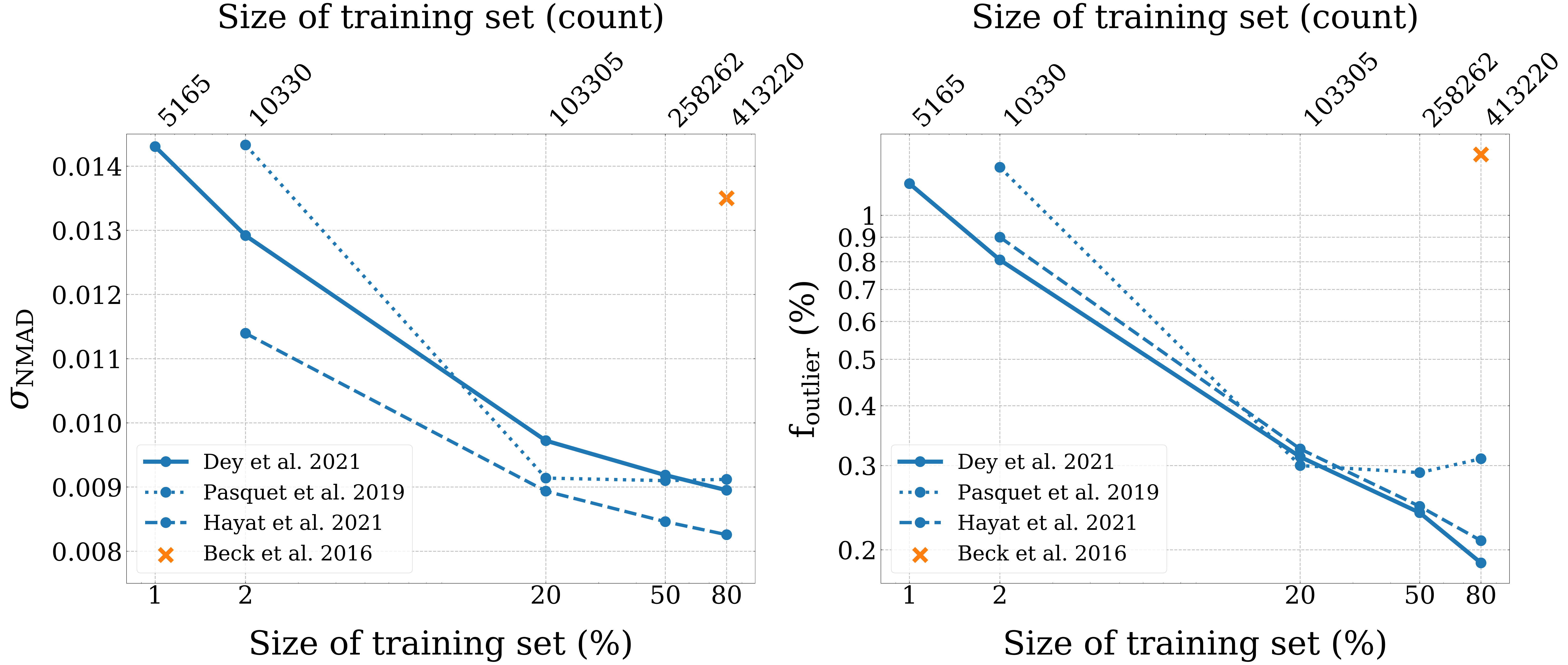}
\caption{An example of the improvements in photometric redshift performance that incorporation of morphological information can enable.   Blue curves show the results from the deep neural network-based methods applied by \citet{Pasquet2019}, \citet{Beck2016}, and Dey et al. (in prep.), which use galaxy images as inputs, for training sets of different sizes; the red symbol shows the performance from the algorithm of \cite{Beck2016}, which employs only galaxy magnitude measurements to predict redshift.  When large training sets are available, morphological information can be exploited to yield better performance, with smaller normalized median absolute deviation between photometric and spectroscopic redshifts, $\sigma_{\mathrm{NMAD}}$, as shown in the left panel, as well as smaller catastrophic outlier rates, $\mathrm{f}_{\mathrm{outlier}}$, as shown at right.  Figure provided by Biprateep Dey (priv. comm.).
}
\label{fig:morphology}
\end{figure}

Deep neural network-based methods that use images of a galaxy in multiple bands as inputs can exploit all of these phenomena, and have made considerable progress since their first application in \citet{Hoyle2016}).  Such methods are now outperforming even the best algorithms which use color information alone when applied to the SDSS Main Galaxy Sample, with $\sim 40\%$ reduction in photometric redshift scatter and even larger gains in catastrophic outlier rates \citep{Pasquet2019,Hayat2020,Henghes2021}, as can be seen in \autoref{fig:morphology}.  This clearly demonstrates that additional redshift information is available in galaxy images beyond what integrated photometry can capture.

However, many open questions remain.  First and foremost is whether such methods can continue to yield gains for fainter objects at higher redshifts.  In that domain, galaxies will only be marginally resolved from the ground, greatly reducing the information available, and training sets will be much smaller and sparser (generally a challenge for deep learning methods). It is clear that at least some improvements are possible when morphological information is utilized even in this domain; for instance, photo-$z$ errors at $z \sim 0.2-1$ are $\sim 10\%$ lower when basic morphological information is provided to a random forest algorithm than when only integrated photometry is used \citep{Zhou2021}.  

The Nancy Grace Roman Space Telescope will provide multi-band resolved images even for higher-redshift galaxies in the future, with sufficient angular resolution to probe similar physical scales at $z \sim 2$ as SDSS imaging does at $z \sim 0.1$. However, at such redshifts the relationship between galaxy color and morphology can be very different from today.  For instance, low star-formation rate but massive galaxies generally show clear evidence for disk components at $z\sim 2$, but that is rarely seen in quiescent objects locally \citep{Chevance2012,Chang2013}. As a result, it is not yet clear whether deep learning methods could offer similar gains at higher redshifts as they do for nearby galaxies from SDSS, even with the high resolution of space-based imaging.

It also remains uncertain whether summary statistics can be identified that could measure morphological characteristics with minimal information loss compared to processing the entire image of each galaxy via a deep neural network, such that a wider array of machine learning methods can be used.  How best to incorporate morphological information into template-based methods (e.g., via morphology-dependent priors on template type or size-dependent priors on redshift) has also not yet been determined.  However, even with these unresolved questions, the recent gains in photometric redshift performance that deep learning methods have made possible for nearby galaxies from SDSS, after more than a decade when many different algorithms all yielded results of similar quality,  offer tantalizing hope that it could be possible to make improvements in photometric redshifts for fainter objects as well.

\ifarxiv
\begin{marginnote}
\entry{Take-away}{Exploiting morphological information has yielded significant improvements to photometric redshift performance at low redshift; it remains to be seen whether similar gains can be achieved at higher $z$.}
\end{marginnote}
\fi

\subsubsection{Photometric Redshifts at Very Low $z$} 

\label{sec:lowz}

\begin{figure}[t]
\centering
    \begin{subfigure}[b]{0.45\textwidth}
        \includegraphics[width=\textwidth]{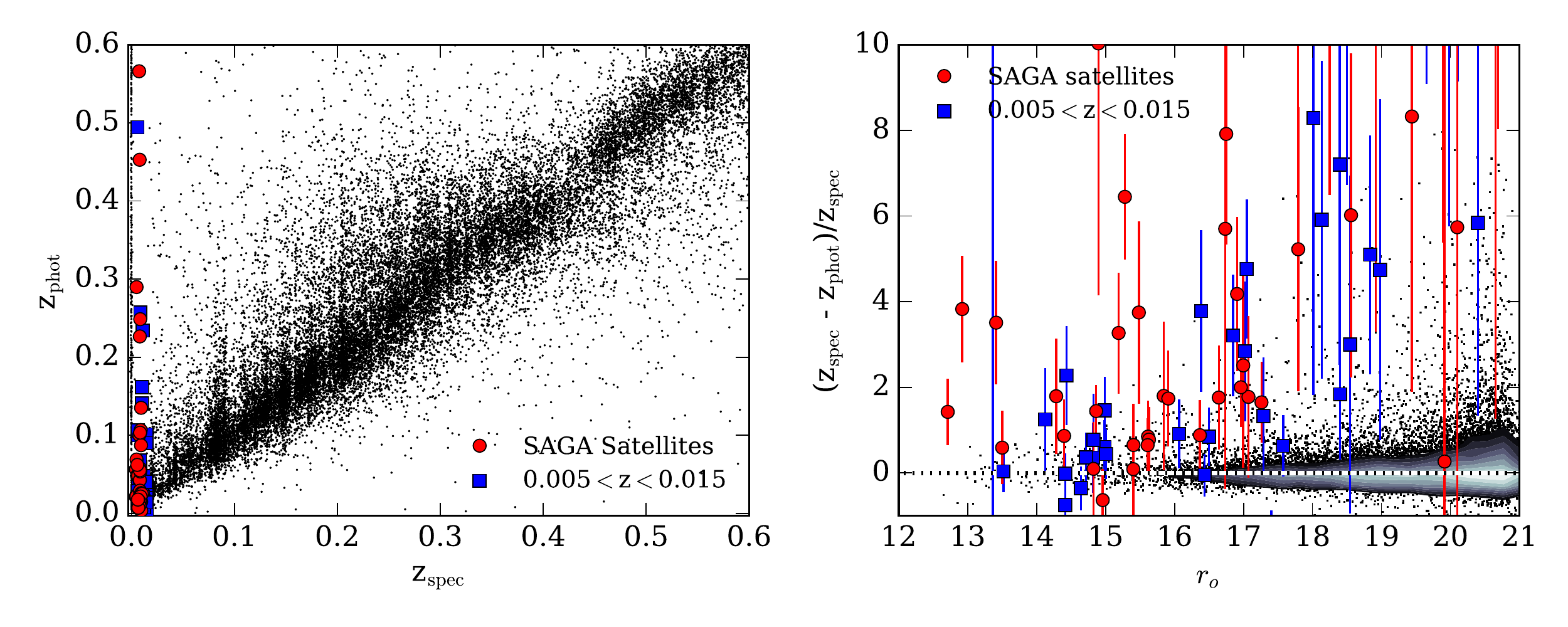}
    \end{subfigure}
    ~ 
      ~
    \begin{subfigure}[b]{0.45\textwidth}
        \includegraphics[width=\textwidth]{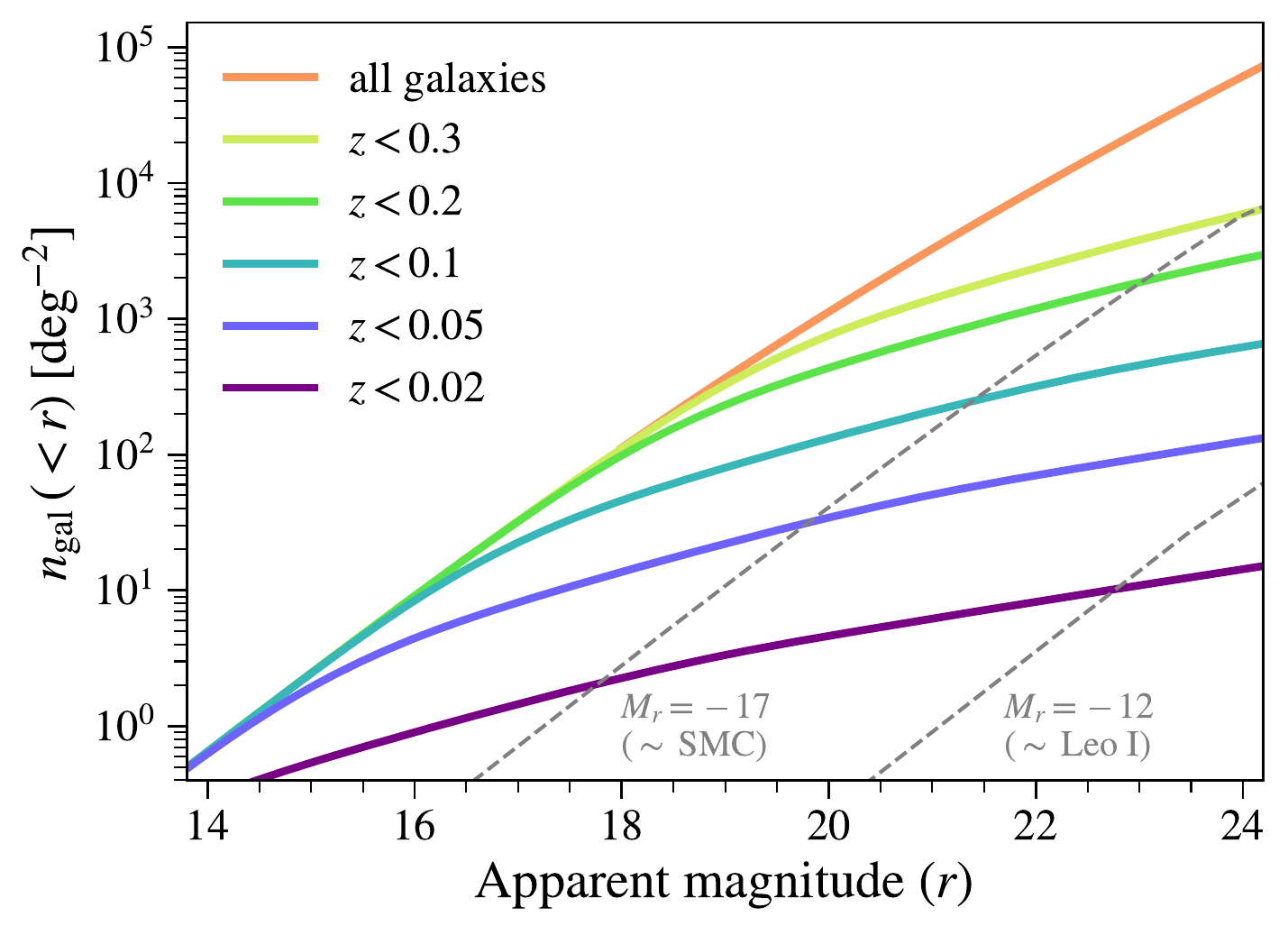}
    \end{subfigure}
\caption{(Left panel) Performance of photometric redshift algorithms at low $z$ can be very poor.  The plot shows the SDSS DR12 photometric redshift estimates from \citet{Beck2016} versus the spectroscopic redshifts of the same objects.  Although in general this algorithm yields good results (with $\sigma_{\rm NMAD} \sim 0.013(1+z)$), many galaxies at $z < 0.015$ (shown by red or blue symbols) are assigned inaccurately high photo-$z$'s.  This plot is adapted from Figure 4 of \cite{Geha2017}. \textcopyright AAS. Reproduced with permission. (Right panel) The low surface density of low-redshift objects makes the collection of training and validation sets of spectroscopic redshifts difficult.  Curves show the number of objects per square degree brighter than a given $r$-band magnitude with redshift below a specified limit.  Only $\sim 10$ galaxies per square degree have $z<0.02$, even when galaxies as faint as $r = 24$ are included.  Adapted from Figure 66 of \cite{MSEScienceBook}, figure courtesy Yao-Yuan Mao.
}
\label{fig:lowz}
\end{figure}

A somewhat surprising challenge that has been encountered in recent work \citep[e.g.,][]{Mao2021} is the comparatively poor performance of both template-based and training-based techniques at the lowest redshifts ($z < 0.05-0.1$).  This problem has many sources.  The first and foremost challenge is that there is very little volume of the Universe at very low $z$, so the lowest-redshift galaxies have a correspondingly low surface density on the sky. There are only $\sim 10$ $z<0.02$ galaxies (or $\sim 100$ $z<0.05$ objects) per square degree down to $r=24$   \citep{MSEScienceBook}, as seen in \autoref{fig:lowz}.  

As a result of their rarity, deep surveys contain very few low-$z$ galaxies, while wide-field surveys contain only the very brightest objects, which tend to have have intrinsically different SEDs than their fainter compatriots.  Flux measurements for bright objects are also subject to a variety of systematic problems that do not affect more compact, fainter galaxies, as photometric pipelines tend to be optimized for the more numerous smaller objects rather than well-resolved ones.  Due to their small numbers, low-redshift galaxies also contribute little to the calculation of losses used to optimize machine learning algorithms, so performance for them may be discarded in favor of improvements at the redshifts which dominate training samples. This effect will be illustrated in the top left panel of \autoref{fig:training_issues}.

This problem, which will tend to cause higher-redshift solutions to be favored, is compounded by the fact that redshifts have a lower bound of zero (modulo small contributions from peculiar velocities).  This violates assumptions commonly made in solving regression problems (including when applying machine learning techniques) that distributions of errors about the predicted value should be normal (or at least symmetric). As a result of this effect, photometric redshift estimates at low $z$ tend to be biased high so that the zero bound will be at one end of the predicted distribution for an object rather than in the middle.

The low volume of the Universe sampled also causes very large fluctuations in redshift distributions (or correspondingly, the occupation of cells in color/magnitude space by galaxies) at low $z$ due to sample/cosmic variance. These fluctuations will then tend to imprint on any training-based redshift distributions.    

Even though they should not be affected directly by these training issues, to date template-based methods which yield good performance at higher redshifts have still provided poorer results at low $z$. One possible cause is that, if $u$-band photometry is not available or is noisy, it can be extremely difficult to localize the 4000 Angstrom break, compromising the ability of an algorithm to determine the redshift with precision. A further challenge is that low-$z$ photometric redshift estimates will rely on the longest-wavelength portions of the templates used, which may be only poorly tested given their irrelevance for most objects.

There are multiple approaches that may lead to improvement in this area.  One is expanding spectroscopic training samples at the lowest redshifts; for instance, the DESI Bright Galaxy Sample will help by surveying galaxies down to a limit roughly two magnitudes fainter than SDSS \citep{Ruiz2020}, and the SAGA survey is obtaining many redshifts for low-$z$ galaxies as part of its search for satellites around Milky Way-mass objects \citep{Mao2021}.  Even for template-based methods, this will allow better understanding of the sources of the problem and testing of solutions.  Incorporating size or surface brightness information into photometric redshifts (cf. \S \ref{sec:morphology}) has also yielded substantial improvements at low $z$ \citep{Mao2021}.

For machine learning methods, gains can also be made by changing the loss function used to optimize redshift predictions.  For instance, penalizing deviations in ${\log{z}}$, rather than the raw redshift value, will give greater weight in the training to the low-redshift regime, as $\Delta(\log{z}) \sim \frac{\Delta(z)} {z}$.  However, so long as a tiny fraction of training samples are at low $z$, higher-redshift objects will dominate the loss calculation due to their much greater numbers in the training sets, unless the nearby objects are provided additional weight; but doing so would degrade performance for the bulk of the sample.  As a result, photometric redshift applications which demand small errors even at low redshifts will likely require bespoke solutions if they employ machine learning-based methods.  

\ifarxiv
\begin{marginnote}
\entry{Take-away}{Current photo-$z$ methods tend to perform very poorly at very low redshift ($z < 0.05)$.}
\end{marginnote}
\fi

\subsection{Challenges for Improving Photometric Redshift Characterization}
\label{sec:characterization}

In this section, we describe a number of effects that significantly affect redshift characterization.
Several, if not all of them, can cause errors exceeding the requirements of near-future cosmological
experiments. Careful calibration and accounting for their effects are therefore necessary for them
not to limit the cosmological insights to be gained. However, this is not always achievable with
currently available methods, so further research in these areas is needed. It is possible that other
systematic issues associated with photo-zs that have not yet been identified could also compromise
cosmological inference at a comparable level to the effects discussed here.

\subsubsection{Spectroscopic Incompleteness}

\label{sec:incompleteness}

Any sample of galaxies can be characterized by a selection function that determines the probability with which an object will be included based on its relevant physical characteristics (e.g., its redshift, observed spectral energy distribution, and surface brightness profile). For any galaxy $i$, we can compute the ratio $r_i=\frac{p^{\rm spec-z}_i}{p^{\rm phot}_i}$ of the probability of obtaining a successful spectroscopic redshift for such an object in available surveys divided by the probability that object has for being placed within some sample of interest (e.g., a particular redshift bin in a cosmological analysis).  We will refer to such a sample, which can be defined based only on the photometric properties of the included objects (possibly incorporating photometric redshift estimates), as a target sample. A direct estimate of the redshift distribution for a target sample can then be written as
\begin{equation}
p^{\rm phot}(z) \propto \sum_{j\in\mathrm{spec-z}} r_j^{-1} \delta(z-z_j) \; ,
\label{eqn:incompleteness}
\end{equation} 
where the sum runs over all galaxies $j$ with successful spectroscopic redshifts $z_j$.

\ifarxiv
\begin{marginnote}
\entry{Impact of spectroscopic incompleteness}{\includegraphics[width=3.5cm]{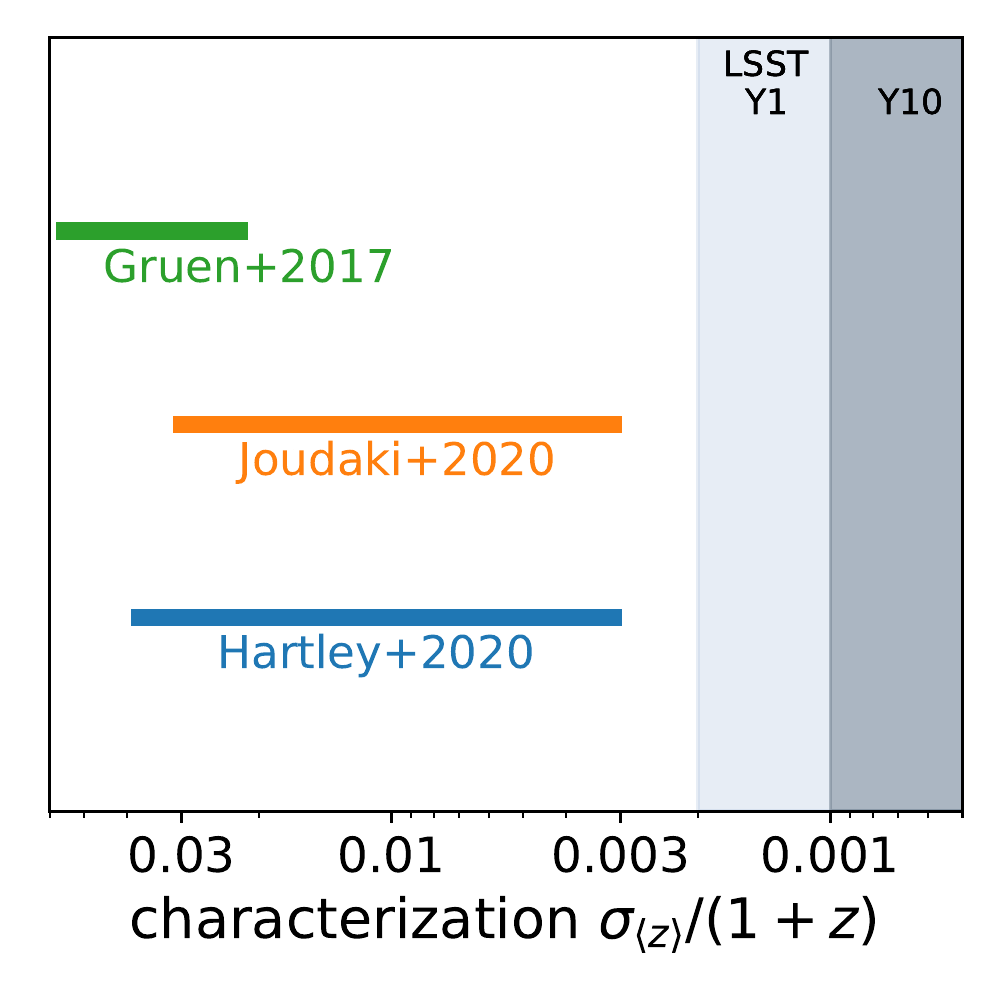}}
\end{marginnote}
\fi

Difficulties can arise when $r$ depends on the \emph{redshift} of a galaxy, which is very common. Two galaxies may have indistinguishable magnitude and colors in a given imaging dataset, yet still have different true redshifts and restframe SEDs (e.g., due to confusion of one spectral break with another, or because of the nearly featureless, power-law-like spectra exhibited by highly star-forming galaxies over a broad wavelength range). Photometrically indistinguishable objects that are at different redshifts can have different probabilities of yielding secure redshifts (e.g., due to strong emission lines passing beyond the optical window at higher $z$). In such a situation, $p^{\rm spec-z}$ remains a crucial factor in redshift characterization but cannot be determined from the photometric data alone. As a simple example, if objects at a given point in color space have two possible redshifts, $z_1$ and $z_2$, which yield secure redshifts at rates of 100\% and 0\%, respectively, the objects at $z_1$ whose spectroscopic redshifts were measured would receive an enhanced weight in calculating the redshift distribution via \autoref{eqn:incompleteness}, whereas $z_2$ would not contribute to the calculation of $p^{\rm phot}(z)$ at all. 
As a result of this effect, so long as spectroscopic success varies strongly across subsets of galaxies distinguished by $>0.1$ in redshift (as it does in real samples), spectroscopic completeness -- i.e., the fraction of targets that yield secure redshifts -- will need to be extremely high ($>99\%$ or even $>99.9\%$; G. Bernstein, private communication) for cosmological measurements from present and future surveys not to be degraded when direct calibration is performed. If such high completeness is not achieved, the selection of spectroscopic samples and the true nature of objects which failed to yield secure redshifts must be modeled carefully and understood well for direct calibration to yield accurate results.

The probability of obtaining a successful redshift, $p^{\rm spec-z}$, within a particular spectroscopic survey will be subject to both intentional and unintended selection effects. \emph{Intentional} selections can include cuts on colors, magnitudes, or surface brightness; these are commonly used in deep spectroscopic samples (e.g., DEEP2 \citealt{Newman2013}, VIPERS \citealt{Guzzo2014}, and zCOSMOS-Deep \citealt{Lilly2007}). They can ensure high spectroscopic success rates for targets of interest within feasible exposure times (e.g., by preferentially selecting blue star-forming galaxies with strong emission features) or to target galaxies in a particular redshift range of interest. The impact of a color space selection is illustrated in the top right panel of \autoref{fig:training_issues}. 

If equivalent photometric data are available for the spectroscopic samples used and the target sample one wishes to characterize, one can determine the intended $r_i$ for each object.  However, in many cases its value may be zero for a subset of the target sample (even when there are no color cuts, one might wish to characterize redshift distributions for objects that go fainter than the spectroscopic samples available). Such galaxies would then have to be excluded from the target sample for direct spectroscopic characterization of their $n(z)$ to be possible. If uniform photometric data are \emph{not} available for both the spectroscopic and target samples, one commonly finds that a non-zero but unknown fraction of galaxies which pass any useful target sample selection are missing from the spectroscopic sample, making direct spectroscopic calibration fundamentally unreliable. In the common case where one combines a variety of spectroscopic surveys, even if no subset of the target sample is excluded by \emph{all} spectroscopic selections, they would have to be correctly reweighted for the joint selection function to be correct; if even \emph{one} of the spectroscopic selections cannot be reproduced on the target sample's photometric data, the procedure becomes ill-defined. 

The value of $p^{\rm spec-z}$ can also be less than one for reasons that were \emph{not intended} by the spectroscopic survey design. At the faint end of deep spectroscopic samples, the rate at which highly-secure redshifts are measured for targeted galaxies is rarely above $75\%$ \citep{Newman2013,LeFevre2013,Bonnett2016}. Whether a secure redshift will be obtained from spectroscopy depends sensitively on the properties of a galaxy and the observations. When spectral features are weak, fall outside the wavelength coverage of a particular instrument, or are at wavelengths affected by atmospheric emission or absorption, the probability of obtaining a successful redshift measurement may be reduced or even eliminated entirely.  This can cause galaxies at some redshifts to be missing entirely in training samples.  The incompleteness of spectroscopic samples is likely worse in regions of high galaxy surface density where blending is more common, further biasing the redshift distributions recovered from direct characterization.  The impact of systematic incompleteness on our ability to map the relationship of colors to redshift is illustrated in the bottom left panel of \autoref{fig:training_issues}, and is apparent in the $z>0.9$ tail of \autoref{fig:zdist}.

\begin{figure}[t]
\centering
        \includegraphics[width=\textwidth]{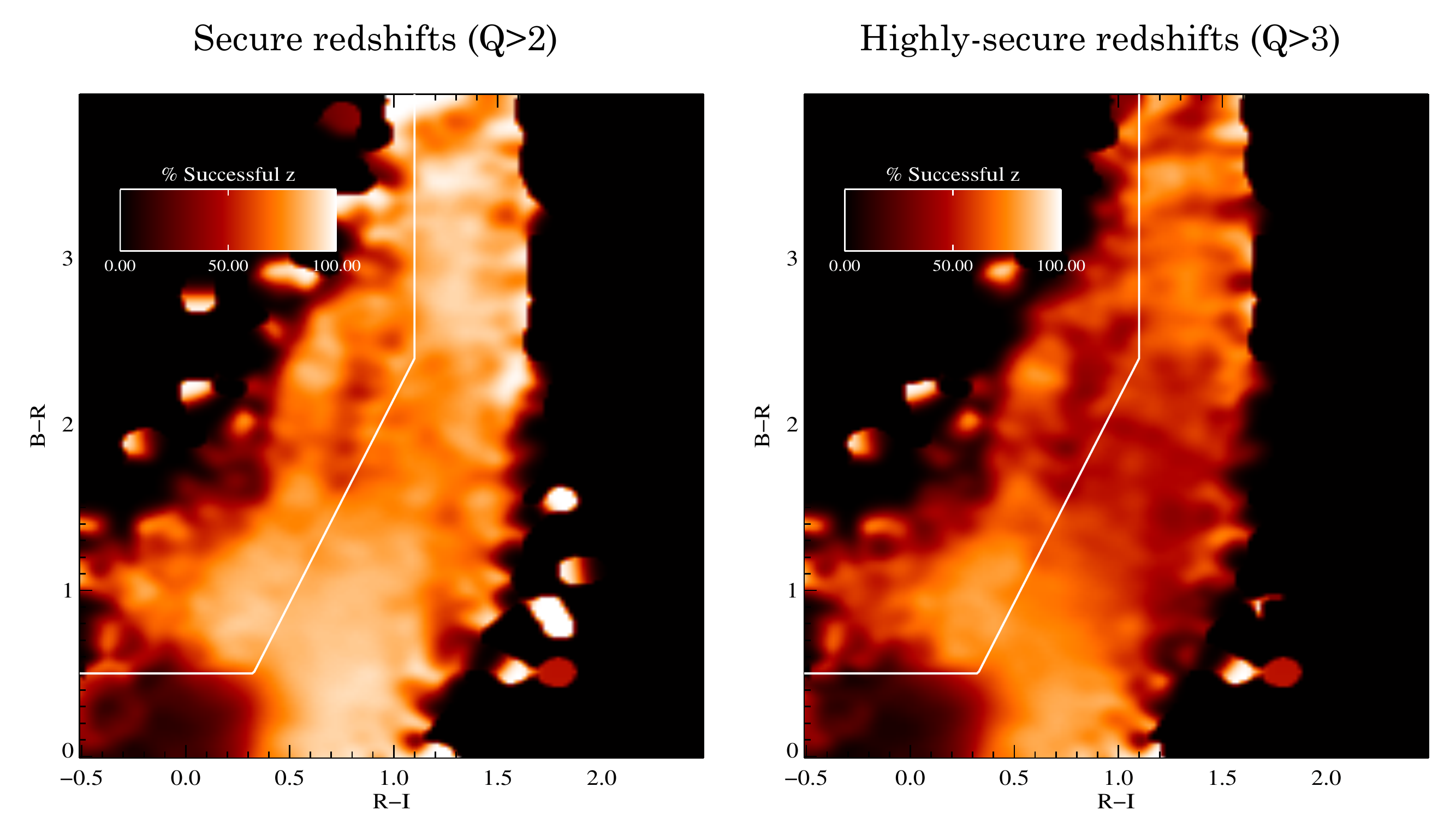}
\caption{
The fraction of objects which spectroscopic surveys of faint galaxies obtain secure redshifts for varies with observed galaxy color, but in parameter spaces defined by the deepest optical bands it nowhere approaches 100\%.  The panel at left shows the fraction of magnitude $R < 24.1$ galaxies targeted by the DEEP2 Galaxy Redshift Survey which delivered secure redshifts ($Q=3$, corresponding to $>95\%$ confidence, or $Q = 4$, corresponding to $>99\%$ confidence), as a function of observed optical $B-R$ and $R-I$ colors.  The white line in each panel shows the color cut used to select targets for DEEP2 in three of the four survey fields, excluding the Extended Groth Strip whose data were used to produce this plot.  At right is shown the fraction of objects which yielded the most secure, $Q=4$ redshifts; although incorrect-redshift rates as low as achieved for this sample, or even lower, may be required for future cosmology applications, requiring purer redshift measurements only increases the challenge of systematic incompleteness in spectroscopic samples.  This figure is adapted from Figures 44 and 45 of \cite{Newman2013}.  \textcopyright AAS. Reproduced with permission.
}
\label{fig:zsuccess}
\end{figure}

All these causes for unintended incompleteness depend on the redshift of a galaxy and are complex to model, posing substantial challenges for redshift characterization.  As can be seen in \autoref{fig:zsuccess}, the problem of incompleteness is further compounded if samples are limited to the most secure redshifts to prevent contamination of characterization by outliers. 

The effects of spectroscopic incompleteness on photo-$z$ characterization in few-band surveys are large compared to current and future requirements. Based upon both many-band template fitting photo-$z$'s \citep{Gruen2017} and simulated spectroscopic observations \citep{Hartley2020}, recent work has found biases in the mean redshift of up to $|\delta z|\approx0.05$ for direct calibration from the combination of intended and unintended spectroscopic selection effects. Similar levels of disagreement have been found by \citet{Hildebrandt2020} when comparing characterization of redshifts performed with spectroscopic samples with different selection functions using the more photometrically  constraining KiDS-VIKING data set and by \citet{Joudaki2020} using mock analyses of few-band DES-like data that include intended spectroscopic sample selections. 

The impact of spectroscopic incompleteness can be reduced if many- and/or narrow-band photometric information is used to select photometric samples (or for reweighting spectroscopic samples, if there are not regions with zero probability due to unintended systematic effects; \citealt{Buchs2019,Myles2021}). Intended selection effects will lead to regions which are devoid of spectroscopic data in such spaces \citep{Masters2015,Hildebrandt2019}. Subsets of the target sample with poor spectroscopic information can be removed based on their position in the many-band color space, or dedicated surveys can  be performed that could potentially  fill in any gaps in coverage \citep{Stanford2021}. 

If at all points covered by a target sample in a high-dimensional color space the width of the redshift distribution is small and the redshift success rate is relatively uniform with $z$, there is little room for unintended selection to bias the characterization. However, the converse is also true. If there are regions of the many-dimensional space which correspond to multiple, significantly-separated redshift values and redshift success rates vary between the different possible solutions, direct characterization will remain biased. 

\ifarxiv
\begin{marginnote}
\entry{Take-away}{Mitigating the effects of systematic incompleteness in the deep spectroscopic samples used to train photo-$z$ algorithms and characterize redshift distributions will be a key challenge for upcoming surveys.}
\end{marginnote}
\fi

\subsubsection{Outliers and Biases in Calibration Redshifts}

\label{sec:outliers}

Training-based photometric redshift methods (as well as direct spectroscopic calibration techniques such as the ``SOM'' method used in \citealt{Hildebrandt2021}) typically assume that all redshifts in a training set are correct. 
However, real-world datasets do not fulfill this assumption.  Deep spectroscopic surveys generally must use low-signal-to-noise spectra to measure redshifts, due to the very long exposure times needed for faint objects. Occasionally, features in a spectrum that are in fact due to noise (particularly in regions affected by night sky emission lines) may match templates at some false redshift, leading to the misattribution of the redshift of an object.  When only a single emission line or spectral break has been clearly detected,  misidentification of that feature will result in a systematically incorrect redshift.  

The rates at which these failures occur are substantial. Current deep surveys generally have assigned quality flags to spectroscopic redshifts based on visual inspection of redshift fits, with quality $Q=3$ corresponding to 95\% certainty that a redshift is correct and $Q=4$ corresponding to $>99\%$ certainty.  In the DEEP2 Galaxy Redshift Survey, $\sim 17\%$ of secure redshifts were assigned $Q=3$, with the remainder receiving $Q=4$ (the high resolution of the DEEP2 spectroscopy, enables splitting of the [OII] doublet, causing most redshifts obtained to be unambiguous).  More than 1000 DEEP2 objects were observed twice, allowing the repeatability of redshifts to be tested; for this sample, the best estimates of the failure rate are $0.75\%/0.15\%$ (with upper limits of $\sim 2.2\%/0.3\%$) for $Q=3/4$, respectively \citep{Newman2013}.  For the zCOSMOS-bright survey, $Q=3$ redshifts dominate ($\sim 58\%$ of secure redshifts) due to the lower spectroscopic resolution used.  Based on objects with repeated spectra, failure rates for zCOSMOS $Q=3$ and $Q=4$ are estimated to be 1\% and 0.2\%, respectively \citep{Lilly2007,zCOSMOSDR}.

Similarly, many-band photo-$z$ suffer from significant catastrophic outlier rates, particularly in regions of parameter space which are poorly characterized by training spectroscopy.  Catastrophic outlier rates ranged from $\sim 0.6\%$ for quiescent objects with $i \sim 22.5$ to $>10\%$ for star-forming objects with $i \sim 24.5$ in the COSMOS2015 catalog of \cite{Laigle2016}.  

Even with perfect spectroscopy, part of this problem  is irreducible due to blending of galaxies, see \S \ref{sec:blending}. Additionally, some fraction of objects will have photometry contaminated by artifacts or failures of measurement algorithms; both effects cause the mapping of color to redshift to be inappropriate for some portion of the training sample.

\ifarxiv
\begin{marginnote}
\entry{Impact of biases in calibration redshifts}{\includegraphics[width=3.5cm]{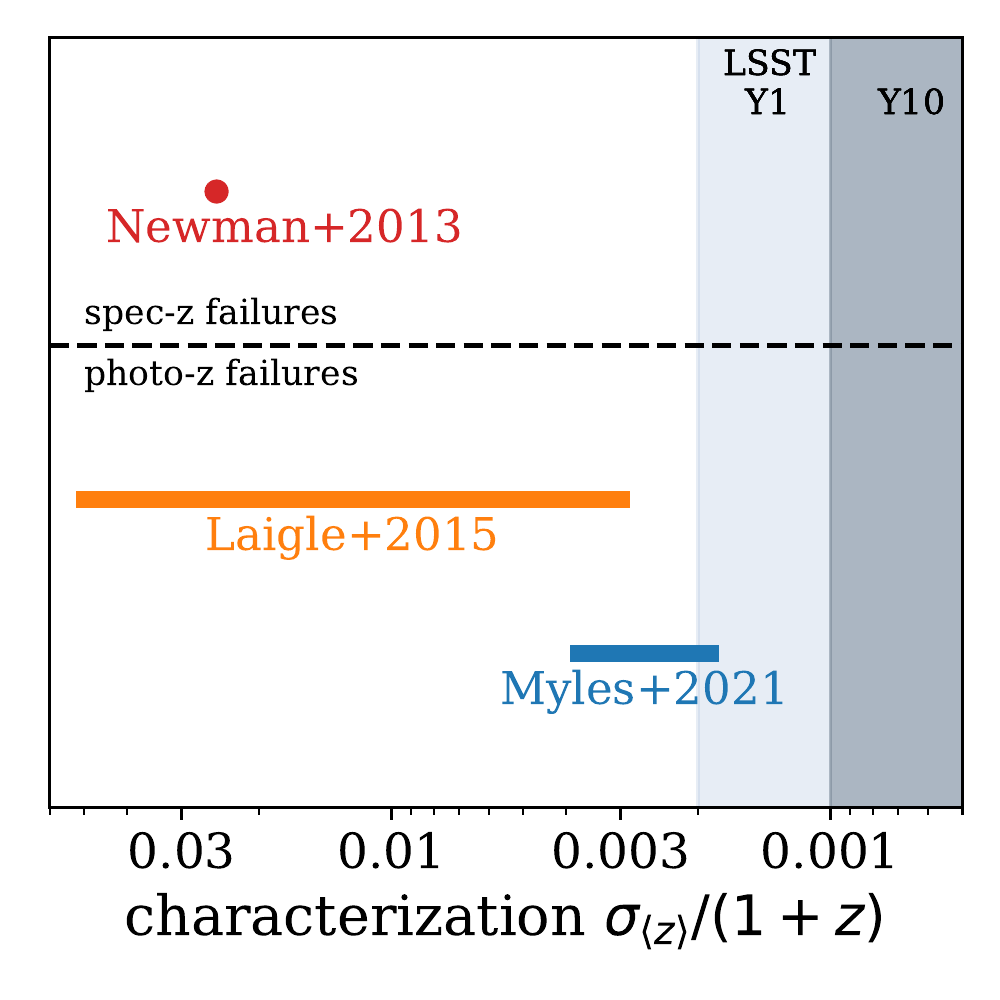}}
\end{marginnote}
\fi

Incorrect training data impact the performance of photo-$z$ algorithms by introducing an unphysical spread of redshift at given photometry, as illustrated in the bottom right panel of \autoref{fig:training_issues}.  However, a greater problem is that we rely on spectroscopic redshifts (or very high-quality photometric redshifts) for characterization; if the redshifts used for this purpose are incorrect, the derived redshift distributions will be too.  The impact  can be large if not accounted for.  

A simple toy model is sufficient to assess the magnitude of this problem.  If a fraction $f_{\rm inc}$ of redshifts for a sample described by a Gaussian distribution of standard deviation $\sigma$ are systematically off from the true mean redshift by $\Delta$ (emulating the common situation where one spectral feature is mistaken for another, leading to an incorrect $z$), the obtained mean redshift for the sample will be shifted by $\delta \langle z\rangle = f_{\rm inc} \Delta$, and the standard deviation will be shifted by 
$$\delta \sigma = \sqrt{(1-f_{\rm inc})\sigma^2+ f_{\rm inc} \Delta^2} - \sigma \approx \frac{f_{\rm inc} \sigma}{2}\left(\frac{\Delta^2}{(1-f_{\rm inc})\sigma^2} -1\right) \approx \frac{f_{\rm inc} \Delta^2}{2 \sigma}$$ when $f_{\rm inc}$ is small and $\Delta$ is large compared to $\sigma$.  For instance, for a sample (e.g., a redshift bin) described by a Gaussian redshift distribution with $\sigma=0.1$, a rate of one $\Delta=0.5(1+z)$ redshift error per thousand spectroscopic redshifts would shift the inferred mean redshift by 0.0005(1+z), smaller than LSST cosmology requirements of $\delta <z> < 0.002(1+z)$, but  the inferred $\sigma$ would be too large by 0.0024(1+z), in tension with the $\delta <\sigma> < 0.003(1+z)$ requirement \citep{DESC_SRD}.

\begin{figure}
\includegraphics[width=0.5\textwidth]{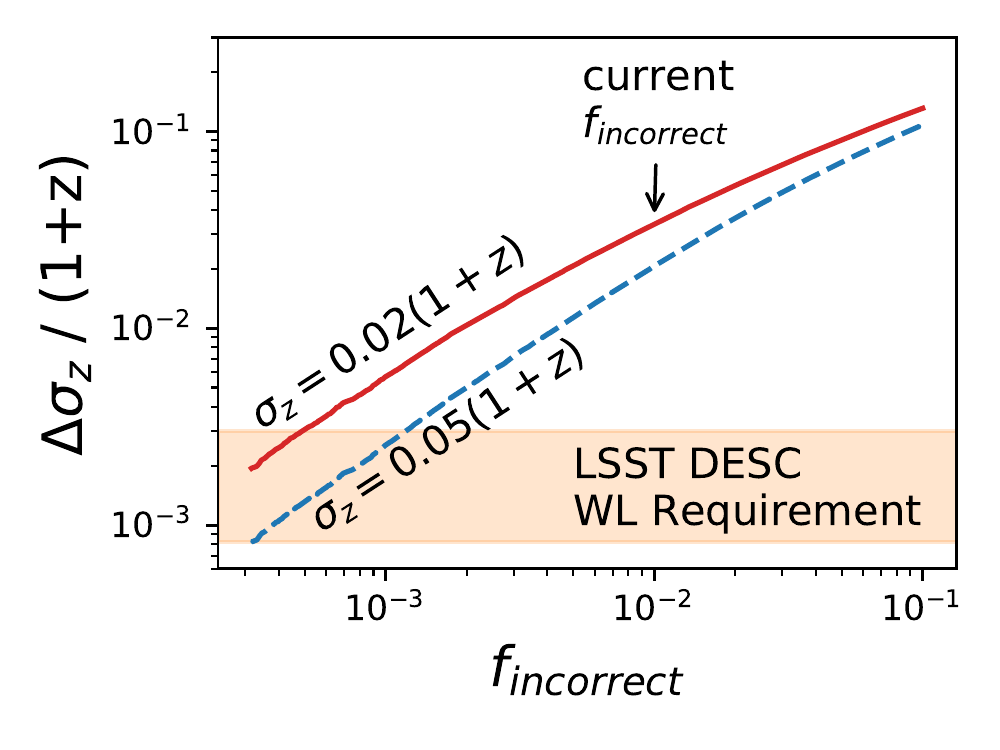}
\caption{Impact of incorrect redshifts in spectroscopic sets used for direct calibration of photometric redshifts.  The red and blue curves show the bias in recovering photo-$z$ uncertainties that results for samples with a true standard deviation $\sigma$ of 0.02 (red solid) or 0.05$(1+z)$ (blue dashed) in Monte Carlo simulations where a fraction $f_\mathrm{\rm inc}$ of redshifts are systematically off by $0.5(1+z)$, a typical value in real spectroscopic datasets.  The shaded region shows the requirement from \cite{DESC_SRD}.  Contamination of spectroscopic calibration sets by incorrect redshifts at the level seen in the best samples from current deep datasets, $\sim 1\%$, would still cause systematic errors in mean $z$ and in photo-$z$ uncertainties that are $\sim$7--10$\times$ larger than the science requirements for Rubin Observatory weak lensing measurements. Requirements for Euclid and Roman weak lensing should be similar.  Figure provided by Brett Andrews (priv. comm.).}
\label{fig:photoz_issues}
\end{figure}

Figure \ref{fig:photoz_issues} illustrates the results from this model. When a fraction $f_{\rm inc}$ of the redshifts used to characterize the distribution of photo-$z$ errors is erroneous, the inferred spread $\sigma_z$ is biased by an amount $\Delta \sigma_z$ from the true value. We here have performed Monte Carlo simulations of scenarios where we measure the standard deviation of Gaussian-distributed photometric redshift errors from a training set where a fraction $f_{\rm inc}$ have their redshift systematically offset by $0.5(1+z)$, emulating the typical effect of line misidentification.  For this toy model, $f_{\rm inc}$ must be  $\sim 10^{-3}$ or less -- an order of magnitude lower than in current deep samples -- for characterization at the level required for Rubin Observatory weak lensing cosmology measurements to be possible.   The bias grows larger, and the requirements on $f_{\rm inc}$ are correspondingly more stringent, when photometric redshift errors are smaller.   If the incorrect redshifts instead had a symmetric, Gaussian distribution about the true value, the inferred $\sigma_z$ would still be biased, though the curves in \autoref{fig:photoz_issues} would differ depending upon assumptions.  
Similar results hold if we consider a bias in the mean redshift 
rather than a bias in the spread.

We might consider three responses to this problem.  The first is to reduce the error rates in spectroscopic training/characterization datasets by roughly an order of magnitude, {which likely would require a combination of much longer integration times, broader wavelength coverage, and more restrictive manual quality control}. Even then, this only solves the problem if we can also reduce all other causes of mismatches between spectroscopic training sets and other objects (e.g., blending and photometric artifacts) to the same level. {One possible path could be to limit the use of spectroscopic redshifts to characterizing the relation between photometry and redshift in a high-dimensional, well-measured color space. If the latter tightly constrains true redshift at given color such that the same color cannot genuinely correspond to multiple redshifts and if the size of the training sample is sufficient, outliers can be identified confidently.}  

A second option is to rely on methods other than direct calibration via deep spectroscopic samples for characterization.   For instance, it is possible to exploit cross-correlations with wide-field surveys of brighter galaxies and quasars, in which we can select only the most secure redshifts and still have a large sample to characterize the redshift distributions of photometric samples, as we discuss in \S \ref{sec:xcorr}.

The third option is to develop methods for characterization that are robust to outliers in the training set, which can be done in a variety of ways. If the potential impact of outliers on the redshift distribution of galaxy bins can be described by a prior, combinations of correlation functions allow for partial self-calibration \citep[e.g.,][]{Zhang2010,Schaan2020}. In hierarchical Bayesian methods that use the full data, biased training sets can potentially be compensated for \citep{Sanchez2019}. 

In principle, hierarchical use of spectroscopic samples could allow for a parameter indicating whether the reported redshift of each spectroscopic galaxy is wrong, whose posterior is constrained by the combination of all photometric and spectroscopic data, that allows to reduce the impact of outliers on redshift characterization.

\ifarxiv
\begin{marginnote}
\entry{Take-away}{The rate of incorrect redshifts in current deep spectroscopic samples is high enough to compromise direct redshift characterization methods for future surveys; the problem is worse for redshifts from many-band photometry or low-resolution spectroscopy.}
\end{marginnote}
\fi

\subsubsection{Impact of Sample Variance on Characterization}

\label{sec:sample_variance}

\begin{figure}[ht]
{\includegraphics[width=0.65\textwidth]{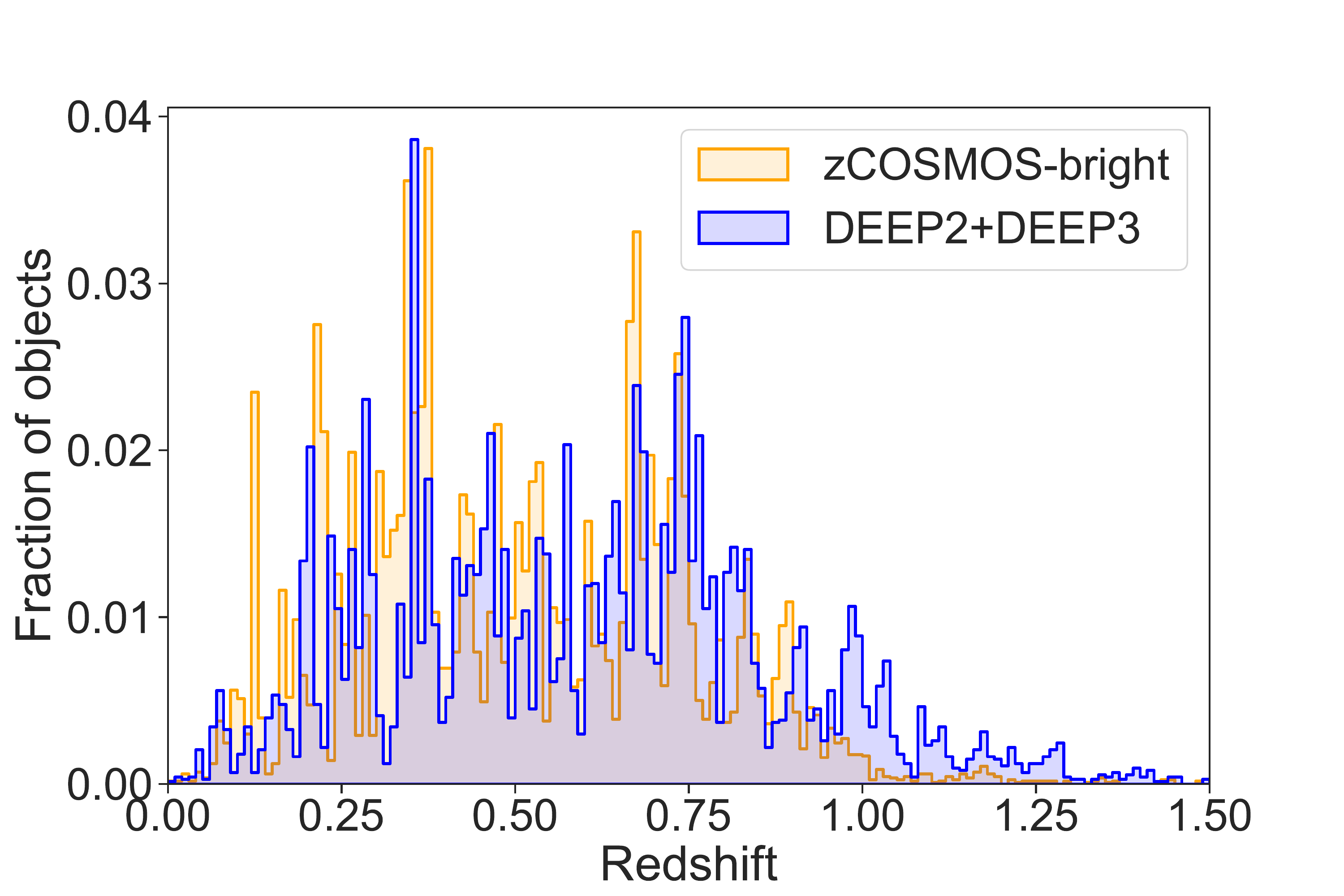}}
{\caption{Redshift distributions of magnitude $i<22.5$ galaxies in the zCOSMOS-bright and DEEP2/DEEP3 surveys \citep{Lilly2007,Newman2013,Zhou2019}.  Histograms show the $z$ distributions of galaxies with secure spectroscopic redshift measurements (confidence class or quality 3 or 4) from each survey, normalized by the total number of $i < 22.5$ objects in each sample. The $i$ is obtained using HST F814W band photometry for zCOSMOS and extinction-corrected CFHTLS $i$-band for DEEP2 and DEEP3 selections.  Even though the zCOSMOS-bright survey spanned 1.7 square degrees of sky, large fluctuations in the number of objects at a given redshift due to sample/cosmic variance are clearly apparent in its redshift distribution.  For DEEP2 and DEEP3, we use only redshifts in the Extended Groth Strip (DEEP2 Field 1) which covered a total of 0.6 square degrees; excursions are even larger in this case.  The redshift distributions differ at $z > 0.9$ largely because of the higher spectral resolution used for DEEP2 and DEEP3; this allows the [OII] 3727 Angstrom doublet to be resolved, enabling higher secure redshift success rates in that regime.   
}
\label{fig:zdist}}
\vskip-0.15in
\end{figure}

\begin{figure}[t]
\centering
        \includegraphics[width=\textwidth]{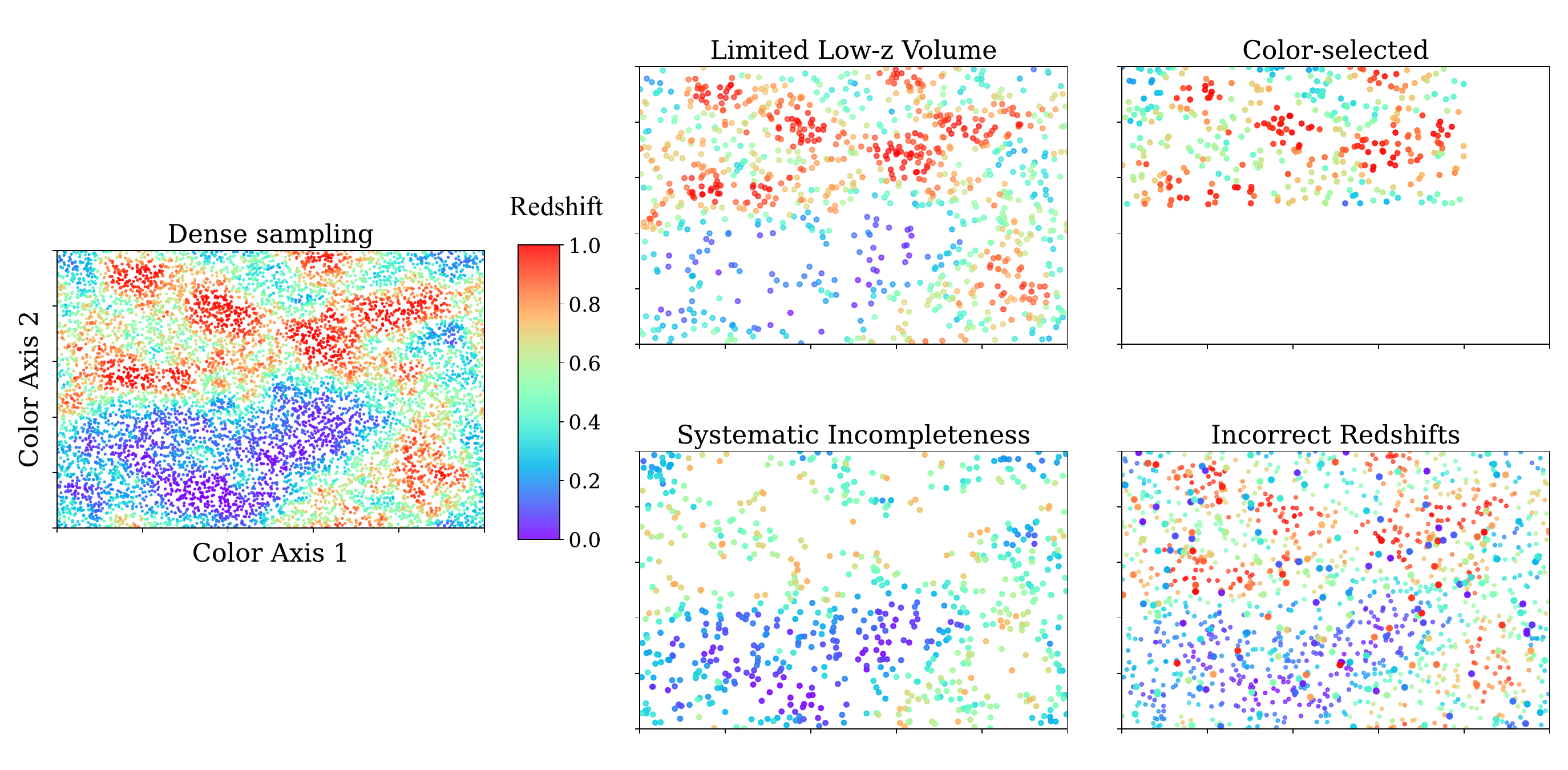}
\caption{Illustration of additional factors which limit the use of spectroscopic training sets to map out relations between color and redshift.  In these figures we continue to use the toy model mapping of galaxy color (or a dimensionality-reduced color space) to redshift that was employed in \autoref{fig:sparsesampling}.  At left we again show the ideal case, where spectroscopic samples cover the color space both densely and uniformly; colors correspond to redshifts ranging from zero to one, as indicated in the color bar. The top middle panel illustrates the impact of the failure of deep spectroscopic training sets to include many objects at low redshift, as a small area of sky will include only a very limited volume at low $z$, and hence correspondingly few objects (cf. \S \ref{sec:lowz}). As a result, deep galaxy samples only sparsely cover color space in that domain, degrading photo-$z$ performance at low redshifts.  The top right panel illustrates the impact of using spectroscopic samples which are restricted to a limited region of parameter space (\textit{intentional} selection effects, as described in \S \ref{sec:incompleteness}. This can provide dense sampling of the relationship between photometry and redshift, but only over a limited region; beyond that range the spectroscopy will have no constraining power.  The bottom middle panel, conversely, shows the impact when spectroscopy systematically fails to obtain secure redshift measurements for objects at high $z$ (corresponding to the \textit{unintended} selection effects in \S \ref{sec:incompleteness}): this again causes gaps in color coverage, leading both to degraded photo-$z$ performance where training samples are lacking and to  systematic biases in redshift characterization.   Finally, the bottom right panel illustrates the impact when incorrect redshifts (here, drawn randomly from a uniform distribution) are assigned to a fraction of targets. As discussed in \S \ref{sec:outliers}, this will cause systematic biases in both any photo-$z$'s that are derived from a training set and in the characterization of redshift distributions. 
}
\label{fig:training_issues}
\end{figure}

As is the case for any cosmological measurement, the limited volume over which it has or even can be made sets a lower bound to the uncertainty of any conclusions that can be drawn from it. In the case of characterizing redshift distributions with spectroscopy, this is fundamentally due to the fact that at a given set of photometric observables, the distribution of redshifts is still broad. When observed with a finite sample of spectra, one retrieves only a sampling of that distribution. When observed over a field of limited area or size, one retrieves such a sampling of only a version of that distribution that is modulated by variations in the mean matter (and thus galaxy) density as a function of redshift within that field. This is a strong effect in current deep samples, as can be seen in \autoref{fig:zdist}.  This phenomenon is commonly referred to as  sample variance or, in some works, as cosmic variance.

It is useful to consider sample variance in redshift calibration as a combination of three effects: 
\begin{itemize}
\item A variation of the observed density $\hat{n}_{ij}$ of galaxies in the calibration field, for instance number per solid angle per color element up to some limiting magnitude, at positions in a high-dimensional color space which we have denoted by a two-dimensional index $i,j$ for notation and illustration purposes. 
\item A variation of the true mean redshift of galaxies of color $i,j$ in the calibration field relative to the true mean redshift of such galaxies over a very large area, due to matter density variations in the finite calibration volume, denoted by $\sigma^z_{\mathrm{CV}, ij}$. Even in the hypothetical case where there are a very large number of galaxies of that color available within a calibration field, their mean redshift would still deviate from the cosmic mean. 
\item An uncertainty in the observed mean redshift of galaxies of color $i,j$ in a calibration field relative to the former hypothetical value due to sampling with a finite number of galaxies. Given a scatter in redshift $\sigma^z_{ij}$ among the sample of galaxies, and a number of galaxies $N_{ij}$ observed, this uncertainty is given by $\sigma^z_{ij}/\sqrt{N_{ij}}$. 
\end{itemize}
The amplitude of the former two uncertainties depends on galaxy type and luminosity (by means of their effect on galaxy clustering bias) and on the field size. The amplitude of the latter uncertainty is set by the intrinsic dispersion of the redshift of galaxies of color $i,j$ and the number of galaxies of that color observed. 

The magnitudes of the three effects depend on not just the size and sampling of the calibration field, but also the photometric information available. Sample variance can have the largest impact for surveys with only a few photometric bands observed, as illustrated in the extreme case of a single magnitude cut in \autoref{fig:zdist}. To see this quantitatively, consider two hypothetical surveys - one where the full photometric information $ij$ is available for all galaxies (both in the calibration sample and in the target sample), and one where only a subset of the photometric bands, $i$, is available. Assume that in both setups, we would like to estimate the mean redshift of the subsample of galaxies with some given color $i$, ${\langle \hat{z}\rangle}_i$.

\ifarxiv
\begin{marginnote}
\entry{Impact of Sample Variance}{\includegraphics[width=3.5cm]{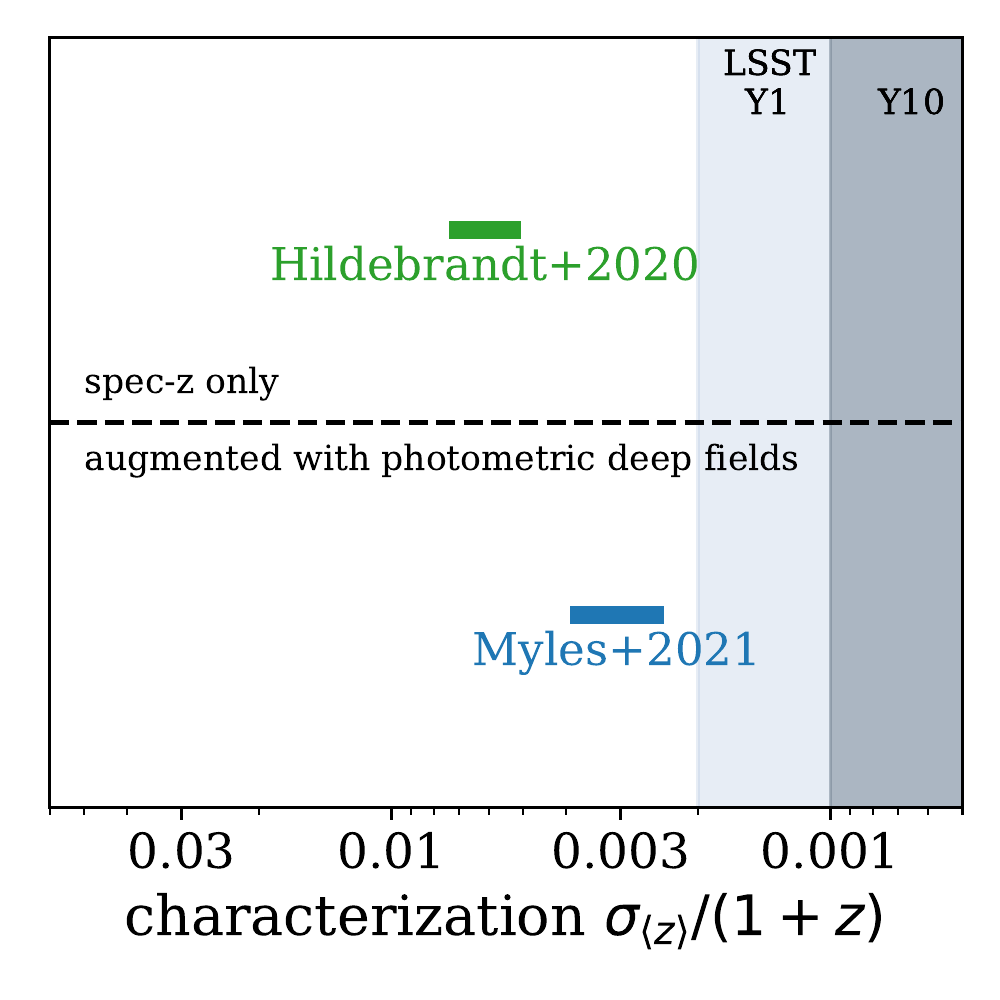}}
\end{marginnote}
\fi

In the former case of a many-band survey, one could write this as a sum over all additional colors $j$,
\begin{equation}
\hat{\langle z\rangle}_i = \frac{\sum_j n_{ij} \hat{z}_{ij}}{\sum_j n_{ij}} \;,
\end{equation} 
where $\hat{z}_{ij}$ is the mean redshift of calibration galaxies of color $ij$, and $n_{ij}$ is the density of galaxies of that color measured over the full survey, essentially to re-weight the calibration sample as a function of color $j$. We can assume $n_{ij}$ to be essentially noiseless for a large-area photometric survey, which leads to an uncertainty of the mean redshift estimate of 
\begin{equation}
\left(\sigma^{\langle z \rangle}_{i}\right)^2 = \frac{\sum_j n^2_{ij} \left[(\sigma^z_{\mathrm{CV}, ij})^2 + (\sigma^z_{ij})^2/N_{ij}\right]}{\left(\sum_j n_{ij}\right)^2} \; .
\end{equation}

Consider instead that the true density $n_{ij}$ is not known because the wide field survey measures only the color $i$, but not $j$. From the calibration sample one would estimate
\begin{equation}
\hat{\langle z\rangle}_i = \frac{\sum_j \hat{n}_{ij} z_{ij}}{\sum_j \hat{n}_{ij}} \;,
\end{equation} 
using the densities of galaxies in the calibration field, $\hat{n}_{ij}$. Note that the r.h.s.~of the above equation is the value of the estimated mean redshift of galaxies of color $i$ regardless of whether or not $j$ is measured in the calibration field, due to the linearity of the expression. The corresponding uncertainty is
\begin{equation}
\left(\sigma^{\langle z \rangle}_{i}\right)^2 = \frac{\sum_j n^2_{ij} \left[(\sigma^z_{\mathrm{CV}, ij})^2 + (\sigma^z_{ij})^2/N_{ij}\right] + \left[\sigma^{\hat{n}}_{ij} (z_{ij}-z_{i})\right]^2}{(\sum_j n_{ij})^2} \; .
\end{equation}
Note the additional term in the nominator, which is due to uncertainty $\sigma^{\hat{n}}_{ij}$ in the number density of galaxies of color $ij$ in the limited volume of the calibration field(s).

The relative importance of the three effects discussed here depends on the survey design. \citet{Bordoloi2010} explore the trade-off between $(\sigma^z_{\mathrm{CV}, ij})^2$ and $(\sigma^z_{ij})^2/N_{ij}$, finding that a hypothetical fully-sampled redshift survey down to faint magnitudes with moderately broad redshift bins would be limited by the former CV term, not by the latter shot noise, and thus that observing a subset of galaxies spread out over a larger area is beneficial. \citet[][their Table 2]{Hoyle2018} confirm this is indeed so for the COSMOS-based calibration of DES Year 1 photometric redshifts, with about $7\times10^{-3}$ and $2\times10^{-3}$ uncertainty in mean redshift due to $(\sigma^z_{\mathrm{CV}, ij})^2$ and $(\sigma^z_{ij})^2/N_{ij}$, respectively. \citet[][their appendix~B]{Gruen2017} show that sample variance can change by an order of magnitude with the same calibration fields, depending on the number of photometric bands used for reweighting. Among current surveys the KiDS-VIKING $ugriZYJK_s$ data most effectively allows to utilize this effect. Re-weighting calibration samples over a high-dimensional color space greatly reduces sample variance and other uncertainties \citep{Hildebrandt2019,Wright2020}. Surveys for which fewer bands are measured over the wide field can still benefit from multi-band photometric deep fields in addition to spectroscopic calibration samples \citep{Buchs2019,Myles2021}. \citet{Buchs2019} show that with $ugrizYJK_s$ photometry, the COSMOS field as a source of redshift calibration is sufficient in terms of galaxy number, but not fully sufficient in terms of volume, to reach total calibration uncertainties of $10^{-3}$ in mean redshift, if larger fields can be used to estimate the density of galaxies in that color space, $n_{ij}$. The limited area over which deep optical and near-infrared photometry is available has a dominant effect on the resulting sample variance.

\ifarxiv
\begin{marginnote}
\entry{Take-away}{Deep spectroscopic and many-band photometric surveys cover only limited areas of sky; as a result, sample variance in redshift distributions is large and can limit direct redshift distribution characterization.}
\end{marginnote}
\fi

\subsubsection{Biases from the Selection of Photometric Samples}

\label{sec:photo_selection}

In parallel to the selection biases in spectroscopic samples discussed in \S \ref{sec:incompleteness}, the samples of galaxies whose redshifts are to be estimated commonly are subject to selections that can bias the estimated redshift distribution at a given point in color space if not accounted for.

Already the presence of Poissonian photometric noise, particularly when noise levels differ between training sample and wide-field photometry, can have a significant impact when methods are not designed to account for it \citep[e.g.,][their table 3]{Wright2020}. The selection of galaxy samples is usually much more complex in reality due to the non-linear interplay between pixel-level noise, detection, star-galaxy separation, model-fitting photometry, and shape measurement for barely resolved galaxies. If these effects are not modeled -- e.g., if using magnitude limited spectroscopic redshift catalogs without further selection -- they introduce biases in redshift characterization of order 0.01 \citep{Gruen2017}. By measuring the detection probability and transfer function of galaxies with image injection \citep{Huang2017,Everett2021}, this bias can be greatly reduced, potentially to levels compatible with Stage IV requirements \citep{Myles2021}.

\ifarxiv
\begin{marginnote}
\entry{Impact of selection in lensing sources}{\includegraphics[width=3.5cm]{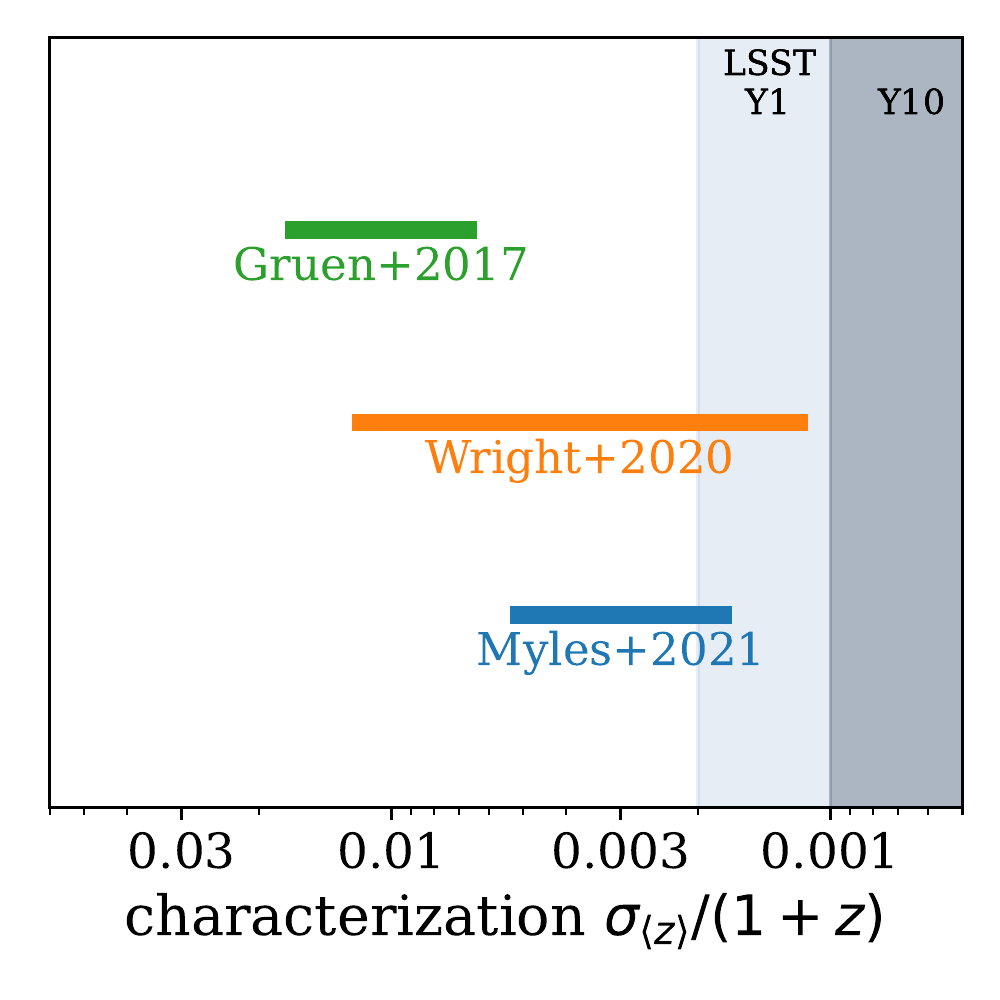}}
\end{marginnote}
\fi

\ifarxiv
\begin{marginnote}
\entry{Take-away}{Selections affecting photometric samples must be taken into account when characterizing photo-$z$'s.}
\end{marginnote}
\fi

\subsubsection{Blending} 

\label{sec:blending}

Multiple galaxies can overlap on a sky image or contribute light to the same fiber or slit of a spectrograph, both because galaxies are intrinsically extended objects and due to blurring by the point-spread function. This blending of light occurs most easily for bright galaxies with intrinsically large angular size, but in those cases it is rarely problematic due to the relative faintness of any overlapping objects. However, it is sufficiently common and severe even for fainter objects for its impact to pose challenges.  For instance, for more than half of the galaxies detected by Rubin Observatory, overlapping galaxies contribute at least 1\% of the total flux within their pixels (\citealt{Sanchez2021}; see also Figure 2 of \citealt{Melchior2021}).

Blending will impact flux and color measurements for some objects within the samples of galaxies with spectroscopy used to train photometric redshift algorithms and characterize redshift distributions. This effect can account for roughly 20\% of persistent photometric redshift outliers in samples with deep many-band photometry \citep{Masters2019}. Worse, when a faint emission line galaxy is blended with another object, the resulting strong line features can dominate the determination of the spectroscopic redshift, even when the broadband colors are primarily determined by another object. 

This occurs in 1\% to 5\% of cases for faint sources \citep{Newman2013,Brinchmann2017,Masters2019}. At $z>1$, $>5\%$ of objects targeted in the DEEP2 Galaxy Redshift survey had multiple counterparts in Hubble Space Telescope imaging within 0.75 arcsec of their nominal position \citep{Newman2013}; at sufficiently small separations blends will not be detectable even from space. 

This will set a floor level of systematic uncertainty in empirically estimated photometric redshifts unless such objects can be excluded with high confidence, e.g., through space-based observations or (in some cases) by visual inspection and re-analysis of spectroscopic data.  Similarly, unless blends can be excluded entirely, blending will limit the performance of photometric redshift estimates for individual objects, even with template-based methods, as the photometry ascribed to an object may not correspond only to its intrinsic properties but rather be altered by contributions from objects at very different $z$.

Blending will also affect cosmological studies by altering the effective $z$ distributions of the redshift bins used in an analysis.  For instance, in the case of tomographic weak lensing, The measured shape of a dominant galaxy at redshift $z_0$ may be affected when a fainter, blended galaxy at $z_1$ is being gravitationally sheared. The observed signal is therefore due to a superposition of lensing of light at $z_0$ and $z_1$, much as if one were looking at a bin of galaxies that contained objects at both redshifts. This effect can be correctly described by modifying the $n(z)$ of each redshift bin accordingly. \citet{MacCrann2021} first formulated this effect and demonstrated it in image simulations, finding impacts $0.003<|\delta z|<0.012$ in the effective mean redshift for samples from DES three-year data.  
The solution to this issue will require extended work on realistic lensing image simulations.

\ifarxiv
\begin{marginnote}
\entry{Impact of blending}{\includegraphics[width=3.5cm]{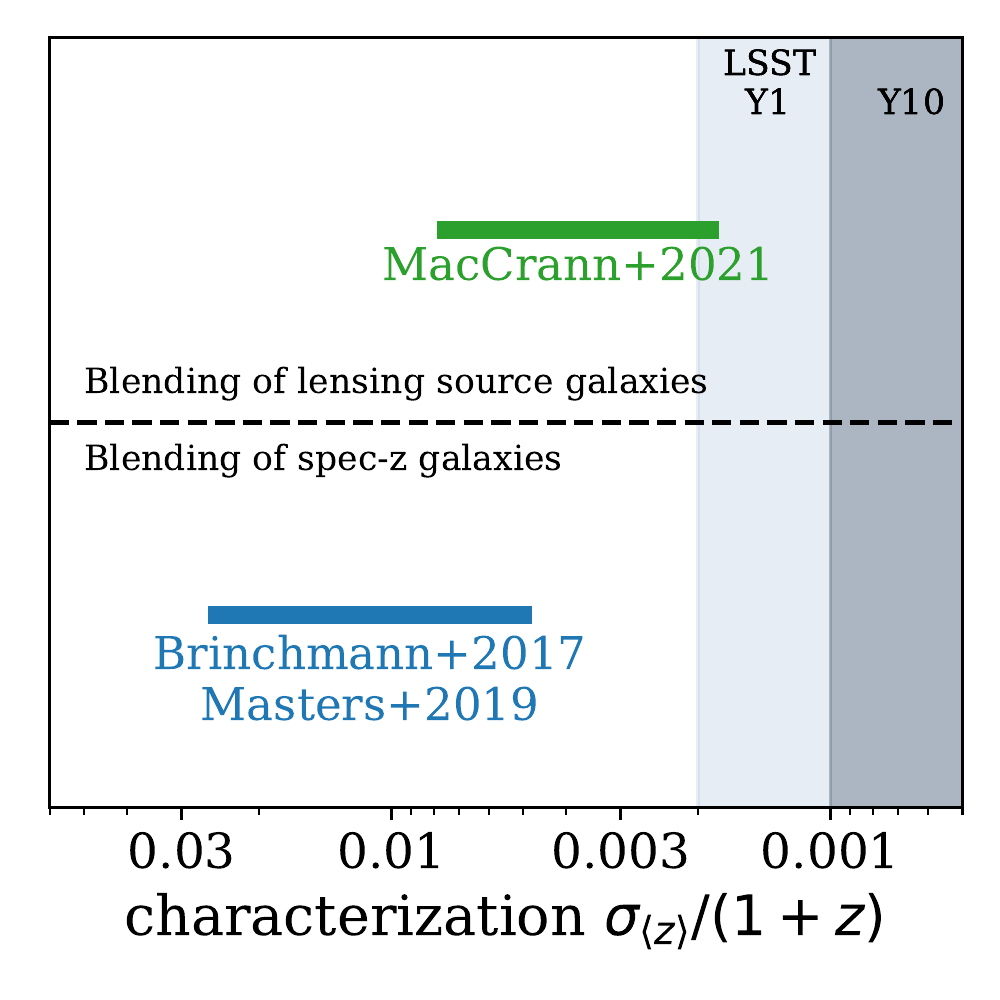}}
\end{marginnote}
\fi

\ifarxiv
\begin{marginnote}
\entry{Take-away}{Undetected blends between galaxies whose images overlap are unavoidable in ground-based data, but make interpretation of photo-$z$'s and spectroscopic samples more difficult.}
\end{marginnote}

\fi

\subsubsection{Astrophysical Systematics on the Cross-correlation Characterization of Redshifts}

\label{sec:xcorr}

In the linear limit of large-scale-structure, the angular cross-correlation $w_{\rm phot,spec}(\theta)$ between objects in a photometric sample and spectroscopic objects of known redshift is proportional to the product of the large-scale-structure bias of each sample (which we will refer to as $b_{\rm phot}$ and $b_{\rm spec}$) and the probability that an object in the photometric sample is at the spectroscopic $z$, $p(z)$: i.e., $w_{\rm phot,spec}(\theta) \propto b_{\rm phot} b_{\rm spec} p(z)$.  These cross-correlations therefore can be exploited to reconstruct the redshift distribution of any photometric sample \citep{Newman2008}, in concert with measurements of the autocorrelation of each sample. The resulting redshift distributions are sometimes referred to as ``clustering redshifts'' \citep{Rahman2015}. 

The cross-correlation method has the great advantage of providing accurate characterization of redshift distributions even if spectroscopic samples are systematically highly incomplete, as has been true of all deep surveys to date (cf. \S \ref{sec:incompleteness}).  In fact, wide-area, shallow surveys of the distant universe such as those that DESI and 4MOST will provide \citep{DESI2016,4MOST2019} will yield much lower errors on cross-correlation measurements than deep, small-area surveys \citep{Matthews2010,Newman2015}.  This makes such methods a promising route for characterizing redshift distributions for samples of faint objects for which it is difficult to obtain statistically complete spectroscopy.

A variant of this idea is to measure the distribution of redshift differences, $|\Delta z|$, between pairs of objects which are near to each other on the sky \citep{Quadri2010,Huang2013}. That distribution will exhibit a peak near zero offsets, the width of which is set by the convolution of the errors on independent objects' photometric redshifts.  This method can accurately characterize the average scale of the core of redshift errors, but does not allow reconstruction of catastrophic error rates or the shape of the redshift distribution of outliers due to the dependence of the measured $|\Delta z|$ distribution on both the strength of galaxy correlations and any evolution of the large-scale-structure bias with redshift.  It is therefore not suitable for the high-precision characterization required for cosmological measurements.

In principle, cross-correlations with upcoming wide-field spectroscopic samples will contain sufficient \emph{information} to reconstruct redshift distributions at the accuracy needed for the next generation of cosmological measurements \citep{Newman2008,Matthews2010}.  However, it has not yet been demonstrated that this can actually be achieved in the presence of astrophysical \emph{systematics}. A number of areas of concern exist which will need to be addressed for these methods to reach the extreme accuracy needed for future imaging experiments \citep{DESC_SRD}.

\begin{figure}[t]
\centering
\includegraphics[width=0.55\textwidth]{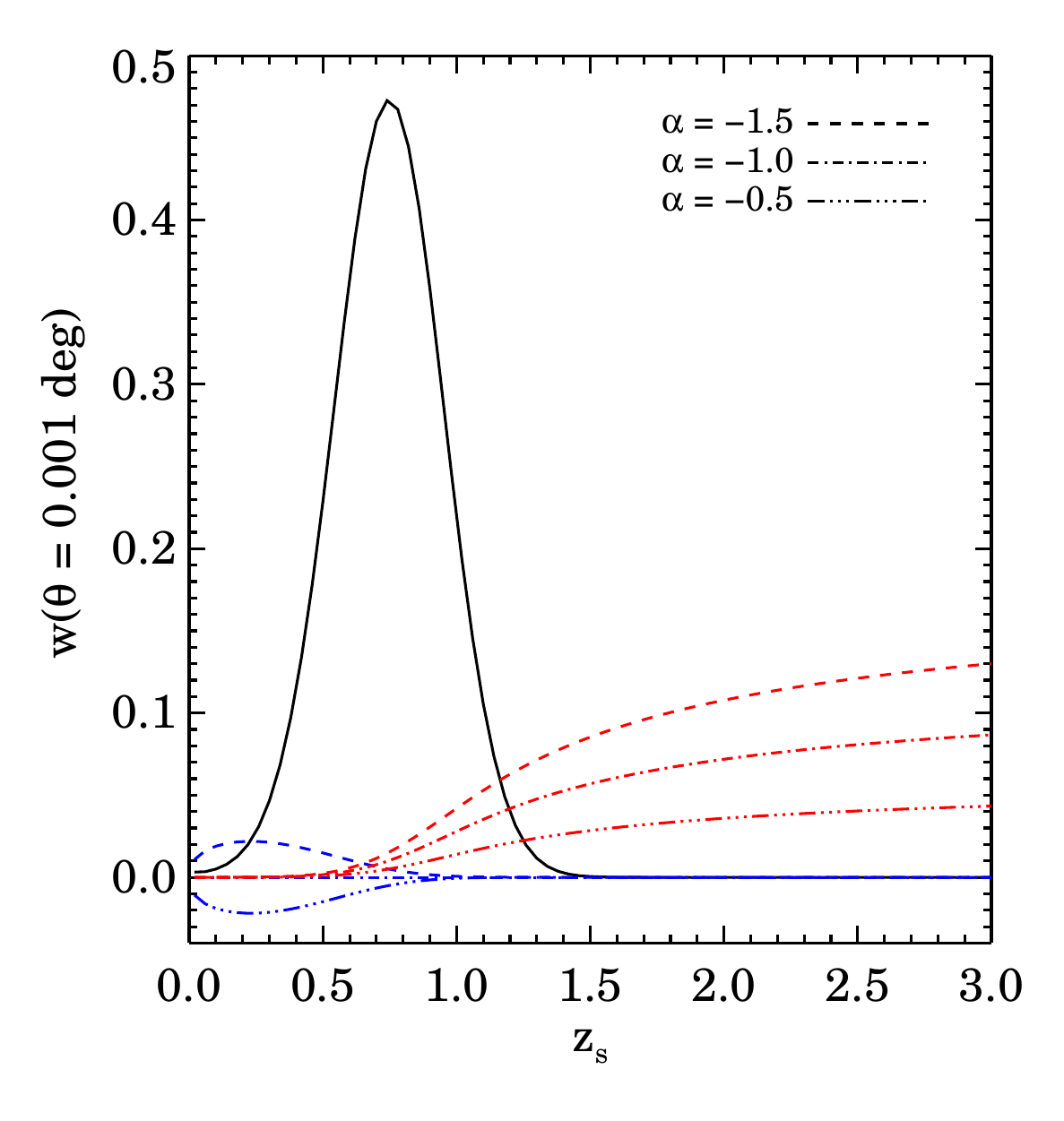}
\caption{An illustration of the impact of weak lensing magnification on cross-correlation measurements that could be used to reconstruct redshift distributions. The black line shows the expected cross-correlation signal, $w_{sp}$, between a spectroscopic sample at redshift $z = z_s$ with a photometric sample having a Gaussian distribution in redshift, plotted as a function of $z_s$ (black line); it is this signal that can be used to reconstruct redshift distributions, for a simple toy model scenario (detailed in \citealt{Matthews2014}).  The cross-correlation signals that result from objects in the photometric sample being lensed by objects in a given spectroscopic redshift bin are shown as blue lines corresponding to different values of the faint-end Schechter function slope $\alpha$.  This signal is comparatively weak, as the lensing effect is null for the typical faint-end slope of $\alpha = -1$. Finally, red curves show the signal resulting from comparatively bright spectroscopic objects being lensed by the photometric objects; this effect is much stronger as the density of bright galaxies drops exponentially with increasing luminosity, making counts at the bright end sensitive to lensing contamination. Figure reproduced with permission from \cite{Matthews2014}.}
\label{fig:magnification}
\end{figure}

One issue that will need to be addressed is magnification of background galaxies due to gravitational lensing \citep{Newman2008,Matthews2014}.  This magnification will cause an excess density of galaxies in regions around foreground objects.  The cross-correlation signal depends on the derivative of the luminosity function of the objects being magnified; lensing has a null effect for a power law slope of -1, as is typical for faint galaxies.  As a result, the  cross-correlation signal from magnification is primarily driven by lensing of intrinsically luminous \textit{spectroscopic} objects by fainter photometric objects in the foreground, rather than lensing of photometric objects by galaxies and quasars with spectroscopic redshift measurements \citep{Moessner1998,Newman2008}, as may be seen in \autoref{fig:magnification}.  There are some hints of this signal in clustering redshift measurements in the literature \citep[e.g.,][]{Hildebrandt2021}, but this will be a much greater issue at the precision required for future experiments.

The strength of this signal should be predictable, given  estimates of the redshift distribution of the lensing objects and the mass associated with them, combined with the luminosity function of the lensed objects.  For the dominant case of lensing by objects in the photometric sample, for instance, the lensing-contaminated redshift distribution inferred from cross correlation measurements can provide an initial guess for $p(z)$, and CMB lensing measurements of the mass associated with that sample can provide the additional information required without imposing a dependence on the optical/IR weak lensing measurements for which we need to characterize the redshift distributions.  \citet{Newman2008} has suggested combining these measurements with the luminosity function of the spectroscopic sample, which would enable the lensing magnification signal to be predicted and  corrected for; this procedure could be iterated until convergence.  However, there has not yet been a demonstration that this method can deliver redshift distributions with the exquisite accuracy required for future cosmological experiments.

\ifarxiv
\begin{marginnote}
\entry{Impact of astrophysical systematics on clustering redshifts}{\includegraphics[width=3.5cm]{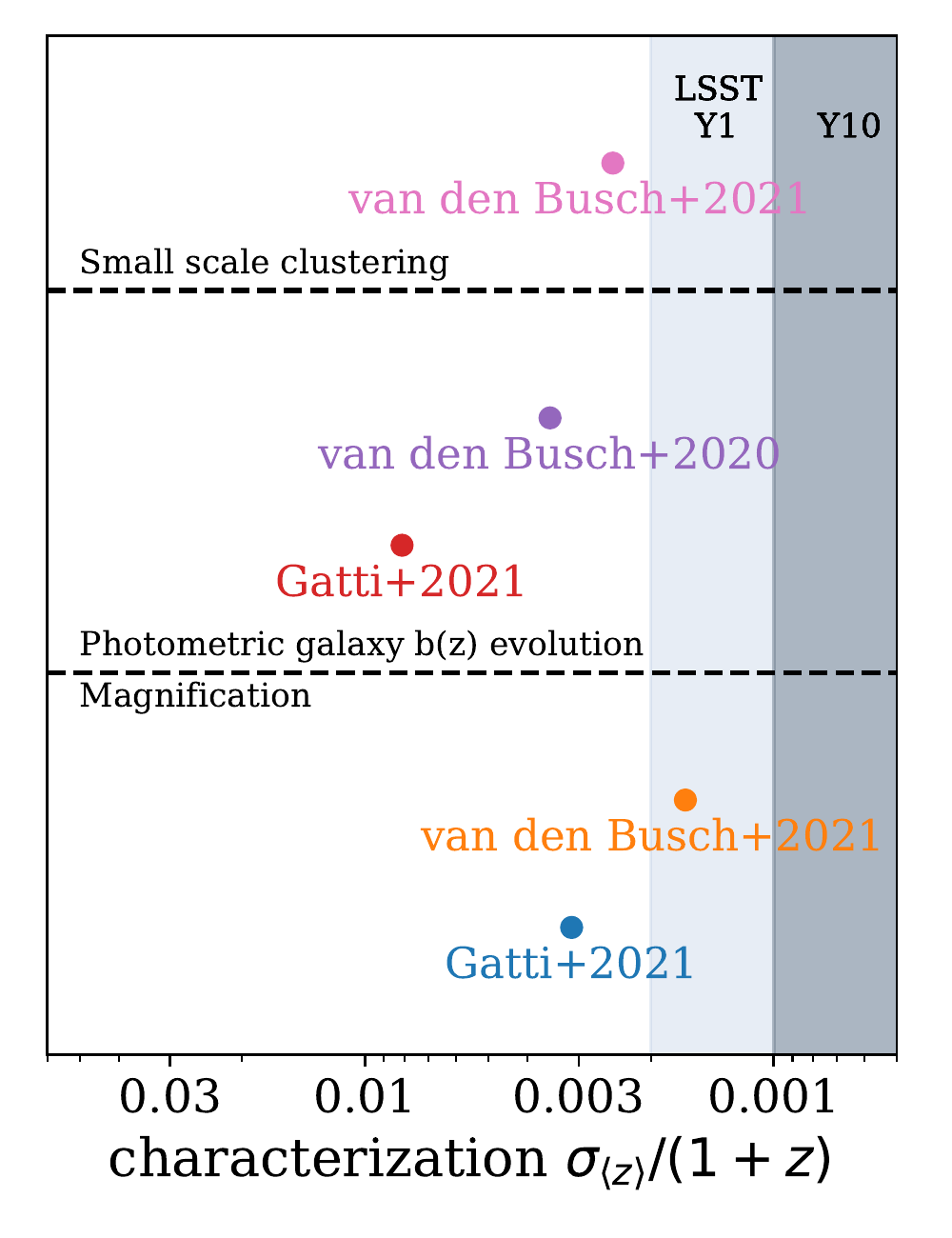}}
\end{marginnote}
\fi
A second area where investigation is needed is tests of whether systematics can be avoided if cross-correlation information from pairs at small separations ($\lesssim 5$ Mpc) is utilized to constrain redshift distributions, and if so, how best to do so.
The signal-to-noise ratio of cross-correlation measurements is maximized at small separations \citep{Newman2008}.  As a result, many applications of clustering redshifts to date have not excluded information from close pairs in order to have the strongest possible constraints on redshift distributions \citep[e.g.,][]{Schmidt2013,Rahman2015}.  However, on those scales structure evolves non-linearly, and the relationship between the clustering of galaxies and that of dark matter, or of one set of galaxies with another, can be complex. In contrast, at the large separations which more theoretical work has focused on \citep[e.g.,][]{Newman2008, McQuinn2013}, the relationship between the clustering of dark matter and that of galaxies is much simpler, and the assumption of linear biasing holds, so that the correlation function of galaxies, $\xi_{gg}$ is simply $b^2 \xi_{mm}$, where $b$ is a constant for a particular galaxy population and $\xi_{mm}$ is the two-point correlation function of matter.  In that case, the intrinsic cross-correlation between the photometric and spectroscopic populations can be treated simply, with the cross-correlation bias determined as $b_{\rm phot,spec}^2 = b_{\rm spec} b_{\rm phot}$; at smaller scales, the relationship can be much more complicated.  

Performing cross-correlation measurements at smaller scales has been invaluable for detecting the signal with currently-available samples; however, the simple analysis methods which can be utilized at large scales may have systematics at unacceptable levels for future experiments when applied for smaller separations.  One example of a cause for concern is the quasar proximity effect \citep{Efstathiou1992}: the large amount of ionizing radiation released by quasars may affect the evolution of galaxies in some volume around each one.  As a result, cross-correlation measurements that use quasars as a spectroscopic tracer of large-scale-structure could yield incorrect results if the scales where pairs between photometric objects and quasars are measured are those where galaxies have been prevented from developing by the proximity effect.  This would not be an issue at larger separations, where the clustering between quasars and photometric objects is driven purely by the underlying web of dark matter.

Given these concerns, it is important to test how well clustering redshifts can perform when small-separation pairs are exploited, in order to determine whether they can be used in a simple manner to characterize redshift distributions for next-generation experiments.  If not, more complicated modeling of the relationship between both photometric and spectroscopic galaxies and dark matter halos will be necessary for accurate characterization to be possible.

A third area where work remains to be done is the handling of redshift evolution of the clustering bias of the photometric galaxy sample. This includes the extreme case where photometric redshift outliers have very different bias from more typical objects, but also the more typical type-redshift degeneracies common with few-band data, which leads to cases where two populations of galaxies with different bias and redshift are indistinguishable photometrically \citep[e.g.,][]{Dunlop2007}. Since the cross-correlation is proportional to the product of the bias and the fraction of objects at a given redshift, if differences in bias are not handled correctly, the redshift distribution inferred will be incorrect. Recent studies show the impact of bias evolution with current methods to be of order $|\Delta z|\approx0.01$ \citep{vandenBusch2020,Gatti2021}.

However, it could be possible to address this issue by exploiting the astrophysical relationships between galaxy color and luminosity and the large-scale structure bias: both locally and at $z\sim 1$, the clustering strength of galaxies is primarily a monotonic function of color, with redder galaxies clustering more strongly, and a weaker function of luminosity \citep{Hogg2003,Cooper2006}.  Although it may be difficult to determine whether an individual galaxy is at either one of two possible redshifts, it is straightforward to determine the restframe color and luminosity of a galaxy, and hence bias as predicted from analyses of broader populations, \textit{conditioned on} the redshift; that is, we can accurately determine $p(C_{\rm restframe}, L | P_{\rm observed}, z)$, where $C_{\rm restframe}$ is some restframe color, $L$ is luminosity, and $P_{\rm observed}$ is the measured photometry for an object.  If we characterize the dependence of bias on restframe properties and redshift, $b(C_{\rm restframe}, L, z)$, either via auto- or cross-correlation analyses, then we have sufficient information to predict $b(z)$ for a sample.  Again, this is a method that should in principle be effective, given the simple behavior of galaxy biasing; however, more work is needed to demonstrate that the level of characterization needed for future surveys can be achieved.

\ifarxiv
\begin{marginnote}
\entry{Take-away}{Cross-correlations between photometric and spectroscopic samples have the potential to characterize redshift distributions even if deep spectroscopy remains systematically incomplete, but a number of astrophysical systematics could limit their power.}
\end{marginnote}
\fi

\section{A VISION FOR THE FUTURE OF PHOTOMETRIC REDSHIFTS}

\label{sec:vision}

\begin{figure}[htp!]
{\includegraphics[width=0.95\textwidth]{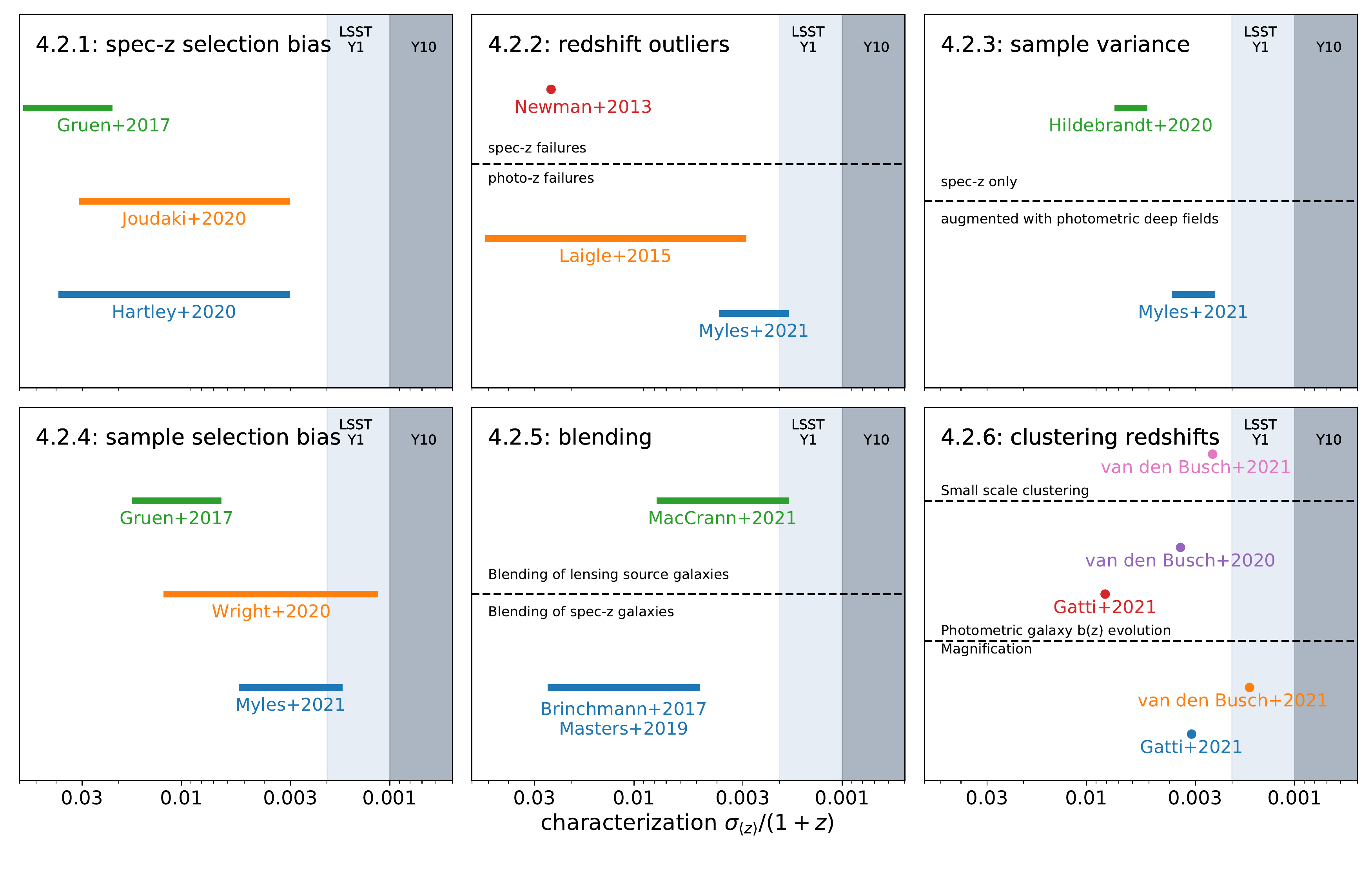}}
{\caption{{Recent estimates of the impact of challenges on the characterization of state-of-the-art photometric redshifts, summarized as the resulting systematic uncertainty of the estimated mean redshift of ensembles of galaxies. Horizontal bars indicate the ranges of impacts consistent with a given study; points correspond to the best estimate of an impact. Grey vertical bands indicate the requirements for cosmological analyses with Year 1 or Year 10 Vera Rubin Observatory LSST data \citep{DESC_SRD}.  At top left we show the impact of biases from the selection of spectroscopic redshift samples \citep{Gruen2017, Joudaki2020, Hartley2020}; cf. \S \ref{sec:incompleteness}.  At top middle we show the impact of incorrect redshifts in spectroscopic or many-band photometric training sets \citep{Newman2013,Laigle2016,Myles2021}, as described in \S \ref{sec:outliers}.  At top right is shown the impact of sample variance due to the limited volume of spectroscopic surveys \citep{Hildebrandt2020, Myles2021}; cf. \S \ref{sec:sample_variance}.  At bottom left is shown the impact of biases in the selection of objects within photometric sample bins, as discussed in \S \ref{sec:photo_selection} \citep{Gruen2017,Wright2020,Myles2021}.  In the bottom middle panel we show the impact of blending between multiple objects, either through its effect on the lensed galaxies \citep{MacCrann2021} or on the spectroscopic samples used for characterization \citep{Brinchmann2017, Masters2019}; cf. \S \ref{sec:blending}.  At bottom right we show the estimated impact of a variety of astrophysical uncertainties on the characterization of redshift distributions via cross-correlations, also known as clustering redshifts \citep{vandenBusch2020,Gatti2021}, as described in \S \ref{sec:xcorr}. Here we consider only sources of uncertainty in the first moment of the redshift distribution, as those are best-studied to date; however, higher moments will need to be characterized to similarly stringent levels \citep{DESC_SRD}. 
\label{fig:char_error}
}
}
}
\vskip-0.15in
\end{figure}

In the previous section, we have considered challenges that may affect applications of photometric redshifts to near-future projects.  Figure \ref{fig:char_error} summarizes the current state of the art on those challenges which will impact the characterization of redshift distributions, a major source of systematic uncertainty for dark energy experiments.  Considerable progress will need to be made in many areas for future experiments to reach their full potential.

In this section, we conclude by evaluating how we might expect photometric algorithms to develop by the time of the final analyses of the next generation of imaging surveys, in the mid-2030s, to help address these challenges.  At that time, we might hope that our understanding of the physics of galaxy evolution has become developed enough that photometric redshift systematics will be inseparably linked to that science, not just to cosmology.  We can also expect that some progress will be made on spectroscopic training samples for photo-$z$ analyses, but that completeness and sample sizes will remain limited; the sorts of large-aperture ($\gtrsim 8$m), highly-multiplexed, wide-field-of-view capabilities that are optimal for photometric redshift training samples are unlikely to be available much before that time \citep{Newman2019}.

\subsection{Potential Developments in Photometric Redshift Algorithms}

\label{sec:pz_vision}

The issues of spectroscopic incompleteness, outliers and biases in calibration redshifts, and sample variance in calibration redshifts discussed in \S \ref{sec:characterization} limit traditional machine-learning photometric redshift approaches at levels significantly exceeding the requirements of Stage IV surveys. At the same time, template-based photometric redshift methods have delivered poorer performance, and often poorer characterization, than machine learning methods within the range of coverage of training sets of spectroscopic redshifts.  Ultimately, we might hope to unify the advantages of both techniques.

A model for the ensemble of rest-frame galaxy SEDs and luminosity functions, as well as the redshift evolution thereof, would allow more meaningful interpolation between the sparse sampling of galaxies targeted by deep spectroscopic surveys in the presence of sample variance. Such a model could also be used to meaningfully extrapolate beyond the limits of the training data, {e.g.,~to emulate intrinsically fainter galaxies}. At the same time, at present no model derived from \emph{a priori} principles has achieved sufficient fidelity; improvements both to stellar population synthesis models and to our understanding of the underlying galaxy population are needed. Rather, there will have to be an interplay of empirical modeling based on artificial intelligence with physics-driven components, incorporating the notion of a rest-frame SED, parameterizations for spectral features that vary from galaxy to galaxy as well as for the characteristics of the underlying galaxy population, flexible models of reddening, and the incorporation of the characteristics of the instrument used for observations. 

Such a model will have to be refined using all available data. In this, spectroscopy and deep multi-band photometry will continue to play decisive roles due to their ability to break degeneracies that are irreducible in shallow broad-band photometry. Narrow-band photometry, if it can be obtained with much greater depths and areas than currently available, could be an important, highly constraining element. The large impact of selection biases both in spectroscopic and in photometric galaxy samples, and their complex and non-linear dependence on observing conditions and analysis algorithms is an additional concern. It can be overcome only if such a scheme includes careful simulation or emulation of observations for interpreting the spectroscopic and photometric data. 

In principle such a model could be used to generate representative artificial training sets of arbitrary size and full redshift coverage as input for machine learning algorithms. Ultimately, it is more desirable that all data be interpreted simultaneously to characterize the likelihood distribution $p({\rm data} | {\rm model})$ -- i.e., the probability of the full set of photometric and spectroscopic observations obtained as a function of the parameters of the underlying model for the galaxy population and SEDs. The most universally useful outputs from this process would be a set of samples from the probability distribution of the parameters of the model (as would be obtained from probabilistic inference methods such as Markov Chain Monte Carlo sampling). Each such sample would then be associated with a set of posterior probability distributions for the redshift of each individual object and/or for ensembles of galaxies; by averaging the posteriors corresponding to different samples, one can obtain the posterior PDF marginalizing over all parameters of the model.

\ifarxiv
\begin{marginnote}
\entry{Take-away}{Flexible, observationally-constrained models of the underlying population of galaxy spectral energy distributions could enable significant improvements in photometric redshift performance and characterization.}
\end{marginnote}
\fi

\subsection{Potential Developments in Spectroscopic Training} 

\label{sec:spec_vision}

As discussed in Sections \ref{sec:incompleteness} and \ref{sec:outliers}, sets of objects with spectroscopic redshifts at the full depth of future imaging surveys from the Rubin Observatory, Euclid, and Roman Space Telescope will be valuable for optimizing the performance of photo-$z$ algorithms and, if the impact of systematic incompleteness and outliers can be avoided, could potentially provide the exquisite characterization needed for precision cosmological measurements. If incompleteness and outlier rates are improved but do not reach the level required for direct characterization, simultaneous forward-modeling of photometric and spectroscopic survey data (e.g., in extension of \citealt{Fagioli2020}) may provide a path forward.  However, such improved samples would still be highly expensive to obtain with instruments that currently exist or are in construction.  It will thus be important both to develop new spectroscopic resources as well as to optimize how we will use existing ones.  

New telescopes and instruments optimized for wide-field, highly-multiplexed spectroscopy on $\gtrsim 8$m diameter apertures can vastly reduce the time required to obtain photometric redshift training samples.  \citet{Newman2015} defined a fiducial photometric redshift spectroscopy sample consisting of 30,000 objects down to magnitude $i=25.3$ distributed over 15 fields of at least 20 arcminute diameter (in order to mitigate and characterize the effects of sample/cosmic variance), with sufficient depth per pointing to obtain redshifts for at least 75\% of targets, and sufficient spectral resolution to split the [OII] 3727 Angstrom doublet at $z \gtrsim 0.7$, where other emission-line spectral features are redshifted beyond optical wavelengths.  \citealt{Newman2019} found that the most efficient currently-available options, the DESI instrument at the Mayall Telescope \citep{DESIInstrument} and the DEIMOS spectrograph at Keck Observatory \citep{Faber2003}, would take more than 1800 dark nights to conduct this survey. In contrast, the upcoming Subaru/PFS \citep{Tamura2016} would take roughly 400 dark nights, while the proposed Maunakea Spectroscopic Explorer \citep{MSE2018}, ESO SpecTel project \citep{Spectel2019}, or the MANIFEST fiber-feed for the GMACS instrument on the Giant Magellan Telescope \citep{Lawrence2020} could potentially complete this fiducial survey in less than 200.

Given the large investment required for spectroscopic surveys in support of photometric redshift measurements, it would be desirable to optimize our use of such datasets in order both to minimize the resources needed and maximize the impact of the data that is obtained. Current photo-$z$ training surveys are already utilizing self-organizing maps to identify regions of parameter space where additional spectroscopy is needed \citep[e.g.,][]{Masters2019}.  Future datasets could take advantage of developments in active learning algorithms, which in this application would identify those objects for which obtaining a spectroscopic redshift measurement would most improve the model \citep[e.g.,][]{Vilalta2017}.

Another area of potential gain is better integration of spectroscopic and many-band photometric surveys (or low-resolution grism surveys, which have similar properties).  Obtaining photometry in large numbers of bands (as in COSMOS, \citealt{Laigle2016}), in narrow bands (as in PAU, \citealt{Alarcon2021}), or alternatively obtaining low-resolution spectroscopy (as used by PRIMUS, \citealt{Coil2011} or slitless spectroscopy (as in 3D-HST, \citealt{Momcheva2016}) can provide redshift estimates for complete samples of galaxies, with larger uncertainties and catastrophic error rates than spectroscopic surveys, but lower errors than broadband photo-$z$'s. 
At fixed telescope etendue, survey time, and number of spectral features resolved, the error on redshifts will be proportional to $\frac{1}{\sqrt{R}}$, where the spectral resolution $R \equiv \frac{\lambda}{\Delta \lambda}$.  Given this scaling,
it is rarely feasible to greatly reduce redshift errors over a full imaging survey area down to the depths reached by broadband imaging, but deep campaigns covering tens or hundreds of square degrees to useful depths could be possible. Such multi-band data can boost the benefits gained from spectroscopic samples. To achieve this, full integration of the information gained from higher-resolution spectroscopic surveys and many-band photometric ones would be desirable (some steps toward that have been taken; e.g., in \citealt{Buchs2019}).

The near-infrared grism spectroscopy capabilities of the Euclid mission and the Roman Space Telescope present an appealing opportunity here.  Low-resolution spectroscopy at IR wavelengths is difficult from the ground due to high sky backgrounds, while the space-based missions have only limited optical capabilities. Deep many-band optical imaging from the ground over the same areas covered deeply by space-based grisms would provide detailed SEDs over a broad wavelength range, yielding a rich dataset for photometric redshift training and characterization as well as for galaxy evolution studies.

Deep multi-wavelength photometry can also help by breaking degeneracies between different redshift solutions that few-band optical colors cannot.  We might hope to gain a deeper understanding of the range of galaxy SEDs at somewhat brighter magnitudes via spectroscopy, and extend that information to fainter galaxies via many-band and multi-wavelength photometry, greatly expanding the training sets that may be used for photometric redshifts (similar to suggestions in \citealt{LSSTScienceBook}).  Ideally, this would again be used to help move us towards a full phenomenological model of galaxy spectral evolution.

If we can mitigate the impact of sample/cosmic variance on photometric redshift characterization, this would further reduce the scale of the training sets required and make wider-field spectrographs (e.g., MegaMapper, \citealt{Schlegel2019}) more useful for this work.  The intermediate-precision redshifts provided by many-band and multiwavelength surveys could play a role here, allowing the large-scale-structure fluctuations in the regions used for deep spectroscopic surveys to be measured and characterized.  By surveying wider areas, such surveys could also enable rare populations to be identified at many different redshifts, rather than just those associated with the highest over- or under-densities in small spectroscopic survey fields.  It would be even better to have a model for galaxy SEDs that can be tuned with the objects in spectroscopic fields, but continuously predicts colors at arbitrary redshift; again, the closer we can come to the ideal of a complete phenomenological model of galaxy spectral evolution (or, if one were even more ambitious a full physical model), the better our understanding of photometric redshifts will be.

\ifarxiv
\begin{marginnote}
\entry{Take-away}{In order to obtain optimal photometric redshift performance and characterization, large investments of observing time on wide-field spectrographs on large telescopes are needed, supported by deep narrow-band and multiwavelength imaging and shallower, wide-area spectroscopy.}
\end{marginnote}
\fi

\vspace{5pt}

Of course, the future of photometric redshifts could lie in different directions than we have speculated here. It is clear, however, that several independent effects are currently limiting their performance and characterization at levels that will be prohibitive for the advancement of the field that the data collection of Stage IV galaxy surveys in principle allows. There are thus many broad avenues of research -- both in areas discussed above and beyond -- that, if successful, will allow photometric redshift methods to be improved far beyond the current state of the art. If a substantial investment in both data-taking and method development is made, it is likely to pay off: photometric redshifts will be a critical tool for studies of both galaxy evolution and cosmology with the next generation of imaging surveys, and should continue to be valuable for extragalactic science for the foreseeable future.

\ifarxiv
    \vskip 0.01in
\fi

\ifarxiv
    \vskip 0.01in
\fi

\section*{DISCLOSURE STATEMENT}
The authors are not aware of any affiliations, memberships, funding, or financial holdings that
might be perceived as affecting the objectivity of this review. 

\section*{ACKNOWLEDGMENTS}
We gratefully acknowledge Paulina Contreras, Biprateep Dey, Moritz Gammel, Elisa Legnani, Alex Malz, Yao-Yuan Mao, Justin Myles, Markus Rau, Samuel Schmidt, and Luca Tortorelli for a variety of helpful conversations and for their reviews of versions of this article. We thank Biprateep Dey, Alex Malz, and Yao-Yuan Mao for providing figures for use in this work.  We also gratefully acknowledge many helpful interactions with members of the LSST Dark Energy Science Collaboration Photometric Redshifts Working Group. We also wish to thank Sandra Faber for her continued help, support, and feedback throughout the development of this review, as well as Robert Kennicutt for his assistance with the editorial process in the later stages of its development. This work was supported by grants DE-SC0007914 and DE-SC0020256 from the United States Department of Energy, Office of Science, Office of High Energy Physics, without which it would not have been possible.


\bibliography{literature}

\end{document}